\pdfoutput=1
\documentclass[11pt,twoside,a4paper,cmspaper,final,collab]{cms-tdr}
\def\svnVersion{dadfb02}\def\svnDate{2025/11/26}\def\cmsCernNoTag{CERN-EP-2025-124}\def\cmsCernDate{\today}\def\cmsMessage{Published in Reports on Progress in Physics as \href{http://dx.doi.org/10.1088/1361-6633/ae2207}{\doi{10.1088/1361-6633/ae2207}.}}
\begin{document}\cmsNoteHeader{HIG-22-013}

\ifthenelse{\boolean{cms@external}}{%
    \providecommand{\cmsLeft}{upper\xspace}
    \providecommand{\cmsRight}{lower\xspace}
}{%
    \providecommand{\cmsLeft}{left\xspace}
    \providecommand{\cmsRight}{right\xspace}
}

\newcommand{\PQj}{{\HepParticle{j}{}{}}\xspace}

\newcommand{\ttej}{\ensuremath{\Pe\PQj}\xspace}
\newcommand{\ttmj}{\ensuremath{\PGm\PQj}\xspace}
\newcommand{\ttltj}{\ensuremath{\Pell+3\PQj}\xspace}
\newcommand{\ttlfj}{\ensuremath{\Pell+{\geq}4\PQj}\xspace}
\newcommand{\ttlj}{\ensuremath{\Pell\PQj}\xspace}

\newcommand{\ttee}{\ensuremath{\Pe\Pe}\xspace}
\newcommand{\ttem}{\ensuremath{\Pe\PGm}\xspace}
\newcommand{\ttmm}{\ensuremath{\PGm\PGm}\xspace}
\newcommand{\ttll}{\ensuremath{\Pell\Pell}\xspace}

\newcommand{\mtop}{\ensuremath{m_{\PQt}}\xspace}
\newcommand{\matop}{\ensuremath{m_{\PAQt}}\xspace}
\newcommand{\pttop}{\ensuremath{p_{\mathrm{T},\PQt}}\xspace}
\newcommand{\ptatop}{\ensuremath{p_{\mathrm{T},\PAQt}}\xspace}
\newcommand{\mtt}{\ensuremath{m_{\ttbar}}\xspace}
\newcommand{\mll}{\ensuremath{m_{\ttll}}\xspace}
\newcommand{\chel}{\ensuremath{c_{\text{hel}}}\xspace}
\newcommand{\chan}{\ensuremath{c_{\text{han}}}\xspace}
\newcommand{\thetaStar}{\ensuremath{\theta^\ast_{\PQt_{\Pell}}}\xspace}
\newcommand{\cpTTT}{\ensuremath{\cos\thetaStar}\xspace}
\newcommand{\acpTTT}{\ensuremath{\lvert\cpTTT\rvert}\xspace}
\newcommand{\thsttop}{\ensuremath{\theta^\ast_{\PQt}}\xspace}

\newcommand{\pAorH}{\ensuremath{{\PSA}\text{/}{\PSH}}\xspace}
\newcommand{\pAandH}{\ensuremath{{\PSA}\text{+}{\PSH}}\xspace}
\newcommand{\pntA}[2]{\ensuremath{\PSA(\mathrm{#1,\, #2})}\xspace}
\newcommand{\pntH}[2]{\ensuremath{\PSH(\mathrm{#1,\, #2})}\xspace}
\newcommand{\pntAH}[2]{\ensuremath{\pAorH(\mathrm{#1,\, #2})}\xspace}

\newcommand{\oneSzero}{\ensuremath{{}^1\mathrm{S}_0}\xspace}
\newcommand{\threePzero}{\ensuremath{{}^3\mathrm{P}_0}\xspace}
\newcommand{\oneSone}{\ensuremath{\oneSzero^{[1]}}\xspace}

\newcommand{\etat}{{\HepParticle{\PGh}{\PQt}{}}\xspace}
\newcommand{\etatnodepth}{\ensuremath{\PGh\smash{{}_{\PQt}}}\xspace}
\newcommand{\muetat}{\ensuremath{\mu(\etat)}\xspace}
\newcommand{\gphitt}[1][\PGF]{\ensuremath{g_{#1\ttbar}}\xspace}

\newcommand{\gAtt}[1][\PSA]{\ensuremath{g_{#1\ttbar}}\xspace}
\newcommand{\gHtt}[1][\PSH]{\ensuremath{g_{#1\ttbar}}\xspace}

\newcommand{\Irel}{\ensuremath{I_{\text{rel}}}\xspace}
\newcommand{\Dnu}{\ensuremath{D_{\PGn}}\xspace}
\newcommand{\muF}{\ensuremath{\mu_{\mathrm{F}}}\xspace}
\newcommand{\muR}{\ensuremath{\mu_{\mathrm{R}}}\xspace}
\newcommand{\hdamp}{\ensuremath{h_{\text{damp}}}\xspace}
\newcommand{\Rinout}{\ensuremath{R_{\text{in}/\text{out}}}\xspace}

\newcommand{\WW}{\ensuremath{\PW\PW}\xspace}
\newcommand{\WZ}{\ensuremath{\PW\PZ}\xspace}
\newcommand{\ZZ}{\ensuremath{\PZ\PZ}\xspace}
\newcommand{\ttV}{\ensuremath{\ttbar\PV}\xspace}
\newcommand{\ttW}{\ensuremath{\ttbar\PW}\xspace}
\newcommand{\ttZ}{\ensuremath{\ttbar\PZ}\xspace}
\newcommand{\tW}{\ensuremath{\PQt\PW}\xspace}
\newcommand{\tX}{\ensuremath{\PQt\PX}\xspace}
\newcommand{\Zjets}{\ensuremath{\PZ\text{+jets}}\xspace}
\newcommand{\Wjets}{\ensuremath{\PW\text{+jets}}\xspace}

\newcommand{\vecNuis}{\ensuremath{\boldsymbol{\nu}}\xspace}
\newcommand{\vecNuisHat}{\ensuremath{\boldsymbol{\widehat{\nu}}}\xspace}
\newcommand{\vecParaPhi}{\ensuremath{\boldsymbol{p}_{\PGF}}\xspace}
\newcommand{\vecParaPhiHat}{\ensuremath{\boldsymbol{\widehat{p}}_{\PGF}}\xspace}
\newcommand{\binBkgr}{\ensuremath{B_i}\xspace}
\newcommand{\binPred}{\ensuremath{\lambda_i}\xspace}
\newcommand{\binObsv}{\ensuremath{n_i}\xspace}
\newcommand{\binSignPhiTot}{\ensuremath{S_{i}^{\PGF}}\xspace}
\newcommand{\binSignPhi}[1]{\ensuremath{s_{#1,i}^{\PGF}}\xspace}
\newcommand{\binSignEtatTot}{\ensuremath{S_{i}^{\etatnodepth}}\xspace}
\newcommand{\binSignEtat}{\ensuremath{s_{i}^{\etatnodepth}}\xspace}
\newcommand{\mpphi}{\ensuremath{m_{\PGF}}\xspace}
\newcommand{\mpH}{\ensuremath{m_{\PSH}}\xspace}
\newcommand{\Gpphi}{\ensuremath{\Gamma_{\PGF}}\xspace}
\newcommand{\Gpphirel}{\ensuremath{\Gpphi/\mpphi}\xspace}
\newcommand{\GpH}{\ensuremath{\Gamma_{\PSH}}\xspace}
\newcommand{\sigstr}{\ensuremath{\mu}\xspace}
\newcommand{\sigstrhat}{\ensuremath{\widehat{\mu}}\xspace}
\newcommand{\teststat}{\ensuremath{\widetilde{q}}\xspace}
\newcommand{\teststatmu}{\ensuremath{\teststat_{\sigstr}}\xspace}

\newcommand{\CP}{\ensuremath{CP}\xspace}

\newcommand{\Hellp}{{\HepParticle{\widehat{\Pell}}{\PQt}{+}}\xspace}
\newcommand{\Hellm}{{\HepParticle{\widehat{\Pell}}{\PAQt}{-}}\xspace}
\newcommand{\mlpb}{\ensuremath{m_{\Pellp\PQb}}\xspace}
\newcommand{\mlmb}{\ensuremath{m_{\Pellm\PQb}}\xspace}

\newlength\cmsTabSkip\setlength{\cmsTabSkip}{1ex}

\cmsNoteHeader{HIG-22-013}
\title{Search for heavy pseudoscalar and scalar bosons decaying to a top quark pair in proton-proton collisions at \texorpdfstring{$\sqrt{s}=13\TeV$}{sqrt(s) = 13 TeV}}
\titlerunning{Search for heavy pseudoscalar and scalar bosons decaying to a top quark pair}

\author[cern]{The CMS Collaboration}

\date{\today}

\abstract{A search for pseudoscalar or scalar bosons decaying to a top quark pair (\ttbar) in final states with one or two charged leptons is presented. The analyzed proton-proton collision data was recorded at $\sqrt{s}=13\TeV$ by the CMS experiment at the CERN LHC and corresponds to an integrated luminosity of 138\fbinv. The invariant mass \mtt of the reconstructed \ttbar system and variables sensitive to its spin and parity are used to discriminate against the standard model \ttbar background. Interference between pseudoscalar or scalar boson production and the standard model \ttbar continuum is included, leading to peak-dip structures in the \mtt distribution. An excess of the data above the background prediction, based on perturbative quantum chromodynamics (QCD) calculations, is observed near the kinematic \ttbar production threshold, while good agreement is found for high \mtt. The data are consistent with the background prediction if the contribution from a simplified model of a color-singlet \oneSone\ \ttbar quasi-bound state \etat, inspired by nonrelativistic QCD, is added. Upper limits at 95\% confidence level are set on the coupling between the pseudoscalar or scalar bosons and the top quark for boson masses in the range 365--1000\GeV, relative widths between 0.5 and 25\%, and two background scenarios with or without \etat contribution.}

\hypersetup{%
    pdfauthor={CMS Collaboration},
    pdftitle={Search for heavy pseudoscalar and scalar bosons decaying to a top quark pair in proton-proton collisions at sqrt(s) = 13 TeV},
    pdfsubject={CMS},
    pdfkeywords={CMS, top quark, scalar boson, pseudoscalar boson}
}

\maketitle

\section{Introduction}

The observation of a Higgs boson with mass of 125\GeV by the ATLAS and CMS Collaborations in 2012~\cite{Aad:2012tfa, Chatrchyan:2012xdj, Chatrchyan:2013lba} confirmed the existence of an elementary spin-0 state, a crucial ingredient of the standard model (SM) of particle physics.
While only one such state is required in the SM, many beyond-the-SM (BSM) extensions predict additional spin-0 states,
such as two Higgs doublet models (2HDMs)~\cite{Branco:2011iw} and models predicting a new electroweak pseudoscalar or scalar singlet~\cite{Huitu:2019}, including models with a combination of a Higgs doublet and such singlet(s)~\cite{Muhlleitner:2017dkd}.
These additional bosons may also provide a portal to dark matter by acting as mediators between SM and dark matter particles~\cite{DMLHC:2015, Arina:2016}.
The new states introduced in these BSM extensions usually include pseudoscalar (\CP-odd) neutral bosons, scalar (\CP-even) neutral bosons, and charged bosons.
We use the symbol \PSA to denote pseudoscalar neutral states, \PSH for scalar neutral states not identified as the one with a mass of 125\GeV, and \PGF as a common symbol to refer to either \PSA or \PSH bosons.

As the heaviest elementary particles, top quarks have a pivotal role in the search for BSM physics, serving as sensitive probes.
Provided that additional \PGF bosons couple to fermions via a Yukawa interaction with coupling strength proportional to the fermion mass,
\PGF bosons with mass larger than twice the top quark mass \mtop may decay to a top quark pair (\ttbar) as the dominant channel.
This is true especially for \PSA bosons with suppressed decays to weak vector bosons due to \CP symmetry,
as well as for \PSH bosons in 2HDMs in the vicinity of the alignment limit~\cite{Craig:2013hca}.

In this paper, we consider a Yukawa-like coupling between \PGF bosons and top quarks.
The corresponding terms in the Lagrangian for the two \CP eigenstates are:
\ifthenelse{\boolean{cms@external}}{\begin{equation}\begin{aligned}
    \mathcal{L}_{\text{Yukawa},\PSA} &= i\gphitt[\PSA]\,\frac{\mtop}{v}\,\PAQt\gamma_{5}\PQt\,\PSA, \\
    \mathcal{L}_{\text{Yukawa},\PSH} &= -\gphitt[\PSH]\,\frac{\mtop}{v}\,\PAQt\PQt\,\PSH,
\end{aligned}\end{equation}}{\begin{equation}
    \mathcal{L}_{\text{Yukawa},\PSA} = i\gphitt[\PSA]\,\frac{\mtop}{v}\,\PAQt\gamma_{5}\PQt\,\PSA,
    \qquad
    \mathcal{L}_{\text{Yukawa},\PSH} = -\gphitt[\PSH]\,\frac{\mtop}{v}\,\PAQt\PQt\,\PSH,
\end{equation}}
where $\gphitt\geq0$ is the real-valued coupling strength modifier and $v$ is the vacuum expectation value of the SM Higgs field.
We probe \PGF boson masses \mpphi in the range $365<\mpphi<1000\GeV$ and total widths \Gpphi relative to \mpphi in the range $0.5<\Gpphirel<25\%$.

\begin{figure}[!b]
\centering
\includegraphics[width=0.47\textwidth]{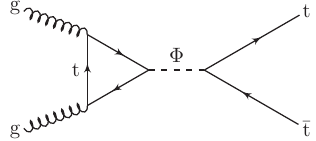}%
\hspace{0.1\textwidth}%
\includegraphics[width=0.25\textwidth]{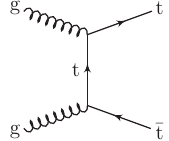}
\caption{%
    Example Feynman diagrams for the signal process (\cmsLeft) and for SM \ttbar production (\cmsRight).
}
\label{fig:feynman}
\end{figure}

\begin{figure}[!ht]
\centering
\includegraphics[width=0.47\textwidth]{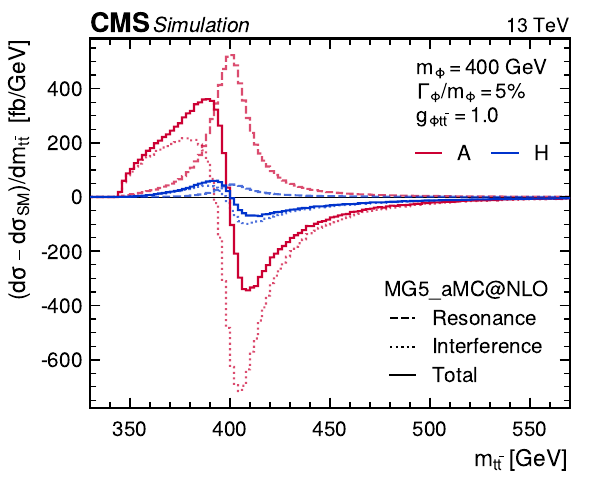}%
\hfill%
\includegraphics[width=0.47\textwidth]{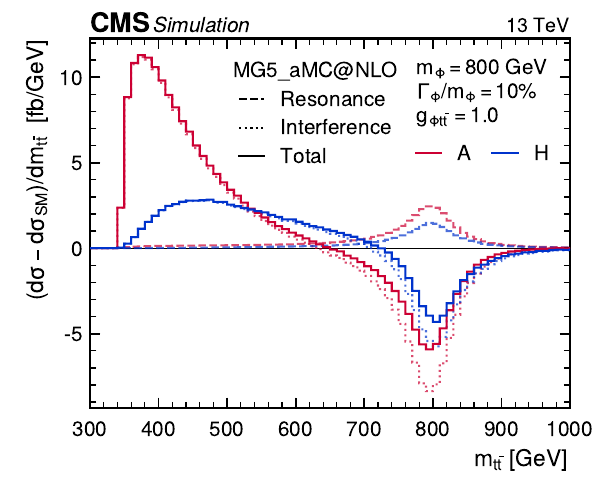}
\caption{%
    Differential cross section of \ttbar production at parton level as a function of \mtt, shown as difference between various BSM scenarios and the SM prediction.
    Shown are the cases of a single \PSA (red) or \PSH (blue) boson for two example configurations: $\mpphi=400\GeV$ and $\Gpphirel=5\%$ (\cmsLeft), or $\mpphi=800\GeV$ and $\Gpphirel=10\%$ (\cmsRight), with $\gphitt=1$ in both cases.
    Separately shown are the cases where only the resonant $\PGF\to\ttbar$ contribution is added to the SM prediction (dashed), where only the interference between SM and \PGF boson contributions is added (dotted), and where both contributions are added (solid).
    The distributions have been calculated using \MGvATNLO as described in Section~\ref{sec:samples}.
}
\label{fig:ahspectrum}
\end{figure}

The production of \PGF bosons is dominated by the gluon fusion process via a top quark loop, followed by a decay into a \ttbar pair, as illustrated in Fig.~\ref{fig:feynman} (\cmsLeft).
This process interferes with SM \ttbar production through gluon fusion, an example of which is depicted in Fig.~\ref{fig:feynman} (\cmsRight).
While the pure \PGF resonance component results in a Breit--Wigner peak in the \ttbar invariant mass (\mtt) distribution,
the interference terms may be either destructive or constructive, with the shape and magnitude of the \mtt distribution depending on the phase space region under consideration, the specific signal model, and the types of particles that appear in the loop of the production diagram~\cite{Carena:2016, Djouadi:2019}.
In general, the sum of the components produce a peak-dip structure in the \mtt distribution~\cite{Gaemers:1984sj, Dicus:1994bm, Bernreuther:1997gs,Bernreuther:2015fts,Bernreuther:2017yhg}, which is shown in Fig.~\ref{fig:ahspectrum} for two example choices of \mpphi and \Gpphi.

Decays of the \PSA and \PSH bosons produce \ttbar systems in the \oneSzero and \threePzero states, respectively~\cite{Bernreuther:1997gs}.
The SM \ttbar production comprises a mixture of spin states, with their relative contributions varying as a function of the partonic center-of-mass energy.
Due to the short lifetime of top quarks,
the information about their spin and polarization states is preserved in the angular distributions of their decay products~\cite{Bernreuther:2004jv, Bernreuther:2008ju, Mahlon:2010gw}.
Therefore, in addition to analyzing the \mtt distribution, we utilize angular observables to investigate the differences in the \ttbar spin states between signal and background processes.

In the SM, \ttbar production is described by quantum chromodynamics (QCD).
State-of-the-art cross section predictions rely on fixed-order (FO) perturbative QCD (pQCD) calculations and include electroweak (EW) corrections.
An additional enhancement of \ttbar production below the kinematic threshold is predicted in nonrelativistic QCD, dominated by the production of color-singlet \ttbar quasi-bound states (toponium)~\cite{Fadin:1987wz, Fadin:1990wx, Hoang:2000yr, Kiyo:2008bv, Ju:2020otc, Fuks:2021xje}.
We account for this effect by using a simplified model of the production of the color-singlet pseudoscalar quasi-bound state \oneSone, referred to as \etat~\cite{Fuks:2021xje}.

This paper describes a search for pseudoscalar and scalar bosons produced in proton-proton collisions at $\sqrt{s}=13\TeV$ and decaying to \ttbar.
The analyzed data were recorded using the CMS detector at the CERN LHC in 2016--2018 and corresponds to an integrated luminosity of 138\fbinv~\cite{CMS:2021xjt, CMS:2018elu, CMS:2019jhq}.
The single-lepton (\ttlj) and dilepton (\ttll) channels are considered, corresponding to the $\ttbar\to\bbbar\PW\PW \to\bbbar\Pell\PGn\PQj\PQj$ and $\bbbar\Pell\PGn\Pell\PGn$ decay chains of the \ttbar system, respectively.
Events in the \ttlj channel are selected with exactly one electron or muon and at least three jets (at least two of which are \PQb tagged), and in the \ttll channel with exactly two oppositely charged leptons (electrons and/or muons) and at least two jets (at least one of which is \PQb tagged).
The top quark four-momenta are estimated using kinematic reconstruction algorithms and the resulting \mtt distribution together with additional observables sensitive to the spin state of the \ttbar system are used to search for \PGF bosons.

The data are interpreted in terms of \PGF boson production, quantified via the coupling \gphitt for given values of \mpphi and \Gpphi, in three signal configurations: one \PSA boson, one \PSH boson, or both of them.
A combined maximum likelihood fit to all channels is used to extract the signals.
Two different background scenarios are investigated: one consists of FO pQCD predictions alone, and the other includes \etat production as part of the background.

Near the \ttbar threshold, there is an excess of the data with respect to the background predicted in FO pQCD alone, with a structure that favors an additional pseudoscalar contribution~\cite{CMS:TOP-24-007}.
In Ref.~\cite{CMS:TOP-24-007}, the companion paper to this publication, we perform an identical analysis to the one presented in this paper in the \ttll channel only, to demonstrate that the observed excess can be explained by \etat production without the need for BSM \PSA boson contributions.
We note that the current experimental resolution does not allow for a significant distinction between the \etat and the \PSA boson production scenarios, nor any potential mixtures of both, if the \PSA boson is produced sufficiently close to the \ttbar production threshold with a width of $\sim$5\% or less.

This search updates a similar analysis performed by the CMS experiment using 35.9\fbinv of data collected in 2016, where a moderate signal-like deviation compatible with \PSA boson production with a mass of 400\GeV was found~\cite{CMS:2019pzc}, without inclusion of any contribution from \ttbar bound states as background.
Searches for $\PGF\to\ttbar$ have also been performed by the ATLAS experiment using 20.3\fbinv of $\sqrt{s}=8\TeV$ data~\cite{ATLAS:2017snw} and 140\fbinv of $\sqrt{s}=13\TeV$ data~\cite{ATLAS:2024vxm}, where no significant deviations from the FO pQCD prediction were observed.
A detailed discussion on the differences of this result and Ref.~\cite{ATLAS:2024vxm} is provided in Ref.~\cite{CMS:TOP-24-007}.

\section{The CMS detector and event reconstruction}
\label{sec:detector}

The central feature of the CMS apparatus is a superconducting solenoid of 6\unit{m} internal diameter, providing a magnetic field of 3.8\unit{T}.
Within the solenoid volume are a silicon pixel and strip tracker, a lead tungstate crystal electromagnetic calorimeter (ECAL), and a brass and scintillator hadron calorimeter (HCAL), each composed of a barrel and two endcap sections.
Forward calorimeters extend the pseudorapidity ($\eta$) coverage provided by the barrel and endcap detectors.
Muons are reconstructed using gas-ionization detectors embedded in the steel flux-return yoke outside the solenoid.
More detailed descriptions of the CMS detector, together with a definition of the coordinate system used and the relevant kinematic variables, can be found in Refs.~\cite{Chatrchyan:2008zzk, CMS:PRF-21-001}.

Events of interest are selected using a two-tiered trigger system.
The first level (L1), composed of custom hardware processors, uses information from the calorimeters and muon detectors to select events at a rate of around 100\unit{kHz} within a fixed latency of about 4\mus~\cite{CMS:2020cmk}.
The second level, known as the high-level trigger, consists of a farm of processors running a version of the full event reconstruction software optimized for fast processing, and reduces the event rate to around 1\unit{kHz} before data storage~\cite{Khachatryan:2016bia, CMS:TRG-19-001}.

The primary vertex (PV) is taken to be the vertex corresponding to the hardest scattering in the event, evaluated using tracking information alone, as described in Section 9.4.1 of Ref.~\cite{CMS-TDR-15-02}.
The particle-flow (PF) algorithm~\cite{CMS-PRF-14-001} aims to reconstruct and identify each individual particle in an event, with an optimized combination of information from the various elements of the CMS detector.
The reconstructed particles are referred to as PF candidates in the following.
The energy of photons is obtained from the ECAL measurement.
The energy of electrons is determined from a combination of the electron momentum at the PV as determined by the tracker, the energy of the corresponding ECAL cluster, and the energy sum of all bremsstrahlung photons spatially compatible with originating from the electron track.
The energy of muons is obtained from the curvature of the corresponding track.
The energy of charged hadrons is determined from a combination of their momentum measured in the tracker and the matching ECAL and HCAL energy deposits, corrected for the response function of the calorimeters to hadronic showers.
Finally, the energy of neutral hadrons is obtained from the corresponding corrected ECAL and HCAL energies.

For each event, hadronic jets are clustered from the PF candidates using the infrared and collinear safe anti-\kt algorithm~\cite{Cacciari:2008gp, Cacciari:2011ma} with a distance parameter of 0.4.
Jet momentum is determined as the vectorial sum of all particle momenta in the jet, and is found from simulation to be, on average, within 5--10\% of the true momentum over the entire transverse momentum (\pt) spectrum and detector acceptance.
Additional proton-proton interactions within the same or nearby bunch crossings (pileup) can contribute additional tracks and calorimetric energy depositions to the jet momentum.
To mitigate this effect, charged particles identified to be originating from pileup vertices are discarded and an offset correction is applied to correct for remaining contributions~\cite{CMS:JME-18-001}.
Jet energy corrections are derived from simulation to bring the measured response of jets to that of particle level jets on average. In situ measurements of the momentum balance in dijet, photon+jet, {\PZ}+jet, and multijet events are used to account for any residual differences in the jet energy scale between data and simulation~\cite{CMS:2016lmd}.
The jet energy resolution amounts typically to 15--20\% at 30\GeV, 10\% at 100\GeV, and 5\% at 1\TeV~\cite{CMS:2016lmd}.
Additional selection criteria are applied to each jet to remove jets potentially dominated by anomalous contributions from various subdetector components or reconstruction failures~\cite{CMS:JME-18-001}.
To be considered in the data analysis, jets are required to satisfy $\abs{\eta}<2.4$, to have $\pt>30$ (20)\GeV in the \ttlj (\ttll) channel, and to be separated by $\DR=\sqrt{\smash[b]{(\Delta\eta)^2+(\Delta\phi)^2}}>0.4$ from any selected lepton, where $\Delta\eta$ and $\Delta\phi$ are the $\eta$ and azimuthal angle differences between the lepton and jet, respectively.

Jets originating from \PQb quarks are identified with the \textsc{DeepJet} algorithm~\cite{CMS:2017wtu, Bols:2020bkb, CMS:DP-2023-005}.
The used working point has a selection efficiency for \PQb quark jets of about 77\%, and a misidentification rate of 15\% for \PQc quark jets and of 2\% for light-quark and gluon jets (considered together and referred to as light jets in the following), as evaluated in simulated \ttbar samples.
Differences between data and simulation in the \PQb tagging efficiency and misidentification rate are accounted for by scale factors that depend on the jet \pt and $\eta$.

Electrons are measured in the range $\abs{\eta}<2.5$ as energy deposits in the ECAL matched to a track.
The momentum resolution for electrons with \pt of around 45\GeV from $\PZ\to\Pe\Pe$ decays ranges from 1.6 to 5\%.
It is generally better in the barrel region than in the endcaps, and also depends on the bremsstrahlung energy emitted by the electron as it traverses the material in front of the ECAL~\cite{CMS:2020uim, CMS-DP-2020-021}.
Only electrons with $\abs{\eta}<2.4$ and $\pt>20\GeV$ are considered in the analysis.
In the \ttll channel, well-identified electron candidates are selected using identification criteria based on boosted decision trees with a working point targeting a 90\% efficiency, with a misidentification rate of 1 and 3\% in the barrel and endcap regions, respectively~\cite{CMS:2020uim}.
In the \ttlj channel, well-identified electrons are selected using the ``tight'' working point of the identification criteria based on sequential requirements,
with an additional requirement of being consistent with originating from the PV~\cite{CMS:2020uim}.
The efficiency of the ``tight'' working point is about 70\%, with a misidentification rate of 1 and 2\% in the barrel and endcap regions, respectively.
Furthermore, the ``veto'' working point of the same sequential-requirements-based identification criteria is used to define a sample of loosely identified electrons used to veto events in the \ttlj channel.
All three employed sets of electron identification criteria are described in detail in Ref.~\cite{CMS:2020uim}.

Muons are measured in the range $\abs{\eta}<2.4$, with detection planes made using three technologies: drift tubes covering the barrel region, cathode strip chambers covering the endcap region, and resistive-plate chambers covering both the barrel and endcap regions.
Matching muons to tracks measured in the silicon tracker results in a relative \pt resolution, for muons with \pt up to 100\GeV, of 1\% in the barrel and 3\% in the endcaps; and of $<$7\% in the barrel for muons with \pt up to 1\TeV.
Only muons with $\pt>20\GeV$ are considered in the analysis.
For use in the main event selection, well-identified muon candidates are required to pass the ``tight'' working point of the identification criteria described in Ref.~\cite{CMS:2018rym}.
The selection efficiency of well-identified muons, together with the isolation requirements described below, is 75--85\%.
The misidentification rate for well-identified muons is 0.1--0.3\%, and the probability to incorrectly label muons within jets as isolated is 5--15\%.
Loosely identified muons are those passing the ``loose'' working point of the identification criteria~\cite{CMS:2018rym}, and are used in the \ttlj channel to veto events.

Lepton candidates are required to be isolated from other activity in the event.
The relative isolation \Irel is calculated as the \pt sum of charged-hadron,
neutral-hadron, and photon PF candidates inside a cone of $\DR=0.4$ around the lepton, divided by the lepton \pt.
An estimated contribution from pileup is subtracted in this calculation~\cite{CMS:2020uim, CMS:2018rym}.
In the \ttlj channel, well-identified muons are required to have $\Irel<0.15$, while loosely identified muons are required to have $\Irel<0.25$.
The same criteria of $\Irel<0.25$ are used for well-identified muons in the \ttll channel, where separate collections of loosely identified leptons are not introduced.
For electrons, isolation requirements are already included in the identification criteria defined in Ref.~\cite{CMS:2020uim}.
Scale factors that depend on the lepton \pt and $\eta$ are used to correct the simulation for small differences in lepton trigger, identification, and isolation efficiency with respect to data.

The missing transverse momentum vector \ptvecmiss is computed as the negative vector sum of the transverse momenta of all the PF candidates in an event, and its magnitude is denoted as \ptmiss~\cite{Sirunyan:2019kia}.
The \ptvecmiss is modified to account for corrections to the energy scale of the reconstructed jets in the event.

\section{Data and simulated event samples}
\label{sec:samples}

The analyzed data were recorded in 2016--2018 using triggers that require the presence of a single isolated electron or muon, or the presence of two such leptons including all possible flavor combinations.
Four independent data-taking eras are considered: 2016pre (19.5\fbinv), 2016post (16.8\fbinv), 2017 (41.5\fbinv), and 2018 (59.8\fbinv).
The 2016 data set is split into two eras because of a modification of the APV strip tracker readout chip settings that affects the efficiency of the track hit reconstruction during the 2016 data-taking period~\cite{CMS-DP-2020-045}, where the identifiers ``pre'' and ``post'' refer to the periods before and after this modification.
The 2016pre, 2016post, and 2017 eras are also affected by an inefficiency caused by the gradual shift in the timing of the inputs to the ECAL L1 trigger in the regions $\abs{\eta}>2.0$~\cite{CMS:2020cmk}.
Correction factors are computed from data and applied to the acceptance evaluated by simulation to account for this effect.

In order to compare the collected data to theoretical predictions, Monte Carlo (MC) samples are produced with events simulating the signal and background processes.
Various programs are used to evaluate matrix elements (MEs) and generate events at parton level.
In all cases, the generators employ the next-to-next-to-leading order (NNLO) NNPDF3.1 parton distribution functions (PDFs)~\cite{Ball:2014uwa} and are interfaced with \PYTHIA8.240~\cite{Sjostrand:2014zea} for fragmentation and hadronization using the CP5 underlying event tune~\cite{Skands:2014pea, CMS:2019csb}.
The nominal value of \mtop is set to 172.5\GeV in all samples involving top quarks, as well as in the computation of theoretical corrections that are applied to them.
The simulated events are processed through the CMS detector simulation based on the \GEANTfour program~\cite{Agostinelli:2002hh}.
Separate MC samples are generated corresponding to the data-taking conditions of each of the four eras.
Pileup interactions are generated with \PYTHIA and overlaid in all samples.
The simulated events are weighted to reproduce the distribution of the number of pileup interactions observed in data, assuming a total inelastic cross section of 69.2\unit{mb}.
On average, there are 23 collisions per bunch crossing in 2016 data and 32 in 2017--2018 data~\cite{CMS:JME-18-001}.

The $\PGF\to\ttbar$ signal process is simulated at leading-order (LO) accuracy using a custom model in the \MGvATNLO2.6.5 event generator~\cite{Alwall:2014hca}.
It implements the full kinematics of the top quark loop of the gluon fusion production, including finite \mtop effects, via a form factor that is implemented as an effective coupling between the \PGF bosons and gluons~\cite{Spira:1995rr}.
Event samples are produced for different \mpphi and \Gpphirel values, such that a good coverage throughout the region of phase space probed in this search is obtained.
They are reweighted to target signal hypotheses by the event-by-event ratios
of the squared MEs of the target signal hypothesis and the one used in the original event simulation.
The target signal hypotheses in this search are \mpphi values of 365, 380, and 400--1000\GeV (in steps of 25\GeV), and
\Gpphirel values of 0.5--3 (in steps of 0.5), 4--8 (in steps of 1), 10, 13, 15, 18, 21, and 25\%.
We use the notation ``\pntAH{400}{3\%}'' to refer to \PGF bosons of a particular \CP eigenstate, \mpphi in \GeVns, and \Gpphirel.
The factorization and renormalization scales, \muF and \muR, are set on an event-by-event basis to $\mtt/2$, following the choice in Ref.~\cite{Hespel:2016qaf}.
The top quarks from the \PGF boson decay are further decayed in \MGvATNLO, preserving their spin correlations.

Separate samples are generated for events corresponding to resonant \PGF boson production, and for events corresponding to interference terms in the ME calculation between \PGF boson and FO pQCD \ttbar background production.
Events in the interference samples can receive negative weights, reflecting the sign of the corresponding part of the squared ME in the presence of a destructive interference.
Since the \PGF boson is produced via gluon fusion with a top quark loop, the $\PGF\ttbar$ coupling appears twice in the ME.
As a result, events originating from the resonance ME terms correspond to a cross section proportional to \smash{$\gphitt^4$}, while those from interference correspond to a cross section proportional to \smash{$\gphitt^2$}.

We calculate cross sections for resonant \PGF boson production at NNLO accuracy in pQCD with the \textsc{SusHi} 1.7.0 program~\cite{Harlander:2012pb, Harlander:2016hcx} in the context of Type-II 2HDM models, where the \textsc{2hdmc} program~\cite{Eriksson:2009ws} is used to calculate the remaining model parameters for a given signal hypothesis.
The coupling modifiers of the \PGF bosons to bottom and charm quarks are set to zero.
The ratio of the NNLO cross section to the LO cross section calculated with \MGvATNLO is used as a $K$ factor to normalize the resonant part of the signal samples, with typical values around 2.

For the interference component of the signal samples, we apply $K$ factors corresponding to the geometric mean of those applied to the resonant signal and the FO pQCD \ttbar process~\cite{Hespel:2016qaf}.
Here, the FO pQCD \ttbar production $K$ factor is calculated as the ratio between the \ttbar cross section at NNLO in pQCD with next-to-next-to-leading logarithmic (NNLL) soft-gluon resummation, as described below, and at LO in pQCD with leading logarithmic resummation. The nominal value of this $K$ factor is 1.49, and is within 1.42 and 1.55 for different \mtop values and scale choices used in the computation.
For the \PSH signal, we have compared the resonance and interference $K$ factors with a recent explicit next-to-LO (NLO) calculation in the scope of a one-Higgs-singlet extension of the SM in Ref.~\cite{Banfi:2023udd}.
We find good agreement for the resonance component and significant differences of about 20\% for the interference component.
However, we have verified that this discrepancy does not significantly alter the conclusions of this work by performing alternative fits using the updated $K$ factors for the interference component, and obtaining compatible exclusion regions as those reported in Section~\ref{sec:results}.
All $K$ factors derived in this analysis are available in Ref.~\cite{hepdata}.

The \etat contribution is implemented as a generic resonance in the \MGvATNLO 2.6.5 event generator at LO accuracy in pQCD using a custom simplified model obtained from Ref.~\cite{Fuks:2021xje}.
The model is similar to the one used for the $\PGF\to\ttbar$ signal generation, although its effective gluon-pseudoscalar coupling is implemented as an effective contact interaction instead of via the top quark loop.
Samples of resonant $\etat\to\PW\PQb\PW\PQb$ events are produced to allow contributions from off-shell top quarks.
The \etat mass and width are set to 343 and 7\GeV, respectively, following Ref.~\cite{Fuks:2021xje}, corresponding to the expectation that the toponium mass is twice \mtop minus a binding energy of about 2\GeV.
A restriction to $\abs{m_{\PW\PQb\PW\PQb}-343\GeV}<6\GeV$ at the generator level is employed, as recommended in Ref.~\cite{Fuks:2021xje}, in order to not influence the high \mtt region which is assumed to be well-described by FO pQCD.
Other simulation parameters are set following the recommendations of the model authors.
The version of the model used here does not include the nonrelativistic Hamiltonian reweighting mentioned in Refs.~\cite{Fuks:2021xje, Fuks:2024yjj}.
This is expected to have a negligible effect on this analysis, which is performed on reconstructed distributions, considering that the reweighting has a very small effect on parton-level distributions~\cite{Sumino:2010bv}.
The \etat sample used in Ref.~\cite{CMS:TOP-24-007} has been updated compared to the one used in this work, with the \etat width set to 2.8\GeV and removing the generator-level requirement on $m_{\PW\PQb\PW\PQb}$~\cite{Maltoni:2024tul}. At the level of precision of this analysis, and comparing reconstructed distributions, both \etat models are in agreement.

The main background contribution originates from the FO pQCD \ttbar production process, and
is simulated at NLO accuracy in pQCD using the \POWHEG~v2 generator~\cite{Nason:2004rx, Frixione:2007vw, Alioli:2010xd, Campbell:2014kua}.
The \muF and \muR scales are set to $\sqrt{\smash[b]{\mtop^2+\pttop^2}}$, where \mtop and \pttop are the mass and \pt of the top quarks in the underlying Born-level configuration.
Decays of the top quarks are performed using the narrow-width approximation~\cite{Gigg:2008yc}.
The sample is normalized to the predicted \ttbar production cross section of $833.9^{+20.5}_{-30.0}\unit{pb}$,
as calculated with the \textsc{top++2.0} program at NNLO in pQCD, and including soft-gluon resummation at NNLL order~\cite{Czakon:2011xx}.
The quoted uncertainty is derived from the independent variations of \muF and \muR, though they are not the only ones that affect the value.
To improve the theoretical description of the FO pQCD \ttbar production process, the sample is further reweighted differentially to account for NNLO pQCD and NLO EW corrections.
The NNLO pQCD prediction is calculated using a private version of the \textsc{matrix} program~\cite{Grazzini:2017mhc}, and the NLO EW prediction is calculated using the \textsc{hathor}~2.1 program~\cite{Bernreuther:2005is,Kuhn:2005it,Bernreuther:2006vg,Kuhn:2006vh,Aliev:2010zk,Kuhn:2013zoa}, both with a nominal scale choice of $0.5\big(\sqrt{\smash[b]{\mtop^2+\pttop^2}}+\sqrt{\smash[b]{\matop^2+\ptatop^2}}\big)$.
Both predictions are computed at the level of stable top quarks, using the same PDF set as the \POWHEG~v2 \ttbar sample.
The weights are applied double-differentially at the generator level as a function of \mtt and the cosine of the angle between the direction of the top quark in the zero-momentum frame (ZMF) of the \ttbar system and the direction of the \ttbar system in the laboratory frame, $\cos\thsttop$.

Other background events originate from single top quark production (\tX), single vector boson production in association with jets including \PQb jets (\Zjets and \Wjets), diboson production (\WW, \WZ, and \ZZ), \ttbar production in association with a vector boson (referred to as \ttV), and events composed uniquely of jets produced through the strong interaction, referred to as QCD multijet processes.
The single top quark production processes, via the $t$ and $s$ channels and as \tW production, are generated at NLO using \POWHEG~v2, \POWHEG, and \MGvATNLO, respectively~\cite{Re:2010bp,Alioli:2009je}.
The samples are normalized using the NLO cross section predictions for the $t$ and $s$ channels~\cite{Aliev:2010zk, singletopNLO2}, and approximate NNLO prediction for the \tW channel~\cite{Kidonakis:2021vob}.
The \Zjets process is generated with the \POWHEG event generator~\cite{Frixione:2007vw, Alioli:2010xd} with a multi-scale-improved NNLO accuracy in pQCD~\cite{Monni:2019whf, Monni:2020nks}, matched with \PYTHIA8 for initial-state radiation (ISR) and the \PHOTOS package~\cite{Barberio:1993qi, Golonka:2005pn} for final-state radiation (FSR).
The \Wjets event samples are generated using \MGvATNLO at LO with up to four additional partons, and the MLM matching scheme~\cite{Alwall:2007fs} is used to combine the different parton multiplicities.
The single vector boson production cross sections are calculated at NNLO~\cite{Melnikov:2006kv, fewz2}.
However, in the \ttll channel, the normalization of the \Zjets contribution is directly determined from a control region in data.
Events simulating the diboson processes are generated using \PYTHIA and normalized to the respective NNLO (\WW)~\cite{wwNNLO} or NLO (\WZ and \ZZ)~\cite{mcfm} cross sections.
For the \WW process, we checked that explicitly simulating nonresonant $\PW\PW\bbbar$ production, which leads to the same final state as \ttbar production, does not change the results of this work.
The \ttV events are generated at NLO with \MGvATNLO, and are normalized using NLO cross section predictions.
The \MCATNLO matching scheme~\cite{Frixione:2002ik} is used for the \ttW samples, while the FxFx matching scheme~\cite{Frederix:2012ps} is used for the \ttZ samples.
Finally, the QCD multijet events are simulated with \PYTHIA.

\section{Data analysis in the \texorpdfstring{\ttlj}{single-lepton} channel}
\label{sec:singlelepanalysis}

Events that contain exactly one well-identified lepton (as defined in Section~\ref{sec:detector}) with $\pt>30\GeV$ are selected for further analysis in the \ttlj channel.
For data recorded during 2018 and most of 2017, except for an early period, a higher threshold of $\pt>34\GeV$ is applied if the lepton is an electron, in order to account for higher trigger-level thresholds.
Events containing additional loosely identified leptons (as defined in Section~\ref{sec:detector}) with $\pt>20\GeV$ are rejected.
Events are required to contain at least three jets with $\pt>30\GeV$, of which at least two are required to be \PQb tagged.
This event selection is referred to as signal region (SR).

\subsection{Kinematic reconstruction}

Each selected event is reconstructed under the assumption of \ttbar pair production with one leptonically and one hadronically decaying \PW boson from the top quark decays.
The first step is to determine the neutrino four-momentum based on the measured \ptmiss, and the second step is to assign jets to the final-state quarks.
Different procedures are followed for events with at least four or exactly three jets, as described below.

The neutrino four-momentum $p^{\PGn}$ is reconstructed with the algorithm described in Ref.~\cite{Betchart:2013nba}, separately using each \PQb jet in the event as candidate for the \PQb accompanying the leptonically decaying \PW boson.
Mass constraints of the \PW boson and leptonically decaying top quark are formulated, and for each \PQb jet candidate the $p^{\PGn}$ that satisfies these constraints and minimizes the distance $\Dnu=\abs{\ptmiss-\pt^{\PGn}}$ is used as the solution~\cite{Betchart:2013nba}.
If no solution is found for any \PQb jet, the event is rejected.

For events with four or more jets, a likelihood function is constructed using the product of the probability density of the minimal \Dnu and the two-dimensional probability density of the invariant masses of the hadronically decaying top quark and \PW boson.
The probability densities are evaluated from simulated events in which all jets are correctly identified.
All possible assignments of jets to the four final-state quarks are evaluated, provided that only \PQb-tagged jets are assigned as \PQb and \PAQb quark candidates.
The best jet assignment is the one that maximizes this likelihood.

For events with exactly three jets, the techniques described in Ref.~\cite{Demina:2013wda} are applied.
The likelihood function is constructed using the product of the probability density of the minimal \Dnu and the probability density of the invariant mass of the two jets assigned as originating from the hadronically decaying top quark.
As with the case of four or more jets, the best assignment is the one that maximizes this likelihood.
There are two typical cases of \ttbar events that only have three jets.
The first and more common case is when one or more quarks from the \ttbar decay lie below the \pt threshold or outside of the detector acceptance, which we refer to as lost-jet events.
The second case typically occurs in the high-momentum regime, where the angular separation between the top quark decay products are lower, leading to multiple quarks being clustered into one jet.
These events are referred to as partially merged events.
Once the best jet assignment is identified, a correction is applied to the four-momentum of the hadronically decaying top quark as a function of its reconstructed mass.
The correction factor, which is derived using simulation as described in Ref.~\cite{Demina:2013wda}, is larger for lost-jet events and is close to one for partially merged events, since a significant energy loss is expected only in the former case.

In events where the required \ttbar decay products, \ie, the lepton and either all four or at least three jets, are inside the detector acceptance and well identified, the correct combination is found in 74\% of events with four or more jets and in 83\% of events with three jets.
With respect to all selected \ttbar events, these correspond to rates of 37 and 61\%, respectively.

The signal is extracted using two-dimensional (2D) templates built using the \mtt and \acpTTT variables.
The angle \thetaStar is defined between the reconstructed leptonically decaying top quark in the ZMF and the direction of the \ttbar system in the laboratory frame, analogously to \thsttop introduced in Section~\ref{sec:samples}.
The spin-0 nature of the signals leads to the top quarks being emitted isotropically in the \ttbar ZMF, resulting in a flat \cpTTT distribution at the generator level in the absence of kinematic selections.
The FO pQCD distribution, on the other hand, peaks toward high values of \acpTTT, due to the contribution from other spin states.
As a result, the \acpTTT distribution will be enriched with signal events at low values.

To assess the precision of the reconstruction algorithm, we compute the relative resolution of \mtt, which is the standard deviation of its relative difference to the generator-level \mtt, evaluated in all selected simulated \ttbar events.
The resolution is in the range of 8\%, for low generator-level \mtt values near the threshold region, to 13\% for high generator-level \mtt values above 1000\GeV, and it does not strongly depend on the number of jets.
Furthermore, the absolute resolution of \acpTTT, defined similarly as the standard deviation of the absolute difference to the generator-level value, is found to be about 0.05 for events with four or more jets and 0.08 for events with three jets.

\subsection{Background estimation}

The background in the \ttlj channel is estimated from MC simulation for FO pQCD \ttbar and single top quark production, as well as for \etat production, as described in Section~\ref{sec:samples}.
QCD multijet production and EW processes (mostly \Wjets and small contributions from \Zjets, diboson, and \ttV production) are estimated using a control region (CR) in data with the same selection criteria as for the SR except for requiring that none of the selected jets is \PQb tagged.
The background distributions are obtained by subtracting the simulated single top quark and \ttbar contributions from the data in the CR.
The ratio of simulated background events in the SR and CR is applied as a normalization factor to the obtained background distributions.
This procedure has been validated in simulation, and the kinematic distributions obtained from the CR are compatible with those in the SR.

\begin{figure}[!t]
\centering
\includegraphics[width=0.47\textwidth]{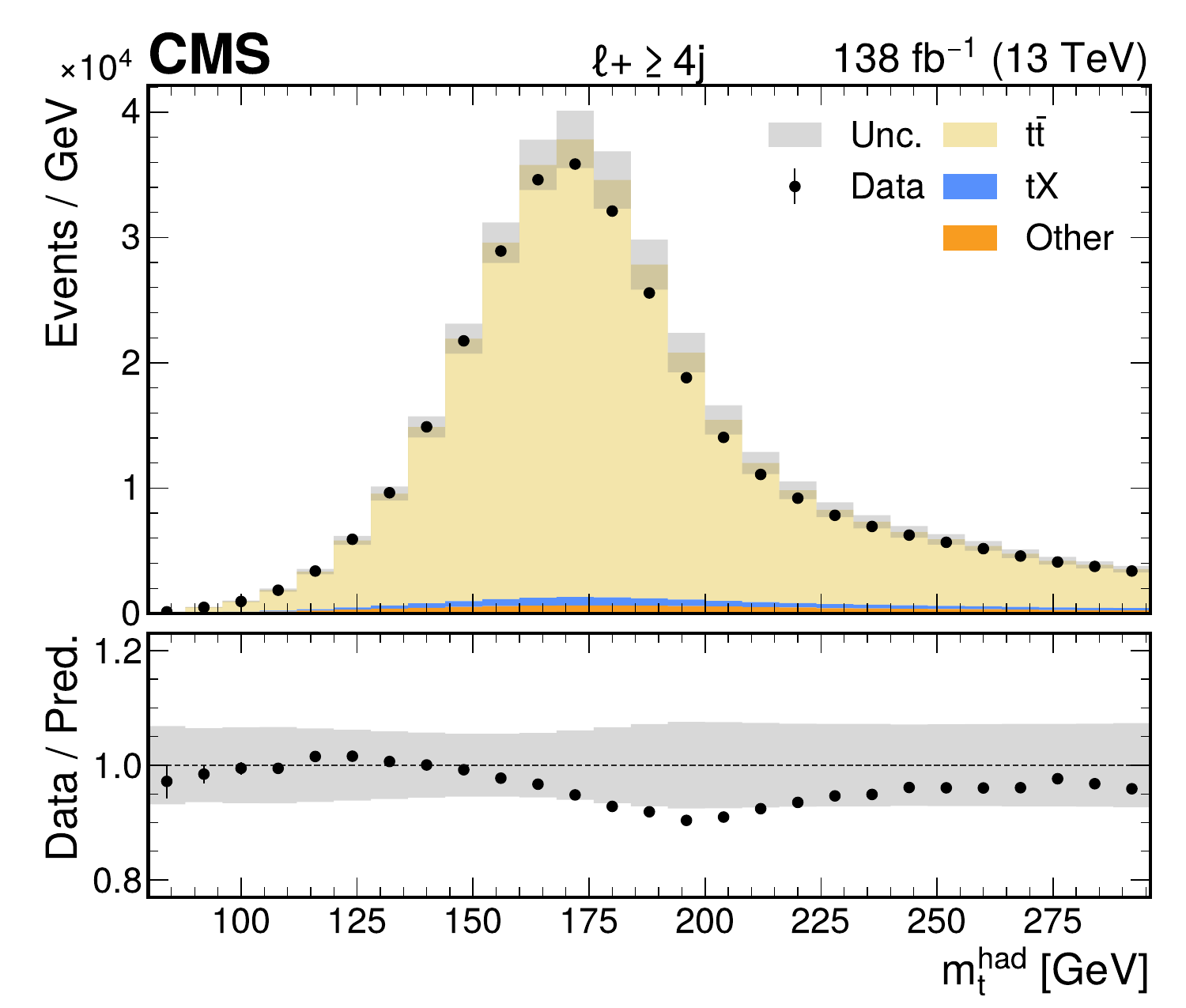}%
\hfill%
\includegraphics[width=0.47\textwidth]{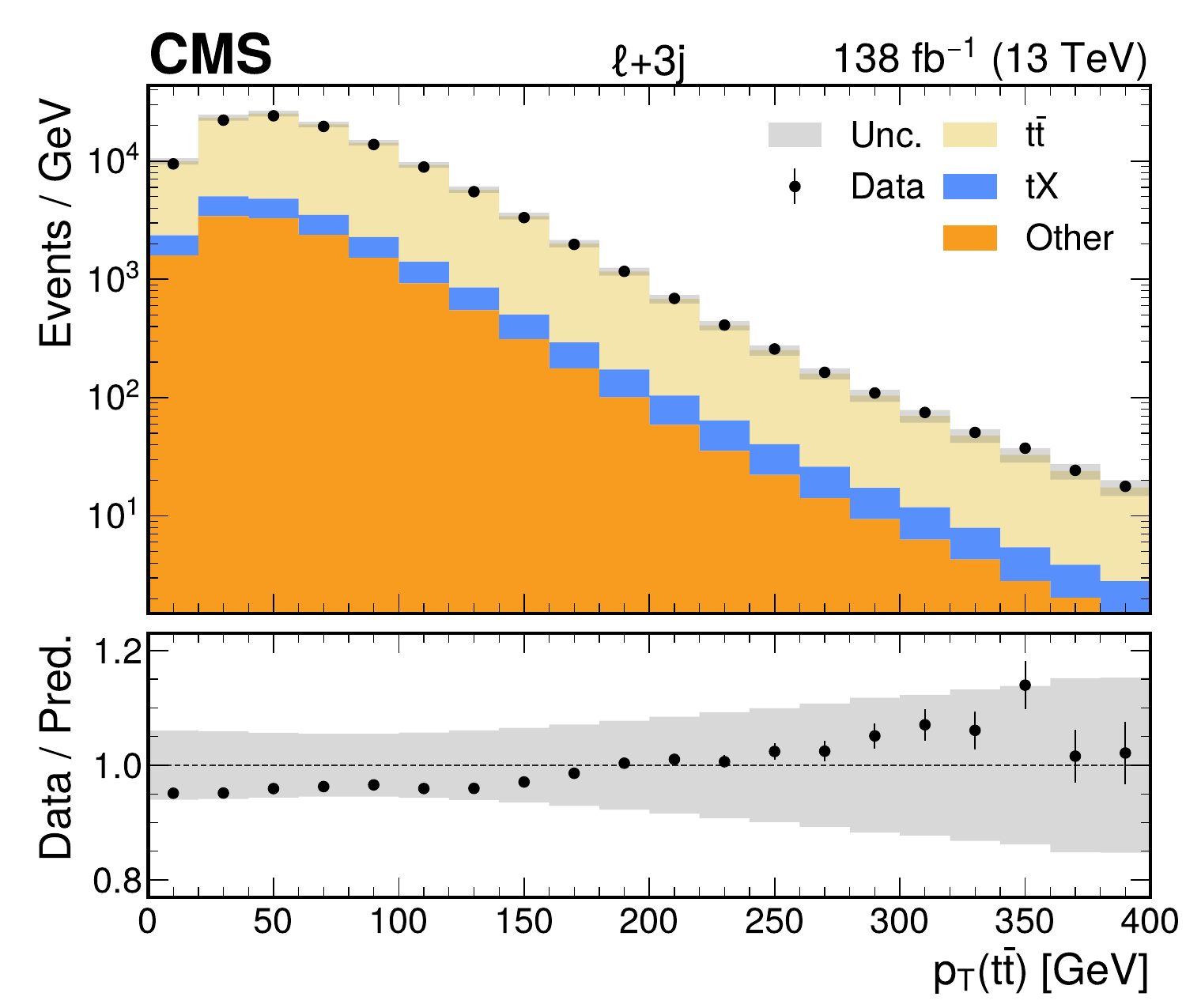}
\caption{%
    Comparison of the number of observed (points) and expected (colored histograms) events in the \ttlj channel after the kinematic reconstruction and background estimation for the distributions of the reconstructed hadronic top quark mass $\mtop^{\text{had}}$ in the region with four or more jets (\cmsLeft) and the \pt of the \ttbar system in the region with exactly three jets (\cmsRight). The ratio to the total prediction is shown in the lower panel, and the total systematic uncertainty is shown as the gray band.
}
\label{fig:controlplots_lj}
\end{figure}

The result of the kinematic reconstruction and background estimation is shown in Fig.~\ref{fig:controlplots_lj}, showing the reconstructed hadronically decaying top quark mass for events with four or more jets as well as the \pt of the \ttbar system for events with exactly three jets.

\section{Data analysis in the \texorpdfstring{\ttll}{dilepton} channel}
\label{sec:dilepanalysis}

In the \ttll channel, events are selected that contain exactly two oppositely charged well-iden\-tified leptons, one with $\pt>25\GeV$ and the other with $\pt>20\GeV$.
Events are rejected if they contain additional well-identified electrons or muons with $\pt>20\GeV$.
Furthermore, the invariant mass \mll of the dilepton pair is required to be larger than 20\GeV, to suppress events from low-mass dilepton resonances, and for same-flavor pairs to be outside of the \PZ boson mass window, $76<\mll<106\GeV$.
To further suppress \Zjets background contributions, events in the \ttee and \ttmm channels are required to have $\ptmiss>40\GeV$.
In all cases, at least two jets with $\pt>30\GeV$ are required, and additional jets with $\pt>20\GeV$ are also considered for further analysis.
At least one of these jets is required to be \PQb tagged.

\subsection{Kinematic reconstruction}

Each selected event is reconstructed under the assumption that the final state consists of a top quark pair that decays into two leptonically decaying \PW bosons.
A kinematic reconstruction algorithm~\cite{CMS-TOP-12-028} consisting of two steps is applied to reconstruct the \ttbar system.
First, of all jets in an event, two are identified as the \PQb and \PAQb quark candidates.
Second, these two candidates, together with the two leptons and \ptmiss, are used to determine the \PQt and \PAQt quark four-momenta by applying mass constraints on the \PW bosons and top quarks, taking into account experimental resolutions.

To find the best assignment of jets to the \PQb and \PAQb quarks, candidate pairs of jets are selected based on the number of \PQb-tagged jets in the event.
For events with two or more \PQb-tagged jets, only those jets are considered as \PQb and \PAQb quark candidates, while for events with exactly one \PQb-tagged jet, this jet is paired with all other jets in the event.
The invariant masses of the visible top quark decay products \mlpb and \mlmb are calculated for each \bbbar candidate pair as well as each assignment to the \PQb and \PAQb quarks, and a likelihood is constructed as the product of the generator-level probability densities of the two invariant masses, evaluated from simulated events.
The candidate pair that maximizes this likelihood is chosen for the next step of reconstruction.

Then, a system of equations for the top quark four-momenta is constructed from energy and momentum conservation as well as additional constraints~\cite{Sonnenschein:2006ud}, namely that:
(i) the top quark mass is equal to 172.5\GeV,
(ii) the \PW boson mass is equal to 80.4\GeV,
(iii) the two neutrinos from the \PW boson decays are the sole source of \ptmiss.
These equations, which are polynomials of fourth order, are solved for the neutrino momenta analytically, and the top quark four-momenta calculated as the vector sum of the decay products.
To resolve ambiguities between the multiple solutions, the one with the lowest reconstructed value of \mtt is used, which minimizes the bias with respect to the true value of \mtt~\cite{CMS:TOP-14-013, Korol:2016wzq}.

In about 55\% of cases, this procedure on its own does not give real solutions for the \ttbar system since it does not take into account the detector resolution.
To remedy this, the system of equations is solved 100 times per event with random smearings applied to the energies and directions of the \bbbar candidates and leptons.
These smearings are sampled, respectively, from distributions of the relative energy difference and angular distance between reconstructed and generator-level objects, as evaluated in simulated events.
The effect of the smearing on the momenta of the \bbbar candidates and leptons is propagated to the measured \ptmiss, by adding to it the opposite of the total change in momenta along the transverse components due to the smearing.
For all samplings that result in a real solution to the system of equations, weighted averages of the \PQt and \PAQt quark four-momenta are computed over all samplings, with the weight given by the same likelihood based on \mlpb and \mlmb as used for the \bbbar quark candidate assignment.
These averages are then considered as the final result of the reconstruction.

The performance of the \ttbar reconstruction algorithm is studied using simulated FO pQCD \ttbar events in the \ttll final state.
The algorithm produces a solution for 90\% of the events.
In 78\% of these events, at least one \PQb quark jet is correctly assigned, while in 61\% both jets are correctly assigned.
The relative \mtt resolution, defined similarly as in Section~\ref{sec:singlelepanalysis}, is in the range of 15\%, achieved at low generator-level \mtt values near the threshold region, to around 30\% at high generator-level \mtt values above 1000\GeV.
The average \mtt resolution is 23\%.

The search is performed by building three-dimensional (3D) templates using \mtt and two observables \chel and \chan that probe the spin correlations of the \ttbar system.
Spin correlation variables have been discussed in detail in Refs.~\cite{Bernreuther:2004jv, Bernreuther:2010ny,Bernreuther:2015yna, Aguilar-Saavedra:2022uye, Maltoni:2024tul, Aguilar-Saavedra:2024mnm}, and we follow the coordinate system and sign convention of Ref.~\cite{Bernreuther:2015yna}.
The observable \chel (referred to as $\cos\varphi$ in Refs.~\cite{Bernreuther:2004jv, Bernreuther:2015yna} and $-\cos\theta_{ab}$ in Ref.~\cite{Aguilar-Saavedra:2022uye}) is defined as the scalar product $\chel=\Hellp\cdot\Hellm$, where \Hellp and \Hellm are the unit vectors of the momenta of the two leptons in the rest frames of their parent \PQt and \PAQt, respectively, obtained by first boosting the leptons into the \ttbar ZMF and then further boosting them into the rest frames of their parent top (anti)quarks.
The observable \chan (identified with $-\cos\theta'_{ab}$ in Ref.~\cite{Aguilar-Saavedra:2022uye}) is obtained by flipping the sign of the component parallel to the top quark direction (the $\widehat{k}$ direction in Ref.~\cite{Bernreuther:2015yna}) for either \Hellp or \Hellm, and then calculating a similar scalar product.
The slopes of both distributions provide sensitivity to the degree of alignment between the \PQt and \PAQt spins.
The absolute resolutions of \chel and \chan as provided by the kinematic reconstruction, defined analogously as for \acpTTT in Section~\ref{sec:singlelepanalysis}, are found to be 0.46 for \chel and 0.60 for \chan.

\begin{figure}[!ht]
\centering
\includegraphics[width=0.47\textwidth]{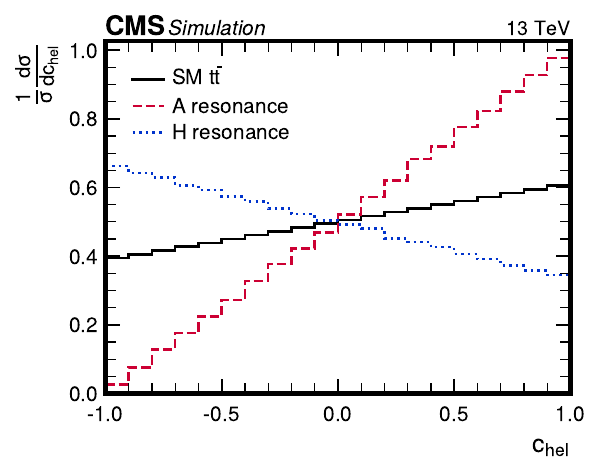}%
\hfill%
\includegraphics[width=0.47\textwidth]{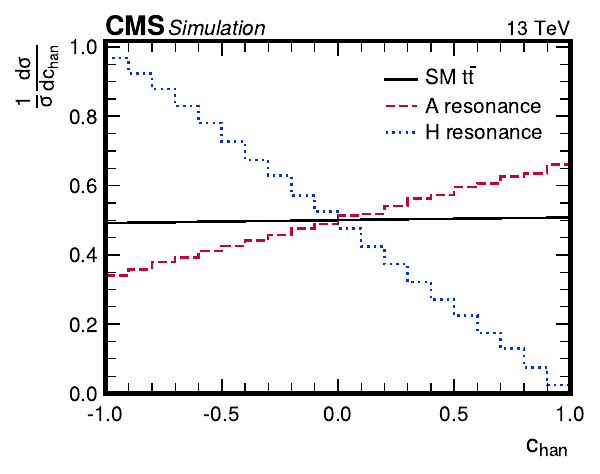}
\caption{%
    Normalized differential cross sections in the spin correlation observables \chel (\cmsLeft) and \chan (\cmsRight) at the parton level in the \ttll channel, with no requirements on acceptance, for SM \ttbar production (black solid), resonant \PSA boson production (red dashed), and resonant \PSH boson production (blue dotted).
    The corresponding distributions for \etat are identical to those of a \PSA boson.
}
\label{fig:chelchan}
\end{figure}

At the generator level and with no requirements on acceptance, the distributions of \chel and \chan, integrated over the phase space of all other variables,
follow a straight line, as shown in Fig.~\ref{fig:chelchan} for SM \ttbar and resonant \PGF boson.
For \chel, the slope is maximally positive for a pseudoscalar resonance, due to the resulting \ttbar system being in the \oneSzero state with anticorrelated \PQt and \PAQt spins.
The slope for the SM \ttbar production is mildly positive, being the weighted average of all possible \ttbar spin states reachable by the initial colliding partons.
Lastly, the slope of the scalar resonance is mildly negative, as a consequence of the \ttbar pair being in the \threePzero spin state.
On the other hand, for \chan, the slope is mildly positive for a pseudoscalar resonance, approximately flat for the SM \ttbar production, and maximally negative for a scalar resonance.
We further remark that, at generator level, the \chel and \chan distributions for \PSA and \PSH resonances have the same slopes regardless of their mass and width values.
The slopes of the SM \ttbar distributions on the other hand are dependent on \mtt---this is because of the change in the relative proportions of the colliding initial partons as well as their helicity combinations.
These features of the \chel and \chan distributions, when combined with \mtt, allow for discrimination between the signal and background processes and between the \PSA and \PSH states in a broad range of phase space.

\subsection{Background estimation}

All background processes in the \ttll channel, namely FO pQCD \ttbar, \etat, single top quark, \Zjets, diboson, and \ttV production, are estimated from simulated event samples.
Both the \ttll and the \ttlj decay channels of \ttbar are considered for the FO pQCD \ttbar sample, and additional misidentified or nonprompt leptons are included.
Contributions from \Wjets events with one additional such lepton or QCD multijet events with two such leptons are found to be small in the \ttll channel and neglected.

In the case of \Zjets production, the total yield of the simulation is corrected using data inside the \PZ boson mass window, which is removed in the main event selection, following a modified version of the procedure described in Ref.~\cite{CMS-TOP-11-002}.
The same selection criteria except for the \mll requirements are applied to the data inside the \PZ boson mass window.
We assume that there, the \Zjets contribution is negligible in the \ttem channel compared to the \ttee and \ttmm channels, and that other backgrounds contribute equally to the three channels up to a combinatorial factor.
Consequently, we can estimate the \Zjets contribution in data inside the \PZ boson mass window by subtracting the data yield in the \ttem channel from the data yield in the \ttee and \ttmm channels while correcting for lepton reconstruction efficiencies, thus subtracting out other backgrounds.

Next, to estimate the ratio of the \Zjets contribution inside and outside the \PZ boson mass window, denoted as \Rinout, we define a second sideband containing events with no \PQb-tagged jets.
The ratio in this region, $\Rinout^{0\PQb}$, can be measured directly by comparing the \Zjets yields in data inside and outside the \PZ boson mass window.
We then assume the ratio of ratios $\Rinout^{\geq1\PQb}/\Rinout^{0\PQb}$ in the regions with $\geq$1 and 0 \PQb tags, respectively, to be well-described by simulation, which is a looser assumption compared to that in Ref.~\cite{CMS-TOP-11-002}.
From this, we can infer $\Rinout^{\geq1\PQb}$, and thus the total \Zjets yield outside the \PZ boson mass window, for events with one or more \PQb tags, as used in the main selection.

\begin{figure}[!ht]
\centering
\includegraphics[width=0.47\textwidth]{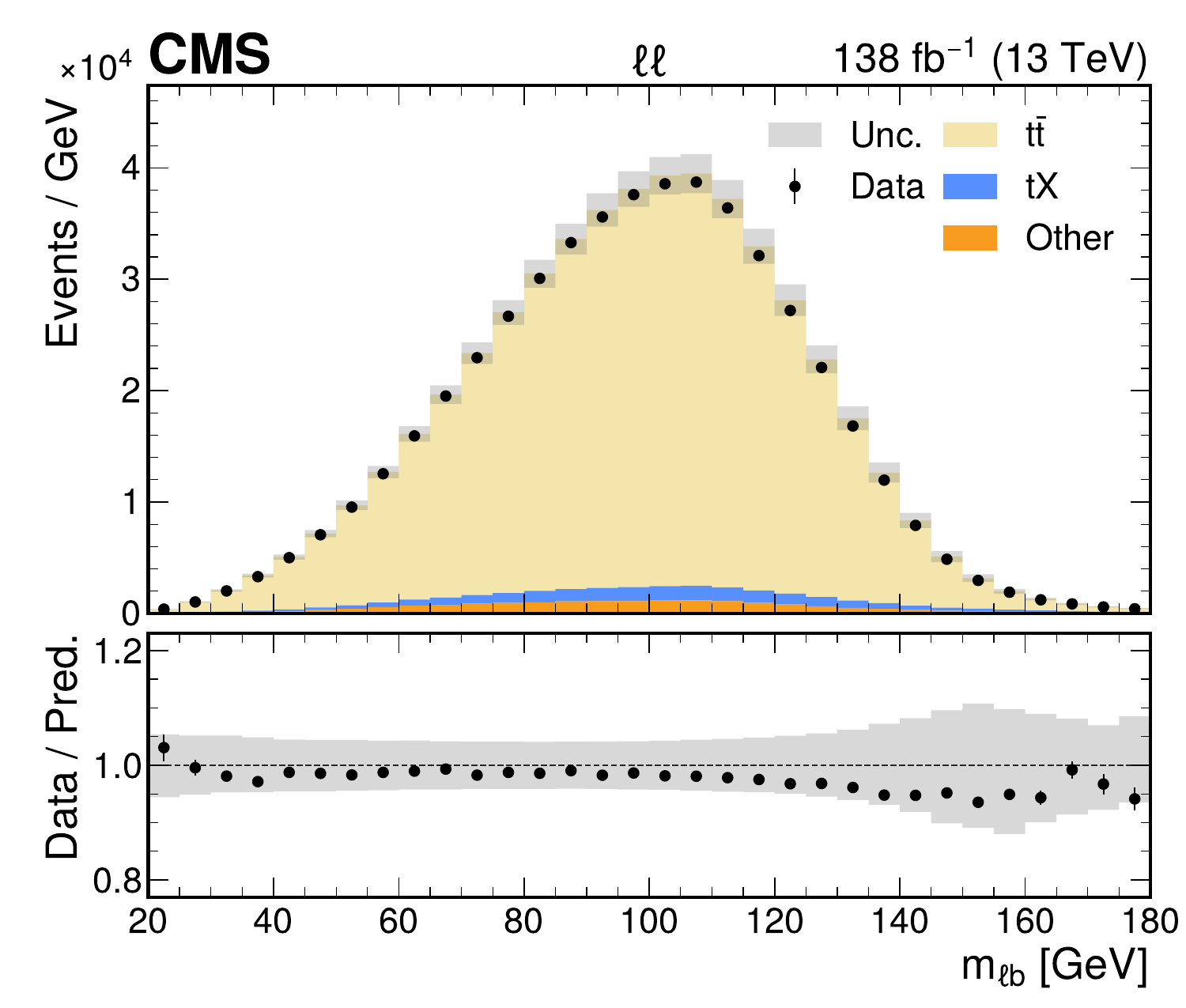}%
\hfill%
\includegraphics[width=0.47\textwidth]{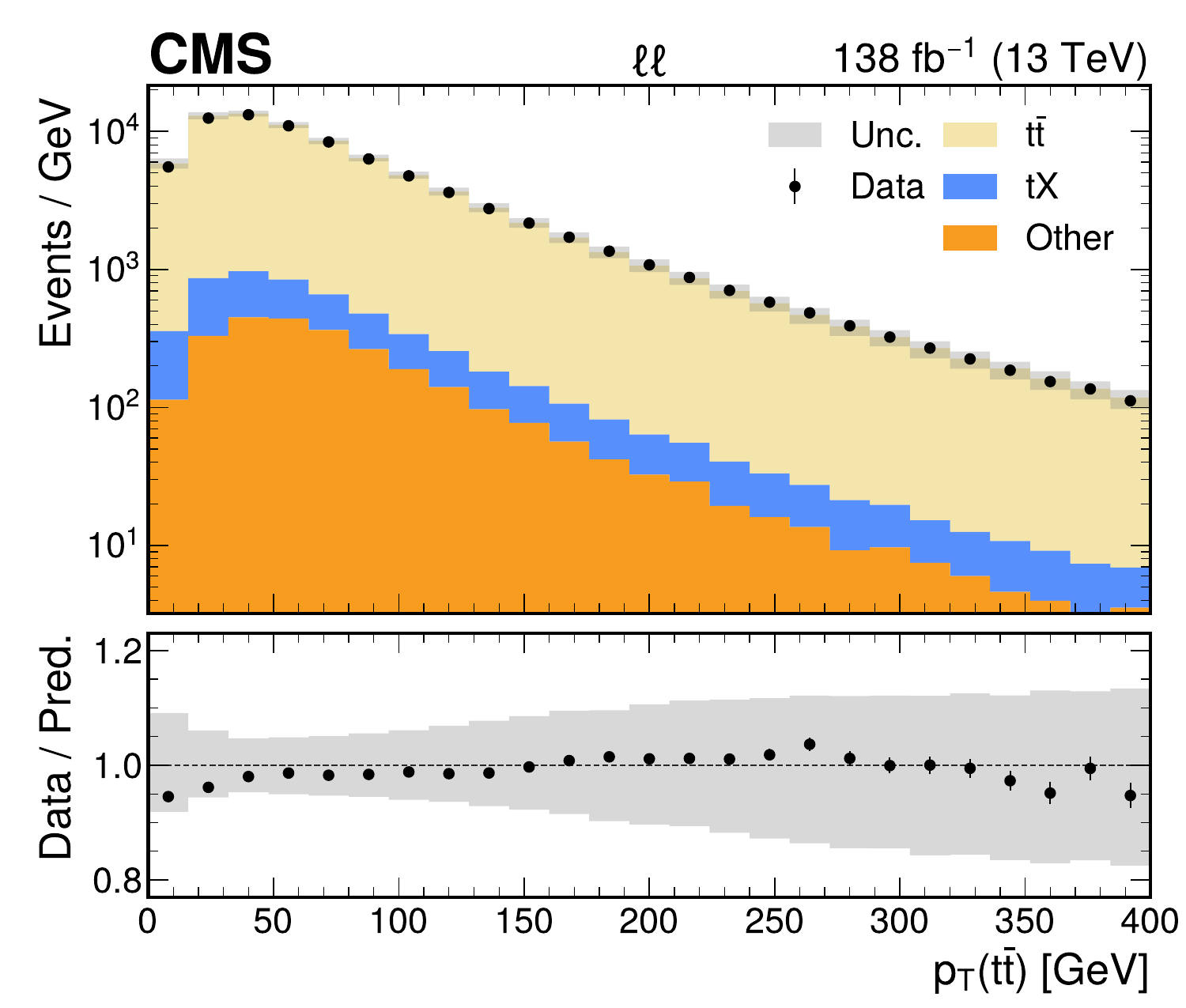}
\caption{%
    Comparison of the number of observed (points) and expected (colored histograms) events in the \ttll channel after the kinematic reconstruction and background estimation for the distributions of the invariant lepton-\PQb jet mass $m_{\Pell\PQb}$ (\cmsLeft) and the \pt of the \ttbar system (\cmsRight). The ratio to the total prediction is shown in the lower panel, and the total systematic uncertainty is shown as the gray band.
}
\label{fig:controlplots_ll}
\end{figure}

The yield is separately estimated for the \ttee and \ttmm channels, and used to normalize the simulated \Zjets contribution.
Compared to the yields predicted by simulation, we find the yield to be 3--12\% lower depending on analysis era and channel.
For the \ttem channel, where the \Zjets contribution is small, the geometric mean of the ratios to simulation is used.
The level of agreement between data and MC simulation after the kinematic reconstruction and background estimation is shown in Fig.~\ref{fig:controlplots_ll} for $m_{\Pell\PQb}$ as well as the reconstructed \pt of the \ttbar system.

\section{Systematic uncertainties}

Various sources of uncertainty affect the distributions of the observables used in this analysis, and are implemented as nuisance parameters in the binned maximum likelihood fit described in Section~\ref{sec:statfit}.
For each considered experimental and theoretical systematic effect, variations of the predicted signal and background distributions are evaluated.
Uncertainties that affect only the normalization of a process are modeled using log-normal constraints, as described in Section~4.2 of Ref.~\cite{CMS:2024onh}.
Gaussian constraints are imposed for all other uncertainties, which are referred to as shape uncertainties and can include a log-normal-constrained variation of the overall normalization, by modifying the product of the event acceptance and the cross sections of the relevant processes.
Unless stated otherwise, all uncertainties are evaluated for signal as well as background processes and treated as fully correlated among the processes, lepton channels, and eras.
The uncertainties are summarized in Table~\ref{tab:systematics}, and described in detail in the following.

\begin{table*}[!t]
\centering
\topcaption{%
    The systematic uncertainties considered in the analysis, indicating in parenthesis the number of corresponding nuisance parameters in the statistical model (if more than one), the type (affecting only normalization or also the shape of the search templates), and the affected physics processes and analysis channels they are applicable to.
}
\label{tab:systematics}
\renewcommand\arraystretch{1.1}
\begin{tabular}{lccc}
\hline
Uncertainty (\# of parameters) & Type & Process & Channel \\
\hline
Jet energy scale (17) & shape & all & all \\
Jet energy resolution (4) & shape & all & all \\
Unclustered \ptmiss (4) & shape & all & all \\
\PQb tagging heavy-flavor jets (20) & shape & all & all \\
\PQb tagging light jets (5) & shape & all & all \\
Single-electron trigger & shape & all & \ttej \\
Single-muon trigger (5) & shape & all & \ttmj \\
Dilepton triggers (12) & shape & all & \ttee, \ttem, \ttmm \\
Electron identification (2) & shape & all & \ttej, \ttee, \ttem \\
Muon identification (10) & shape & all & \ttmj, \ttem, \ttmm \\
ECAL L1 trigger inefficiency (3) & shape & all & all \\
Pileup & shape & all & all \\
Integrated luminosity (5) & norm. & all & all \\[\cmsTabSkip]
Top quark Yukawa coupling & shape & FO pQCD \ttbar & all \\
EW correction scheme & shape & FO pQCD \ttbar & all \\
\mtop & shape & FO pQCD \ttbar, \PGF & all \\
ME \muR (5) & shape & FO pQCD \ttbar, \PGF, single \PQt, \Zjets & all \\
ME \muF (6) & shape & FO pQCD \ttbar, \PGF, \etat, single \PQt, \Zjets & all \\
PS ISR (6) & shape & FO pQCD \ttbar, \PGF, \etat, single \PQt, \Zjets & all \\
PS FSR (6) & shape & FO pQCD \ttbar, \PGF, \etat, single \PQt, \Zjets & all \\
Color reconnection (2) & shape & FO pQCD \ttbar & all \\
\hdamp & shape & FO pQCD \ttbar & all \\
PDF (2) & shape & FO pQCD \ttbar & all \\[\cmsTabSkip]
Single top quark normalization & norm. & Single \PQt & all \\
EW+QCD normalization & norm. & EW+QCD & \ttlj \\
EW+QCD shape (20) & shape & EW+QCD & \ttlj \\
\ttV normalization & norm. & \ttV & \ttll \\
\Zjets normalization & norm. & \Zjets & \ttll \\
Diboson normalization & norm. & Diboson & \ttll \\
\hline
\end{tabular}
\end{table*}

The uncertainty in the jet energy scale~\cite{CMS:2016lmd} is evaluated by varying the corresponding corrections within their uncertainties, resulting in a total of 17 nuisance parameters that correspond to the absolute and relative jet energy scales, calibration uncertainties in specific detector regions, \pt balance in dijet or \Zjets events used in the jet energy calibration, and flavor-dependent jet response split into one source for \PQb quark jets and another for all other.
Of these, 12 nuisance parameters are specific to individual data-taking eras.
The uncertainty in the jet energy resolution measured in calibration data is propagated to the scale correction and smearing of the jet energy resolution in simulation.
An uncertainty in the unclustered component of \ptmiss is computed by shifting the energies of PF candidates not clustered into jets with $\pt>15\GeV$ according to the energy resolution for each type of PF candidate~\cite{Sirunyan:2019kia}.

Uncertainties in the scale factors to correct the \PQb tagging efficiency in simulated events are evaluated by varying them within their respective uncertainties~\cite{CMS:2017wtu}, independently for heavy-flavor (\PQb and \PQc quark) and light jets.
We assign 20 nuisance parameters for the heavy-flavor jet scale factors that correspond to the parton shower (PS) modeling, the presence of leptons within the jet, the jet energy scale, the number of pileup interactions, and differences between different SF estimation methods.
Of these, 4 nuisance parameters affect individual eras.
For the light jet scale factors, 5 nuisance parameters are assigned, of which 4 affect individual eras.

Uncertainties in the trigger, electron identification, and muon identification scale factors are considered~\cite{CMS:2020uim, CMS:2018rym}.
For the single-muon trigger and muon identification scale factors, each uncertainty component is further split into statistical components that are uncorrelated across eras and a correlated systematic component.
The effects of the inefficiency caused by the gradual shift in the timing of the inputs of the ECAL L1 trigger~\cite{CMS:2020cmk} are considered by assigning one nuisance parameter to each era except 2018, where the effect was not present.

The effective inelastic proton-proton cross section used for pileup reweighting in the simulation is varied by 4.6\% from its nominal value.
The uncertainty in the integrated luminosity amounts to 1.6\%~\cite{CMS:2021xjt, CMS:2018elu, CMS:2019jhq} and affects the normalization of all simulated processes.
It is split into 5 nuisance parameters with different correlation assumptions between the eras.

The prediction of the FO pQCD \ttbar production is affected by various sources of theoretical uncertainty.
The computation of the NLO EW correction, discussed in Section~\ref{sec:samples}, depends on the value of the SM top quark Yukawa coupling through interference with diagrams containing virtual SM Higgs bosons.
This coupling is modified with respect to its SM value in many BSM scenarios relevant to this analysis, and its experimental measurement uncertainty is significantly larger than the uncertainty on the top quark mass.
Thus, we consider
an uncertainty in the coupling by varying its value by $1.00\,^{+0.11}_{-0.12}$, where the range is given by the measurement reported in Ref.~\cite{CMS:2018uag}.
Furthermore, the uncertainty in the application scheme of the NLO EW corrections when combined with NNLO pQCD corrections is considered by taking the difference between the multiplicative and additive approaches of about 1--2\%, as recommended in Ref.~\cite{Kuhn:2013zoa}.
The uncertainty in \mtop is considered by shifting its value in simulation by $\pm$3\GeV, and the induced variations are then rescaled by a factor of 1/3 to emulate a more realistic top quark mass uncertainty of 1\GeV~\cite{Khachatryan:2015hba}.
The effect of the choice of \muR and \muF in the ME calculation is evaluated by varying these scales independently by a factor of two up and down.
The effects of the \mtop, \muR, and \muF variations on the acceptance and shape of the search templates are considered at NLO accuracy, while the effects on the overall FO pQCD \ttbar normalization is considered at NNLO+NNLL accuracy~\cite{Czakon:2011xx, Butterworth:2015oua}.
Decoupling the theoretical nuisance parameters based on their effects---one each for the acceptance and shape, and one additional parameter for the overall FO pQCD \ttbar normalization---does not alter the conclusions of this analysis.
Unlike Ref.~\cite{CMS:TOP-24-007}, no additional nuisance parameters comparing the predictions of different ME and PS programs are assigned to the FO pQCD \ttbar background.

The scales used to evaluate the strong coupling constant \alpS in the PS simulation of ISR and FSR are also varied independently by a factor of two up and down.
The effect of the uncertainties in the underlying event tune is estimated by varying the parameters of the CP5 tune~\cite{CMS:2019csb}.
Two uncertainties are assigned for the color reconnection model, with one based on the ``QCD-inspired'' model~\cite{Christiansen:2015yqa}, and the other by switching on the early resonance decay option in \PYTHIA~8.240~\cite{CMS:2022awf}.

The uncertainty in the matching scale between the ME and PS is evaluated
by varying the \POWHEG parameter \hdamp, which controls the suppression of radiation of additional high-\pt jets.
The nominal value of \hdamp in the simulation and its variations are $1.58\,^{+0.66}_{-0.59}\,\mtop$~\cite{CMS-PAS-TOP-16-021}.
The uncertainty arising from the choice of the PDF set is evaluated by
reweighting the simulated \ttbar events using 100 replicas of the NNPDF3.1 set.
A principal component analysis is performed on the variations from the PDF replicas to construct base variations
in the space of the predicted event yields in each bin of the search templates,
from which the base variation with the largest eigenvalue is used as the PDF uncertainty.
The second largest eigenvalue is found to be almost two orders of magnitude smaller than the largest one, thus the base variations corresponding to it and smaller eigenvalues are not considered.
The uncertainty in the \alpS parameter used in the PDF set forms a second independent PDF variation uncertainty.

The \muR and \muF scale uncertainties in the \PGF signal simulation are treated independently for the resonance and interference components.
Compared to the alternative of varying the scales for the two components simultaneously, we found this to be the more conservative option.
The effect on the acceptance and shapes of the search templates is considered at LO accuracy, while the effect on signal cross section is considered at NNLO accuracy.
The scales used in the PS simulation of ISR and FSR are also varied independently by a factor of 2 in each direction and are treated independently for the resonant and interference components.

The uncertainty in \mtop is also considered for the signal by varying its value in simulation
by $\pm$1\GeV.
Its effect on acceptance, shape, and cross section is considered in the same way as \muR and \muF variations.
Given that this is a variation on the same physical parameter, it is treated as fully correlated with the background processes.
Other theoretical uncertainties in the signal, such as the PDF, are neglected as they are small compared to those already considered.

The \etat background simulation, if applied, considers \muF, ISR, FSR, and \mtop uncertainties, affecting only the acceptance and shape.
They are handled identically to the corresponding variations in the \PGF signal simulation. The overall normalization of \etat is always taken to be a free parameter of the fit in this analysis.
Since the used model describes effective \etat production via a contact interaction, without the emission of extra partons at the LO ME level, the model encodes no dependence on \alpS.
Therefore, \muR variations have no effect on the \etat prediction.

The \muR, \muF, ISR, and FSR scale uncertainties are also independently considered for the \Zjets and single top quark production processes.
For these processes, the \muR and \muF uncertainties affect only acceptance and shape, not normalization.

The expected yields for most of the non-\ttbar background processes are derived using theoretical predictions for the cross sections at NLO or higher accuracy.
The uncertainties assumed in the normalization of these processes are conservative and always exceed those of the corresponding theoretical computations.
For single top quark production, we assign an uncertainty of 15\%, based on relevant cross section measurements~\cite{Aaboud:2016ymp, Sirunyan:2018rlu, Sirunyan:2018lcp}.
In the \ttll channels, the uncertainty in the \ttV production is taken to be 30\%~\cite{Sirunyan:2017uzs, Aaboud:2019njj}.
The uncertainty of the \Zjets production is taken to be 5\%~\cite{ATLAS:2016oxs}.
To account for the fact that this search probes a restricted region of the phase space of the corresponding processes, we assign a normalization uncertainty of 30\% for diboson production,
which has little impact on the overall sensitivity due to the small contribution of these processes.
All normalization uncertainties for non-\ttbar background processes are considered uncorrelated between each other.

In the single-lepton channels, the normalization uncertainty of the EW+QCD background estimate evaluated from a CR in data is taken to be 50\%.
Furthermore, to estimate the effect of changing the \PQb tagging requirements on the kinematic distributions, the estimation is repeated for three different selections of the highest allowed \PQb tagging discriminant value in the event.
The shape differences between the central selection and the selections with a higher and lower allowed value of the highest \PQb tagging discriminant are taken into account as uncertainties in the background estimation.
As an additional uncertainty, we take into account a variation of the subtracted single top quark and \ttbar contributions,
in which their expected contributions are scaled by the ratio of observed and expected events in the CR.

{\tolerance=800
The nominal background prediction is affected by the limited size of the simulated MC event samples.
This statistical uncertainty is evaluated using the ``light'' Barlow--Beeston method~\cite{barlowbeeston}, by introducing one additional nuisance parameter for each bin of the search templates.
These parameters are uncorrelated across all channels and eras.
\par}

Several systematic variations, most notably those constructed from dedicated MC samples, are affected by statistical fluctuations.
We suppress these fluctuations with a smoothing procedure, which is described in Ref.~\cite{CMS:2019pzc} and is based on the LOWESS algorithm~\cite{Cleveland79, Cleveland88}.

In general, the relative importance of different systematic uncertainties depends greatly on the signal hypothesis, especially the mass of the scalar bosons.
Close to the \ttbar production threshold, uncertainties due to the modeling of \ttbar dominate the total uncertainty, in particular the top quark Yukawa coupling, the application scheme of the NLO EW corrections, \muR, \mtop, the color reconnection model, and the \etat normalization (if considered).
A further nonnegligible contribution comes from the estimation of the EW+QCD background.
For larger values of \mpphi, the ME-PS matching uncertainty for the FO pQCD \ttbar background as well as experimental uncertainties due to heavy-flavor jet tagging become similarly important, while the effect of the \etat and EW+QCD contributions become small.
In addition, the total MC statistical uncertainties in all bins together often outweigh every other individual uncertainty.

\section{Statistical analysis}
\label{sec:statfit}

To evaluate the consistency of the observed data with the background-only hypothesis and with different signal hypotheses, we perform a statistical analysis using the search templates described in Sections~\ref{sec:singlelepanalysis}--\ref{sec:dilepanalysis}.
The \ttlj and \ttll final states do not overlap as they correspond to orthogonal lepton selection criteria.

The statistical model is defined by the likelihood function
\ifthenelse{\boolean{cms@external}}{\begin{multline*}
    L(\vecParaPhi,\muetat,\vecNuis)= \\
    \Bigg(\prod_i\frac{\binPred(\vecParaPhi,\muetat,\vecNuis)^{\binObsv}}{\binObsv!}\,\re^{\displaystyle-\binPred(\vecParaPhi,\muetat,\vecNuis)}\Bigg) G(\vecNuis),
\end{multline*}
where
\begin{multline}\label{eq:likelihood}
    \binPred(\vecParaPhi,\muetat,\vecNuis)= \\
    \binSignPhiTot(\vecParaPhi,\vecNuis)+\binSignEtatTot(\muetat,\vecNuis)+\binBkgr(\vecNuis),
\end{multline}}{\begin{equation}\begin{aligned}\label{eq:likelihood}
    L(\vecParaPhi,\muetat,\vecNuis)&=\Bigg(\prod_i\frac{\binPred(\vecParaPhi,\muetat,\vecNuis)^{\binObsv}}{\binObsv!}\,\re^{\displaystyle-\binPred(\vecParaPhi,\muetat,\vecNuis)}\Bigg) G(\vecNuis), \quad\text{where} \\
    \binPred(\vecParaPhi,\muetat,\vecNuis)&=\binSignPhiTot(\vecParaPhi,\vecNuis)+\binSignEtatTot(\muetat,\vecNuis)+\binBkgr(\vecNuis),
\end{aligned}\end{equation}}
with \binBkgr denoting the combined FO pQCD background yield in a given bin $i$, \binSignPhiTot the \PGF signal yield dependent on signal model parameters \vecParaPhi, \binSignEtatTot the \etat contribution dependent on the signal strength \muetat, \vecNuis the vector of nuisance parameters on which the signal and background yields generally depend, and \binObsv the observed yield.
The external constraints on the nuisance parameters are taken into account in this likelihood via a product of corresponding probability density functions, $G(\vecNuis)$.

The \PGF signal yield is given by
\ifthenelse{\boolean{cms@external}}{\begin{multline}\label{eq:ahyield}
    \binSignPhiTot(\vecParaPhi,\vecNuis)=\sum_{\PGF=\PSA,\PSH}\Big(\gphitt^4 \, \binSignPhi{R}(\mpphi,\Gpphi,\vecNuis) \\
    +\gphitt^2 \, \binSignPhi{I}(\mpphi,\Gpphi,\vecNuis)\Big),
\end{multline}}{\begin{equation}\label{eq:ahyield}
    \binSignPhiTot(\vecParaPhi,\vecNuis)=\sum_{\PGF=\PSA,\PSH}\Big(\gphitt^4 \, \binSignPhi{R}(\mpphi,\Gpphi,\vecNuis)+\gphitt^2 \, \binSignPhi{I}(\mpphi,\Gpphi,\vecNuis)\Big),
\end{equation}}
where \binSignPhi{R} and \binSignPhi{I} are the yields for the resonant and interference part, respectively.
The vector \vecParaPhi represents the signal model parameters and comprises the \PGF boson mass \mpphi, width \Gpphi, and \gphitt.
Equation~\eqref{eq:ahyield} is kept generic by including contributions from both \PSA and \PSH.
Since there is no interference between them, the corresponding signal distributions are trivially added together.

The yield of the \etat contribution is given by
\begin{equation}
    \binSignEtatTot(\muetat,\vecNuis)=\muetat \, \binSignEtat(\vecNuis),
\end{equation}
where \binSignEtat are the predicted \etat signal yields and \muetat is the signal strength modifier, which is a free parameter of the fit.
There is no additional interference between \PSA and \etat productions~\cite{Djouadi:2024lyv, Aguilar-Saavedra:2024mnm, Maltoni:2024tul}.

The background-only model is constructed by setting $\gphitt=0$.
The compatibility between data and a given hypothesis is determined by performing scans over the parameters of the signal models in different scenarios, using methodologies described in the following.

\subsection{Methodology for single \texorpdfstring{\PGF}{A/H} boson interpretation}

In the single \PGF boson interpretation, constraints on the coupling strength modifier \gphitt are derived as a function of \mpphi for fixed \Gpphirel values, separately for \PSA and \PSH.
This is done while setting the coupling modifier for the other \CP state in Eq.~\eqref{eq:likelihood} to zero, thus excluding it from the statistical model.
The scan is performed for the \mpphi and \Gpphirel values listed in Section~\ref{sec:samples}.
Coupling strength values up to 3 are probed to guarantee that the amplitudes preserve perturbative unitarity for all calculations, in accordance with the lower bound $\tan\beta=1/\gphitt[\PSA]\gtrsim0.3$ given in Ref.~\cite{Branco:2011iw} in the context of 2HDMs, where $\tan\beta$ is the ratio of the vacuum expectation values of the Higgs doublets coupling to the up- and down-type quarks.

A variant of the LHC profile likelihood ratio test statistic \teststatmu equivalent to those described in Refs.~\cite{Cowan:2010js, CMS-NOTE-2011-005} is utilized:
\begin{equation}
    \teststatmu(\vecParaPhi)=-2\ln\frac{L(\sigstr,\vecParaPhi,\vecNuisHat_{\sigstr,\vecParaPhi})}{L(\sigstrhat,\vecParaPhi,\vecNuisHat_{\sigstrhat,\vecParaPhi})},
    \quad
    0\leq\sigstrhat\leq\sigstr.
\end{equation}
Because the \PGF signal scales nonlinearly with the coupling modifiers \gphitt, we introduce an auxiliary overall signal strength modifier \sigstr in terms of which the test statistic is expressed, in the same way as in Ref.~\cite{CMS:2019pzc}.
This facilitates testing different \PGF signal hypotheses in a computationally efficient way.
The auxiliary parameter scales the overall \PGF signal yield in Eq.~\eqref{eq:ahyield}, keeping the other parameters in \vecParaPhi fixed.
The likelihood in the numerator is maximized with respect to the nuisance parameters, and $\vecNuisHat_{\sigstr,\vecParaPhi}$ denotes the vector of their values at the maximum for a given \vecParaPhi.
Depending on the scenario considered, the \etat signal strength is kept as a free parameter of the fit and treated as part of the nuisance parameters, or it is fixed to $\muetat=0$ in both numerator and denominator.
A similar notation is used in the denominator, where the likelihood is maximized with respect to both \sigstr and \vecNuis, under the additional constraint $0\leq \sigstrhat\leq\sigstr$.
The requirement $\sigstrhat\geq0$ excludes cases in which the shape of the overall BSM contribution gets flipped, resulting in a qualitatively different effect from what is targeted in this search.
The condition $\sigstrhat\leq\sigstr$ prevents the exclusion of a signal hypothesis if the data are more compatible with a model that predicts the BSM contribution of a similar shape but a larger overall size.

For each signal hypothesis, we perform a test according to the \CLs criterion~\cite{Junk:1999kv, Read:2002hq}.
An asymptotic approximation~\cite{Cowan:2010js} is employed to efficiently construct the distributions of the adopted test statistic.
We exclude a configuration \vecParaPhi at 95\% confidence level (\CL) if the \CLs value computed for $\sigstr=1$, which reproduces the nominal signal expectation, is smaller than 0.05.

\subsection{Methodology for \texorpdfstring{\pAandH}{A+H} boson interpretation}
\label{sec:2Dinterp}

In the \pAandH boson interpretation, we consider the more general case where two \PGF states exist at the same time.
We confine ourselves to the case with exactly one \PSA boson and exactly one \PSH boson, \ie, the case considered in 2HDMs~\cite{Branco:2011iw}.
Constraints in the \gAtt--\gHtt plane are set using the following test statistic:
\begin{equation}\label{eq:dnll}
    \teststat_{\vecParaPhi}=-2\ln\frac{L(\vecParaPhi,\vecNuisHat_{\vecParaPhi})}{L(\vecParaPhiHat,\vecNuisHat_{\vecParaPhiHat})},
\end{equation}
expressed directly in terms of \gAtt and \gHtt.
In contrast to the single \pAorH interpretation, the asymptotic approximation on the form of the test statistic distribution is not exploited, rendering the auxiliary parameter \sigstr unnecessary.

For each \gphitt configuration under consideration, its compatibility with the data is evaluated with the Feldman--Cousins prescription~\cite{Feldman:1997qc,Cousins:1991qz}.
An iterative procedure is applied to reduce the number of points for which the test statistic needs to be evaluated.
An initially sparse grid of \gphitt configurations are evaluated and refined around the region of the exclusion contour boundary at a given \CL.
The procedure is repeated until the minimum distance of two neighboring \gphitt configurations in the plane is small enough.
Like in the single \pAorH boson interpretation, we scan within the range of $\gphitt\leq3$ in the \pAandH boson interpretation.

\section{Results}
\label{sec:results}

The data are interpreted in the context of \PGF boson production under two background scenarios, one including \etat production and one without.
When \etat is not included, a deviation from the background prediction is observed near the \ttbar production threshold.
In Section~\ref{sec:ahsmvseta_2D}, we compare the two different background scenarios to the signal scenario corresponding to the highest local significance for this deviation.
Next, in Section~\ref{sec:ahsmtt_1D}, limits on the production of a single \PGF boson are presented, assuming that the background prediction is based on FO pQCD calculations alone.
Then, in Section~\ref{sec:ahetat_1D}, the same \PGF boson interpretations are presented, but now with \etat included as part of the background.
Finally, in Section~\ref{sec:ahetat_2D}, we show exclusion contours for the simultaneous presence of \PSA and \PSH bosons for a few examples of \mpphi and \Gpphirel, in the background scenario with \etat production included. Constraints on \gphitt in the single \PGF boson interpretation, as well as exclusion contours in the \pAandH boson interpretation, for mass and width values not included in this paper are provided in the corresponding HEPData entry~\cite{hepdata}.

We refer to the companion paper Ref.~\cite{CMS:TOP-24-007} for an interpretation of the excess around the threshold region in terms of a pseudoscalar \ttbar quasi-bound state without invoking any BSM degrees of freedom, performed in the \ttll channels only. For \mtt values close to the \ttbar threshold and with the chosen analysis strategy without spin correlation observables, the \ttlj channels contribute only subleading sensitivity to a \ttbar quasi-bound state. As a result, including the \ttlj channels in Ref.~\cite{CMS:TOP-24-007} would not significantly change the conclusions of said work.

\subsection{Data compared to background scenarios with and without \texorpdfstring{\etat}{toponium} contribution}
\label{sec:ahsmvseta_2D}

\begin{figure*}[!tp]
\centering
\includegraphics[width=0.87\textwidth]{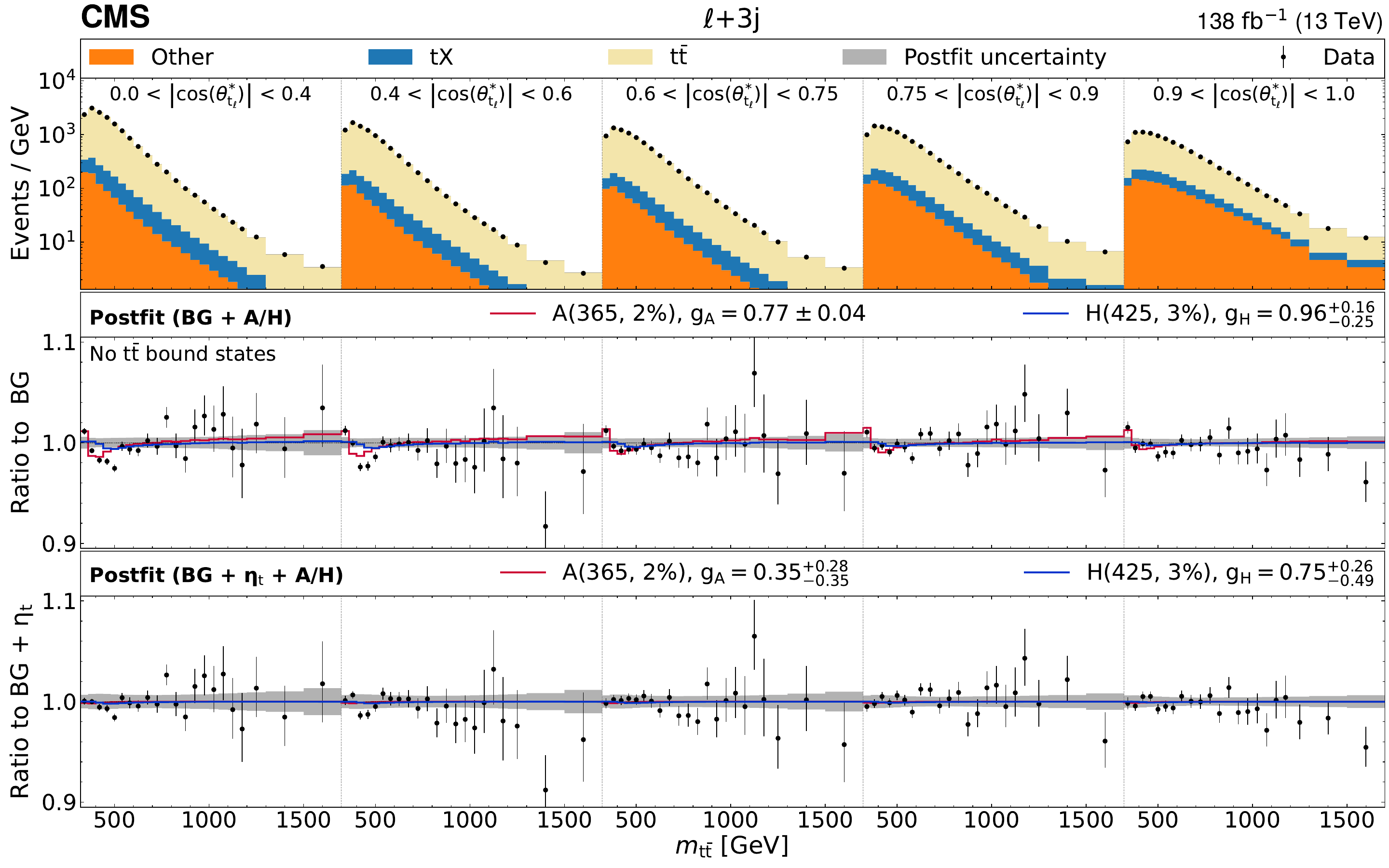}
\caption{%
    Observed and expected \mtt distribution in bins of \acpTTT, shown for the \ttltj channel summed over lepton flavors and eras.
    In the upper panel, the data (points with statistical error bars) are compared to \ttbar production in FO pQCD and other sources of background (colored histograms) after the fit to the data in the \pAandH interpretation.
    The ratio of data to the prediction is shown in the middle panel, where the two signals \pntA{365}{2\%} and \pntH{425}{3\%}, corresponding to the best fit point, are overlaid.
    The lower panel shows the equivalent ratio for the fit where \etat is considered as an additional background, for the same signal points.
    In both cases, the gray band shows the postfit uncertainty, and the respective signals are overlaid with their best fit model parameters.
}
\label{fig:combined_l3j}
\end{figure*}

\begin{figure*}[!tp]
\centering
\includegraphics[width=0.87\textwidth]{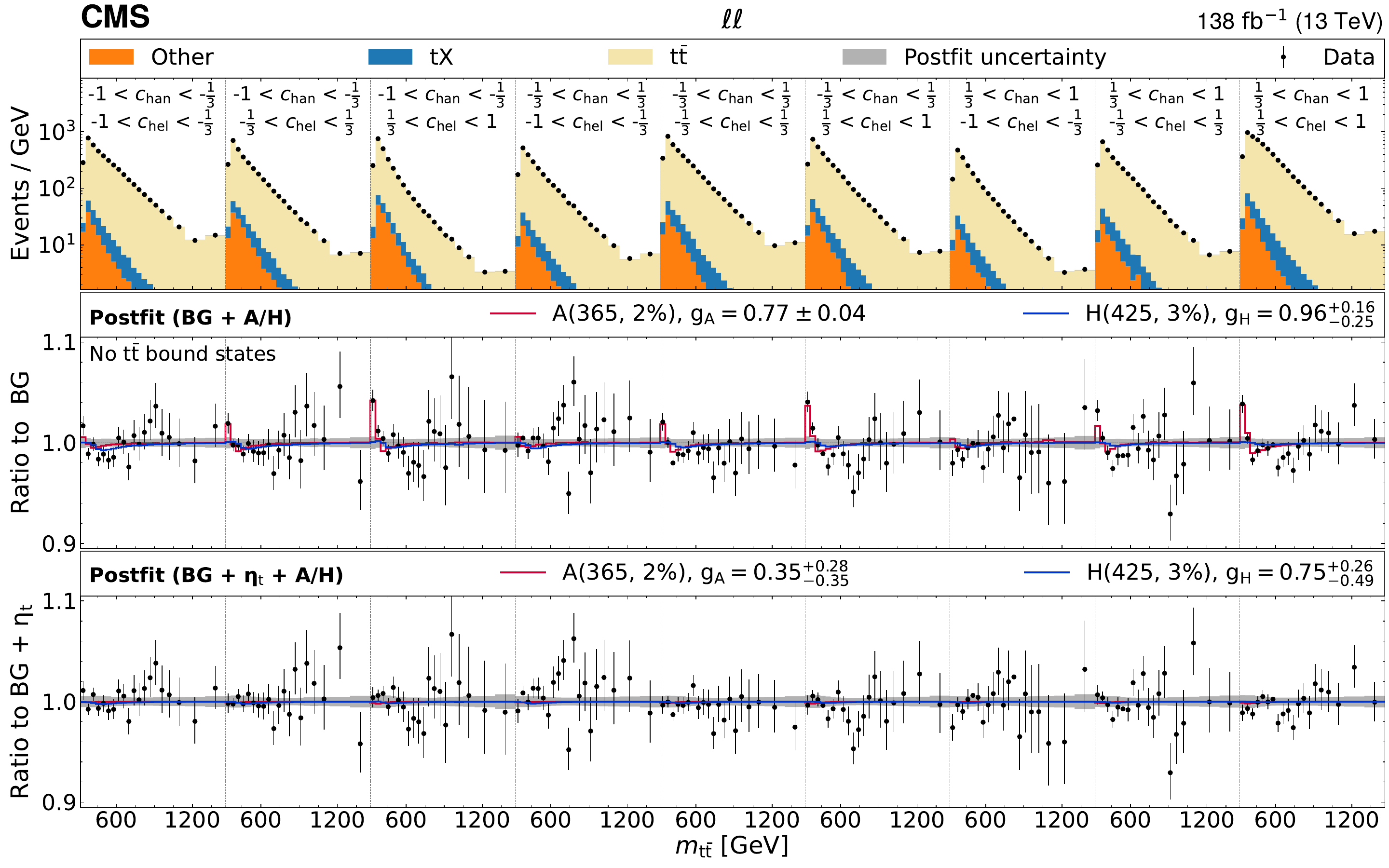}
\caption{%
    Observed and expected \mtt distribution in \chel and \chan bins, shown for the \ttll channel summed over lepton flavors and eras.
    Notations as in Fig.~\ref{fig:combined_l3j}.
}
\label{fig:combined_ll}
\end{figure*}

The expected and observed distributions are shown after the fit in Figs.~\ref{fig:combined_l3j}--\ref{fig:combined_ll} for the three channels considered.
In the middle panels, where no \etat contribution is included,
a deviation from the background prediction can be seen at low values of \mtt.

\begin{figure*}[!tp]
\centering
\includegraphics[width=0.87\textwidth]{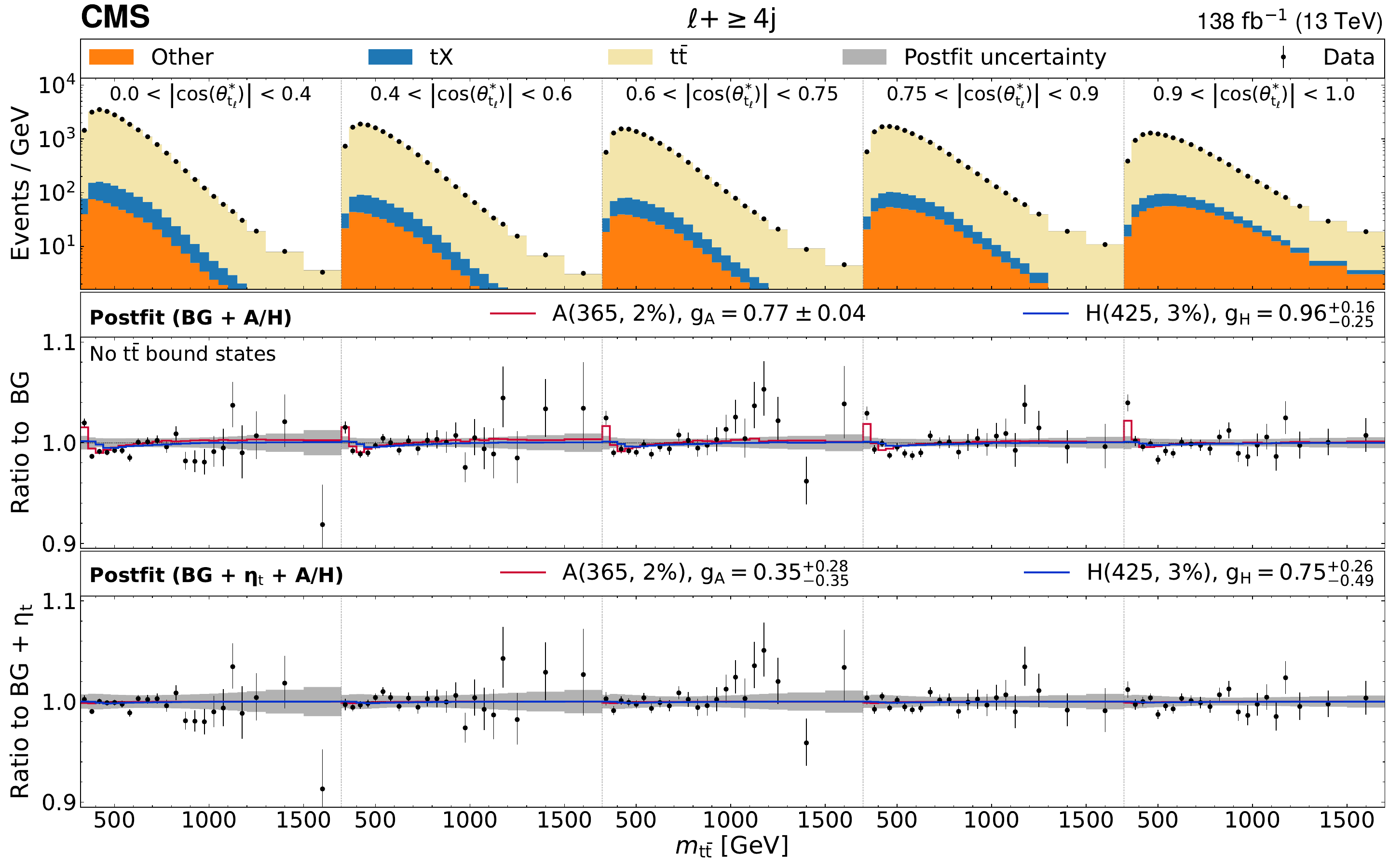}
\caption{%
    Observed and expected \mtt distribution in \acpTTT bins, shown for the \ttlfj channel summed over lepton flavors and eras.
    Notations as in Fig.~\ref{fig:combined_l3j}.
}
\end{figure*}

The shown fit is performed using the signal pair \pntA{365}{2\%}{+}\pntH{425}{3\%}, using the notation introduced in Section~\ref{sec:samples}, which corresponds to the highest observed local significance.
To find this signal pair, the local significance of an \pAandH boson pair is estimated using the square root of the value of the test statistic from Eq.~\eqref{eq:dnll} when fixing $\gphitt=0$, \ie, comparing the case of zero \pAandH contribution to the one that best describes the data, in the background scenario without \etat~\cite{Cowan:2010js}.

It becomes apparent in Fig.~\ref{fig:combined_ll} (middle) that the contributions of \pAorH boson production at the best fit \gphitt values are dependent on \chel and \chan, highlighting their sensitivity to discriminate between the signals.
In general, \PSA boson production is favored by the data over \PSH boson production. Comparing \pntA{365}{2\%} and \pntH{365}{2\%}, corresponding to the best fit mass and width for single \pAorH boson signals, we find a difference in negative log-likelihood of $2\Delta\ln L\approx 53$, indicating a strong preference for the \CP-odd contribution.

For the lower panels of Figs.~\ref{fig:combined_l3j}--\ref{fig:combined_ll}, \etat production was included in the fit as additional background with the normalization being a freely floating parameter of the fit,
as discussed in Section~\ref{sec:statfit}.
In this case, the contributions for \PSA and \PSH boson production vanish, showing that the data prefers \etat production over \PSA or \PSH boson production. However, we note that the considered \pAorH masses are different from the \etat mass of 343\GeV, as described in Section~\ref{sec:samples}, and \etat and \PSA are thus not directly comparable.
A further difference between \etat and \pAorH is the inclusion of SM \ttbar--\pAorH interference, leading to peak-dip structures in \mtt, while \etat is modeled as a pure resonance~\cite{Djouadi:2024lyv}.

\begin{figure}[!tp]
\centering
\includegraphics[width=0.47\textwidth]{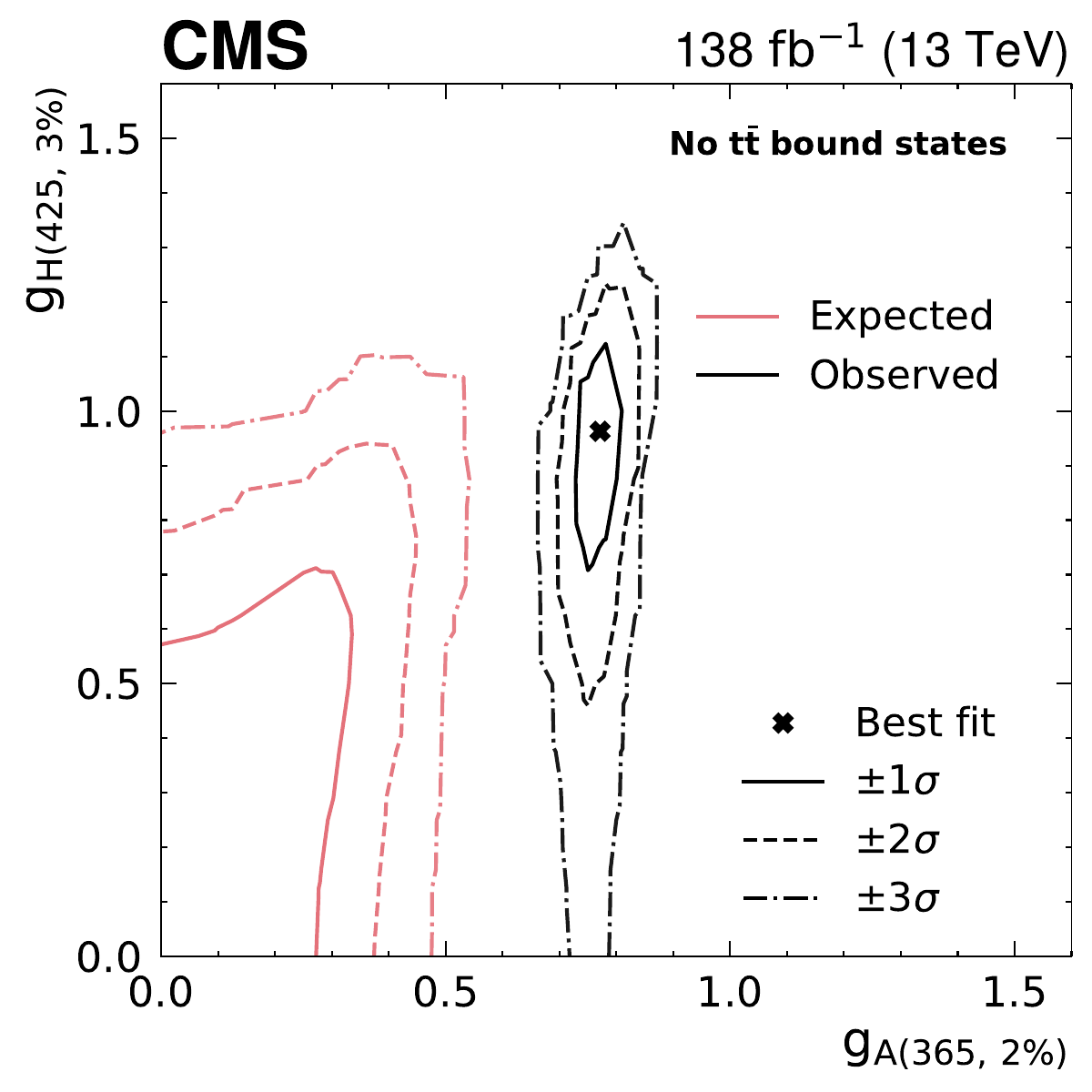}%
\hfill%
\includegraphics[width=0.47\textwidth]{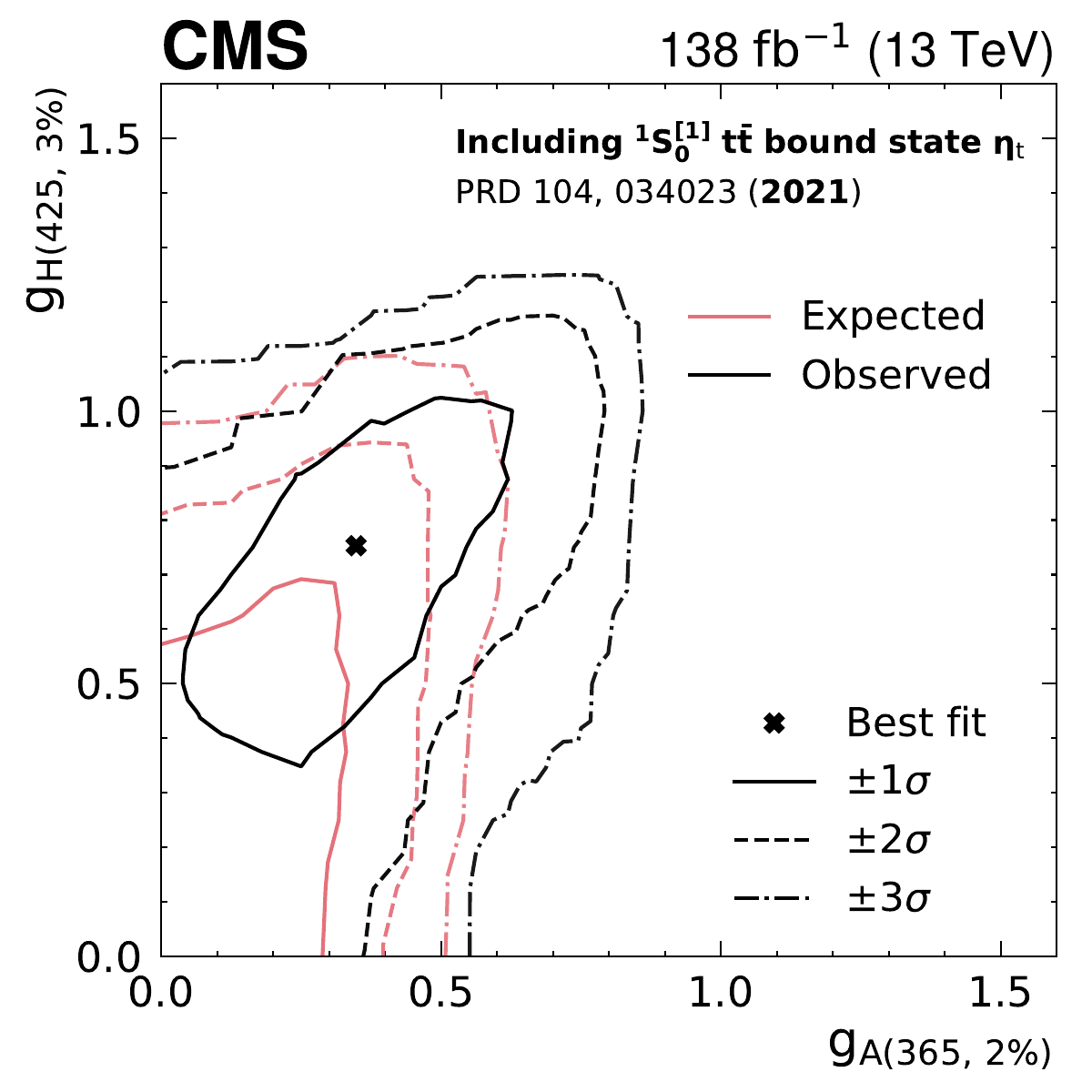}
\caption{%
    Frequentist 2D exclusion contours for \gphitt[\PSA] and \gphitt[\PSH] for the \pntA{365}{2\%}{+}\pntH{425}{3\%} signal point, in the background scenario excluding (\cmsLeft) and including (\cmsRight) \etat production.
    The expected and observed contours, evaluated with the Feldman--Cousins prescription~\cite{Feldman:1997qc,Cousins:1991qz}, are shown in black and pink, respectively, with different line styles denoting progressively higher {\CL}s. The regions outside of the contours are considered excluded.
}
\label{fig:limit_2D_sm_vs_eta}
\end{figure}

In addition, the Feldman--Cousins exclusion contours (as discussed in Section~\ref{sec:2Dinterp}) for the two scenarios are shown in Fig.~\ref{fig:limit_2D_sm_vs_eta}.
The expected contours are similar in shape, though the one in the background scenario including \etat (\cmsRight) is slightly wider.
This is due to the fact that in the regions of \gphitt[\PSA] relevant for the contours, the interference component of the signal dominates, effectively manifesting as a deficit of expected events. Since this occurs at higher \mtt compared to the enhancement predicted by \etat, the addition of \etat to the background does not significantly affect the expected exclusion in \gphitt[\PSA]. Furthermore, since \PSH and \etat can be distinguished based on \chel and \chan, the exclusion in \gphitt[\PSH] is not affected either.

The observed exclusion contours are significantly different for the two background scenarios.
If \etat production is not included as in Fig.~\ref{fig:limit_2D_sm_vs_eta} (\cmsLeft), the observed pseudoscalar-like excess in data manifests as a narrow strip of compatible \gphitt[\PSA] values significantly different from zero.
In contrast, the value of \gphitt[\PSH] for this parameters point is compatible with zero within three standard deviations (SDs).
This demonstrates the pseudoscalar nature of the excess.

In the \etat background scenario as presented in Fig.~\ref{fig:limit_2D_sm_vs_eta} (\cmsRight), the observed allowed values of both \gphitt[\PSA] and \gphitt[\PSH] are compatible with zero within two SDs, and the excess has vanished.

\subsection{Single \texorpdfstring{\PGF}{A/H} boson interpretation without \texorpdfstring{\etat}{toponium} in the background model}
\label{sec:ahsmtt_1D}

\begin{figure*}[!tp]
\centering
\includegraphics[width=0.42\textwidth]{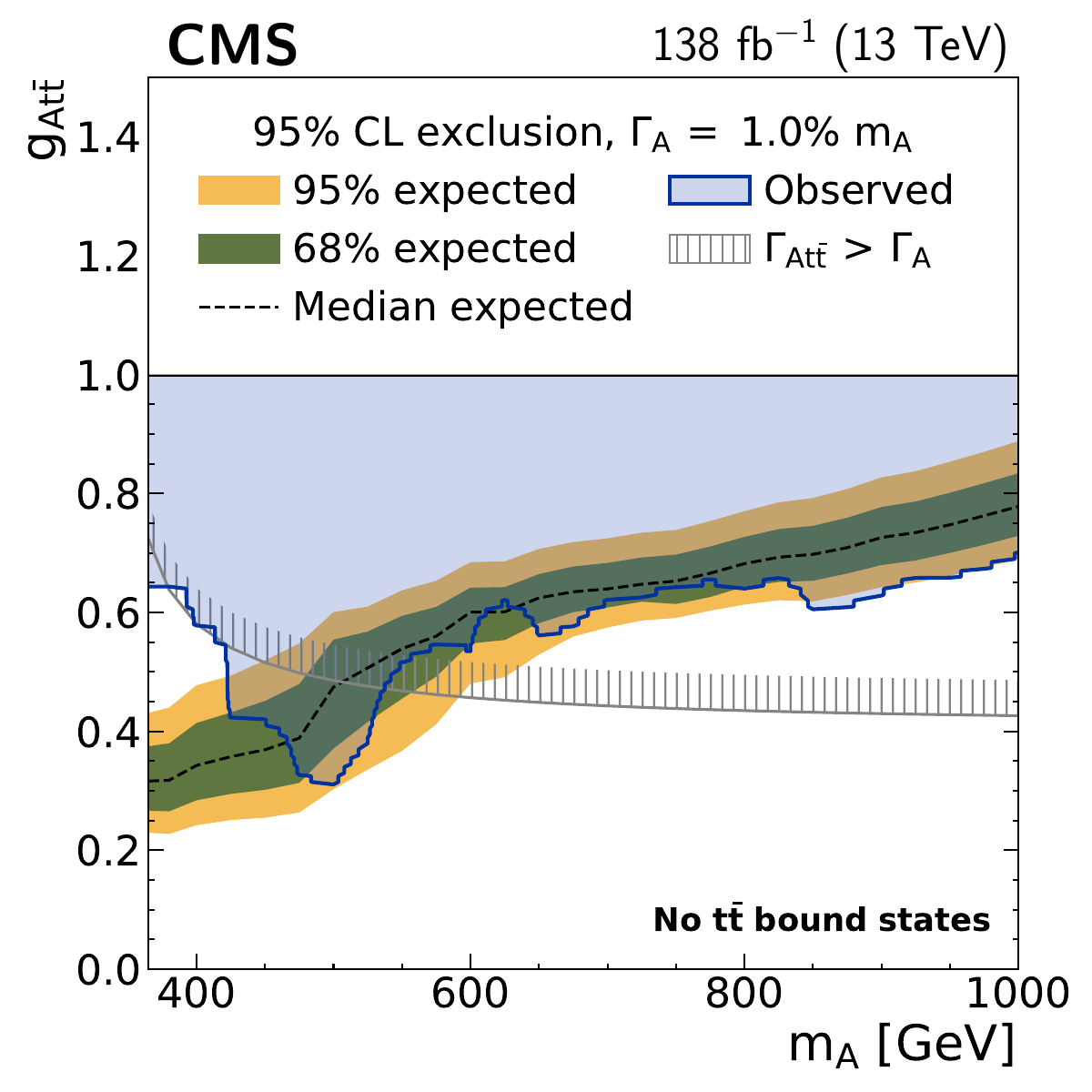}%
\hspace*{0.05\textwidth}%
\includegraphics[width=0.42\textwidth]{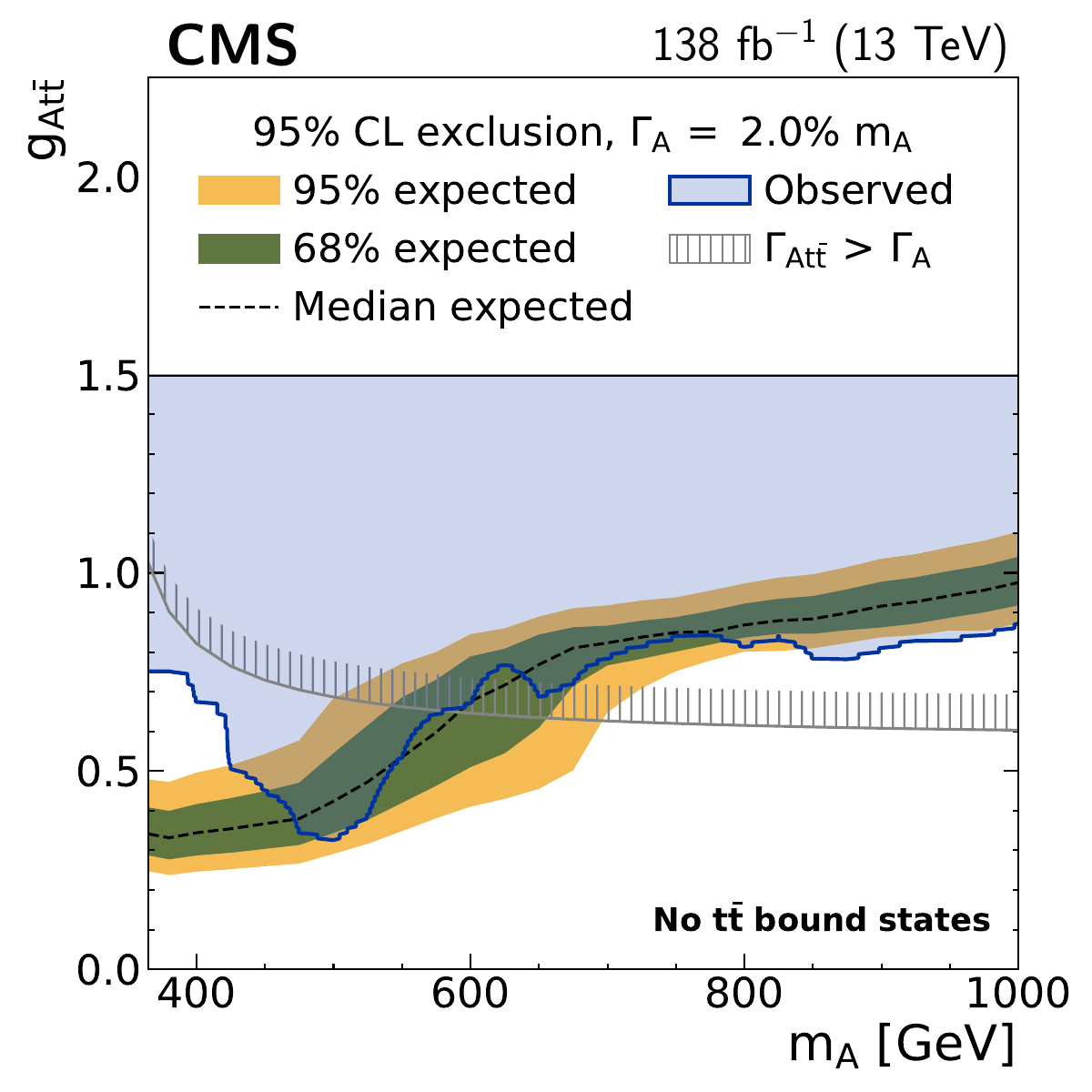} \\
\includegraphics[width=0.42\textwidth]{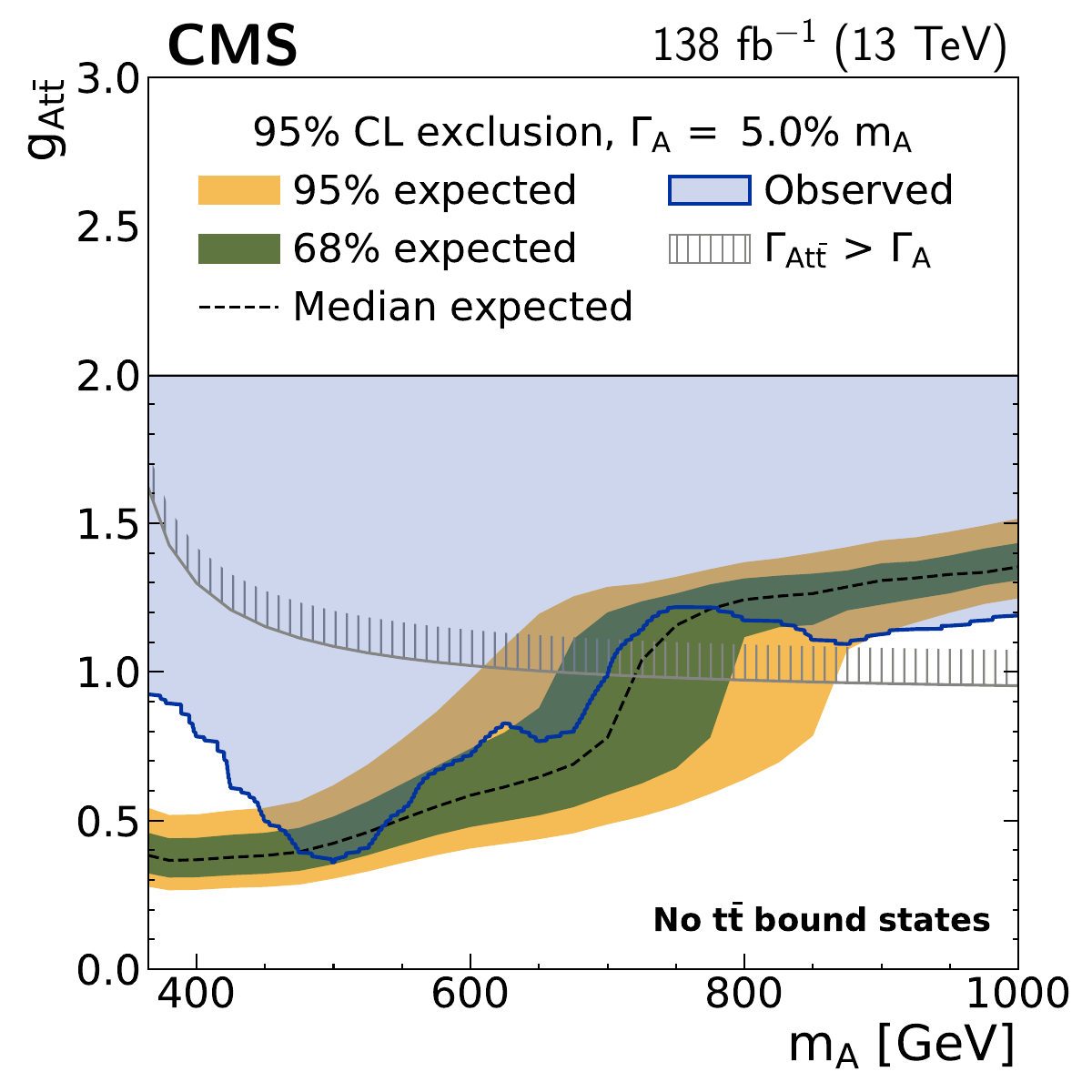}%
\hspace*{0.05\textwidth}%
\includegraphics[width=0.42\textwidth]{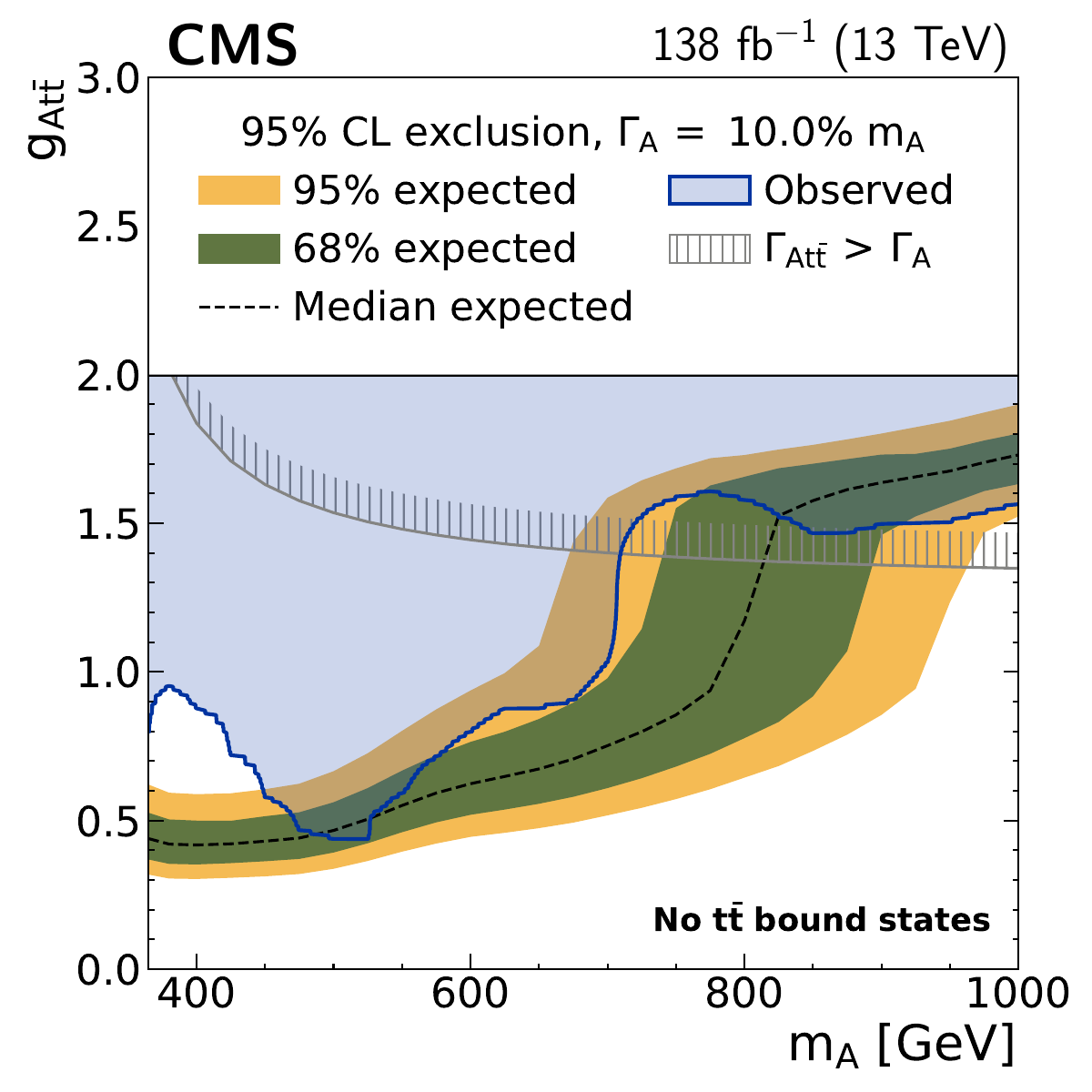}
\\
\includegraphics[width=0.42\textwidth]{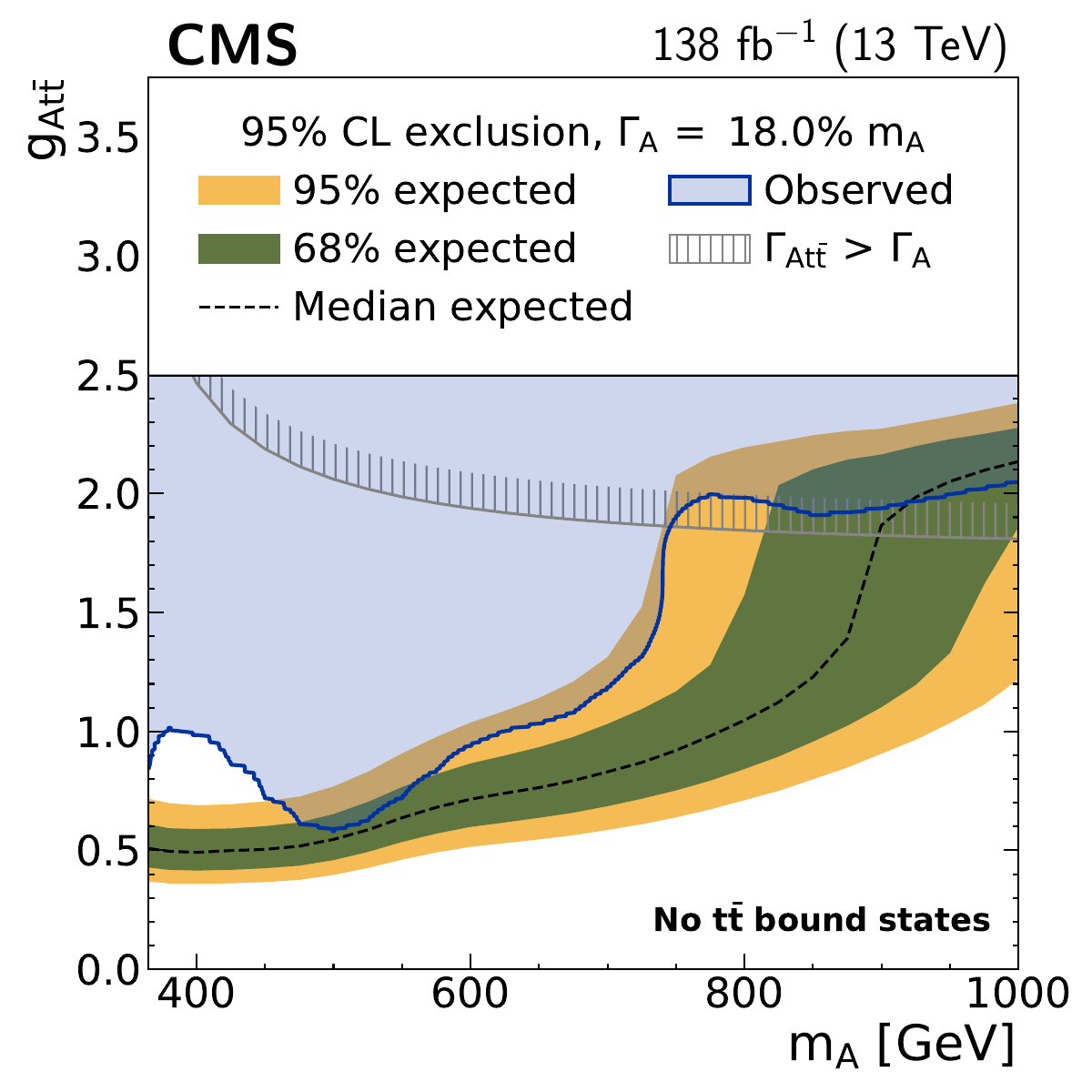}%
\hspace*{0.05\textwidth}%
\includegraphics[width=0.42\textwidth]{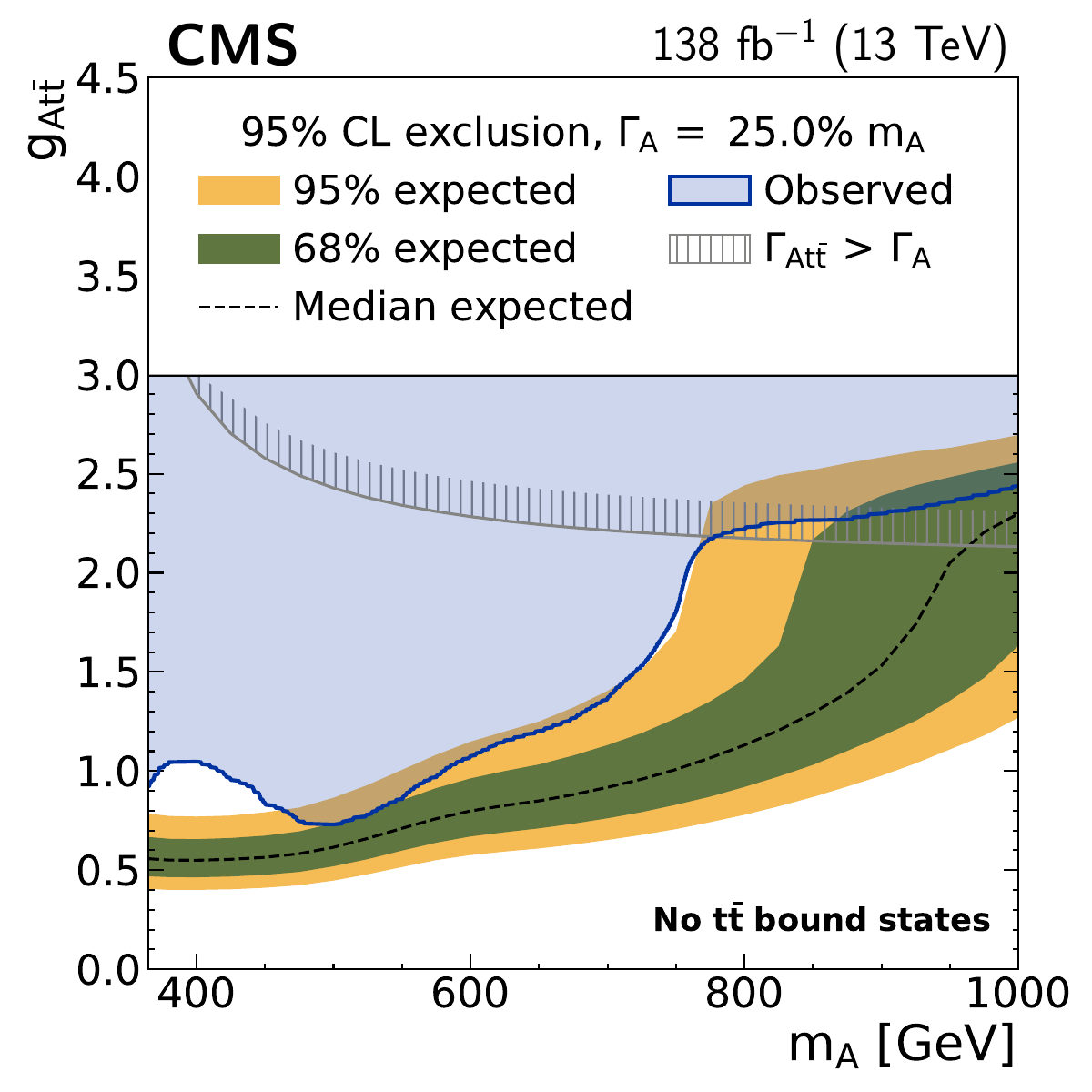}
\caption{%
    Model-independent constraints on \gphitt[\PSA] as functions of the \PSA boson mass in the background scenario without \etat contribution, for \Gpphirel of 1, 2, 5, 10, 18, and 25\% (from upper left to lower right).
    The observed constraints are indicated by the shaded blue area, bounded by the solid blue curve.
    The inner green and outer yellow bands indicate the regions containing 68 and 95\%, respectively, of the distribution of constraints expected under the background-only hypothesis.
    The unphysical region of phase space in which the partial width $\Gamma_{\PSA\to\ttbar}$ becomes larger than the total width of the \PSA boson is indicated by the hatched line.
}
\label{fig:limit_1D_a_smtt}
\end{figure*}

\begin{figure*}[!p]
\centering
\includegraphics[width=0.42\textwidth]{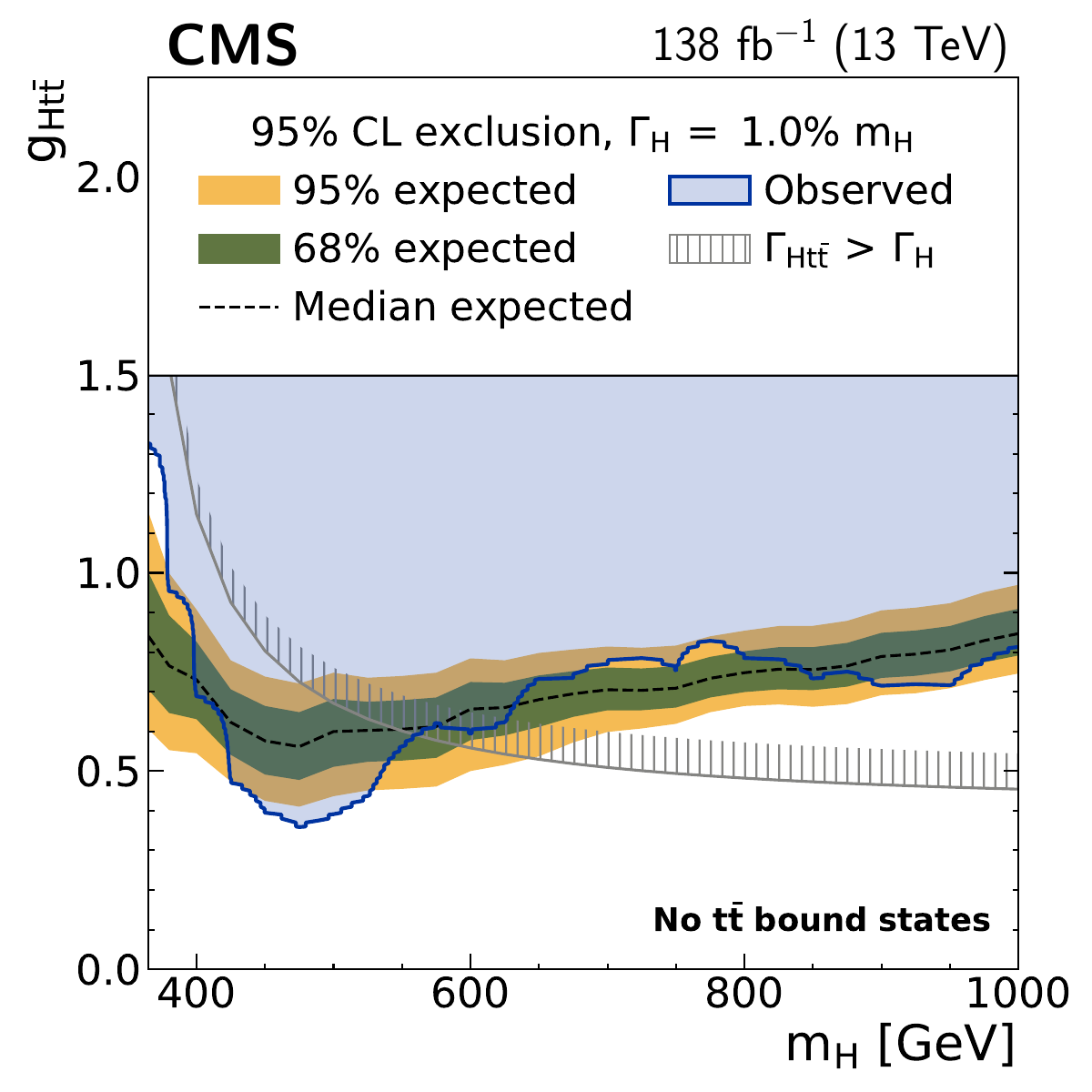}%
\hspace*{0.05\textwidth}%
\includegraphics[width=0.42\textwidth]{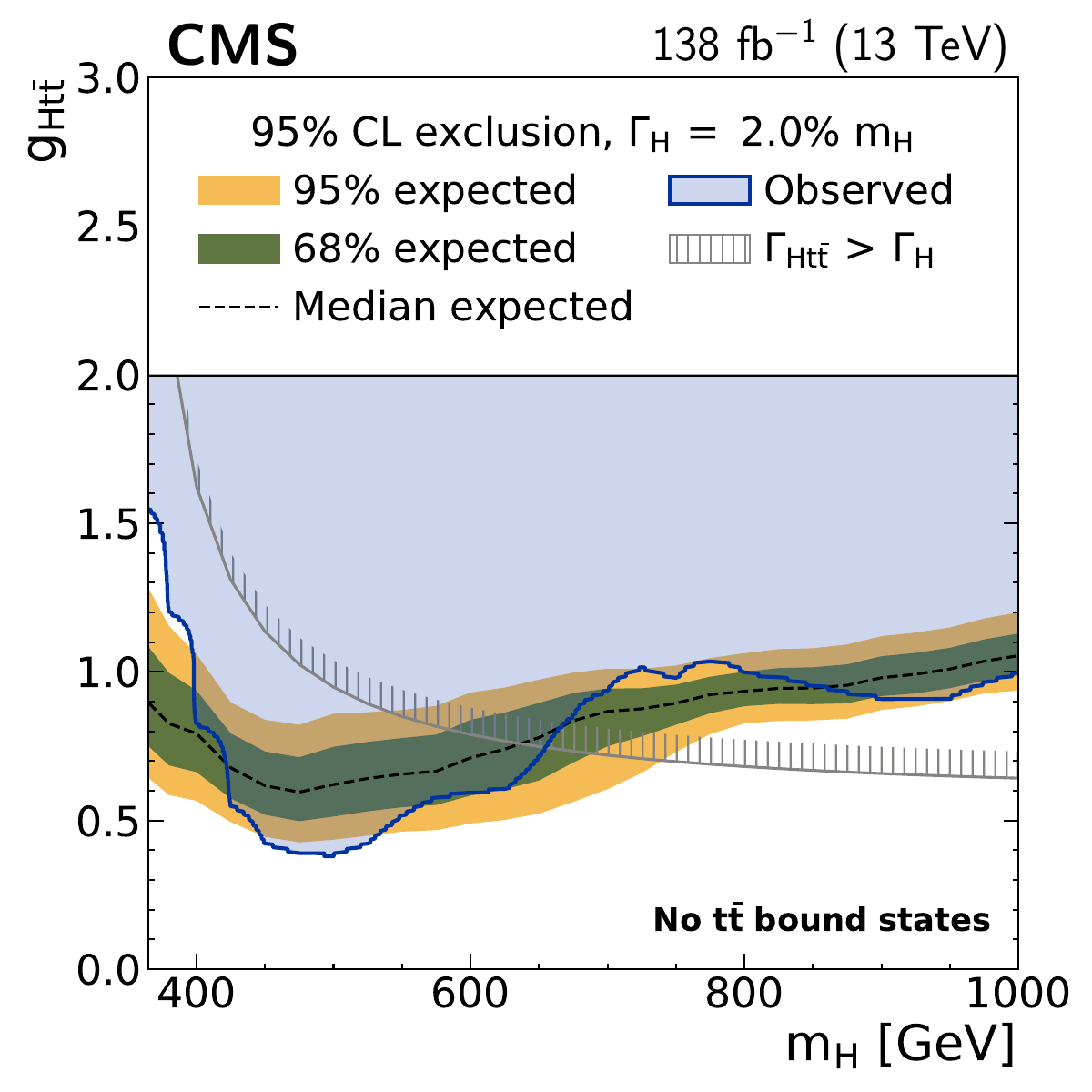} \\
\includegraphics[width=0.42\textwidth]{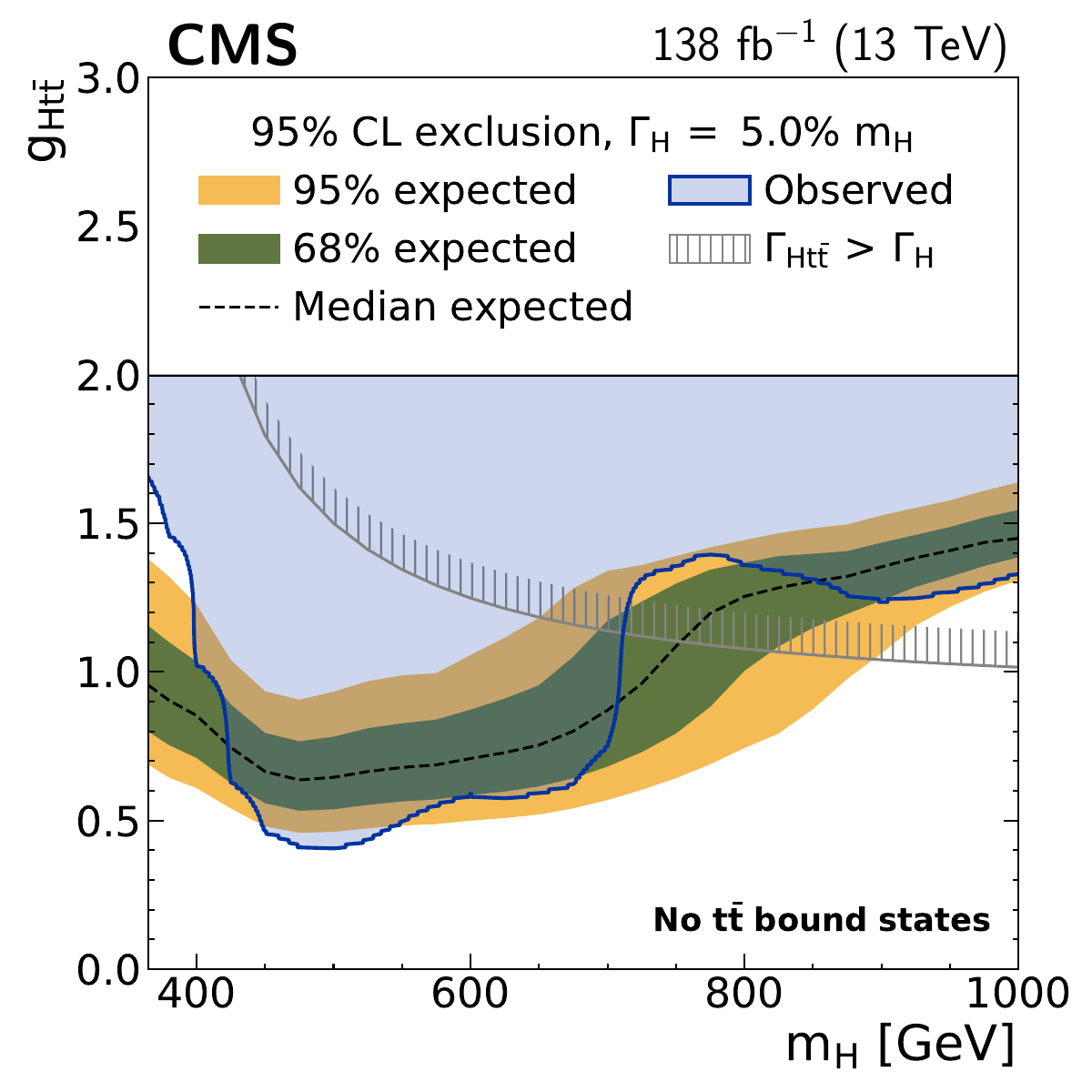}%
\hspace*{0.05\textwidth}%
\includegraphics[width=0.42\textwidth]{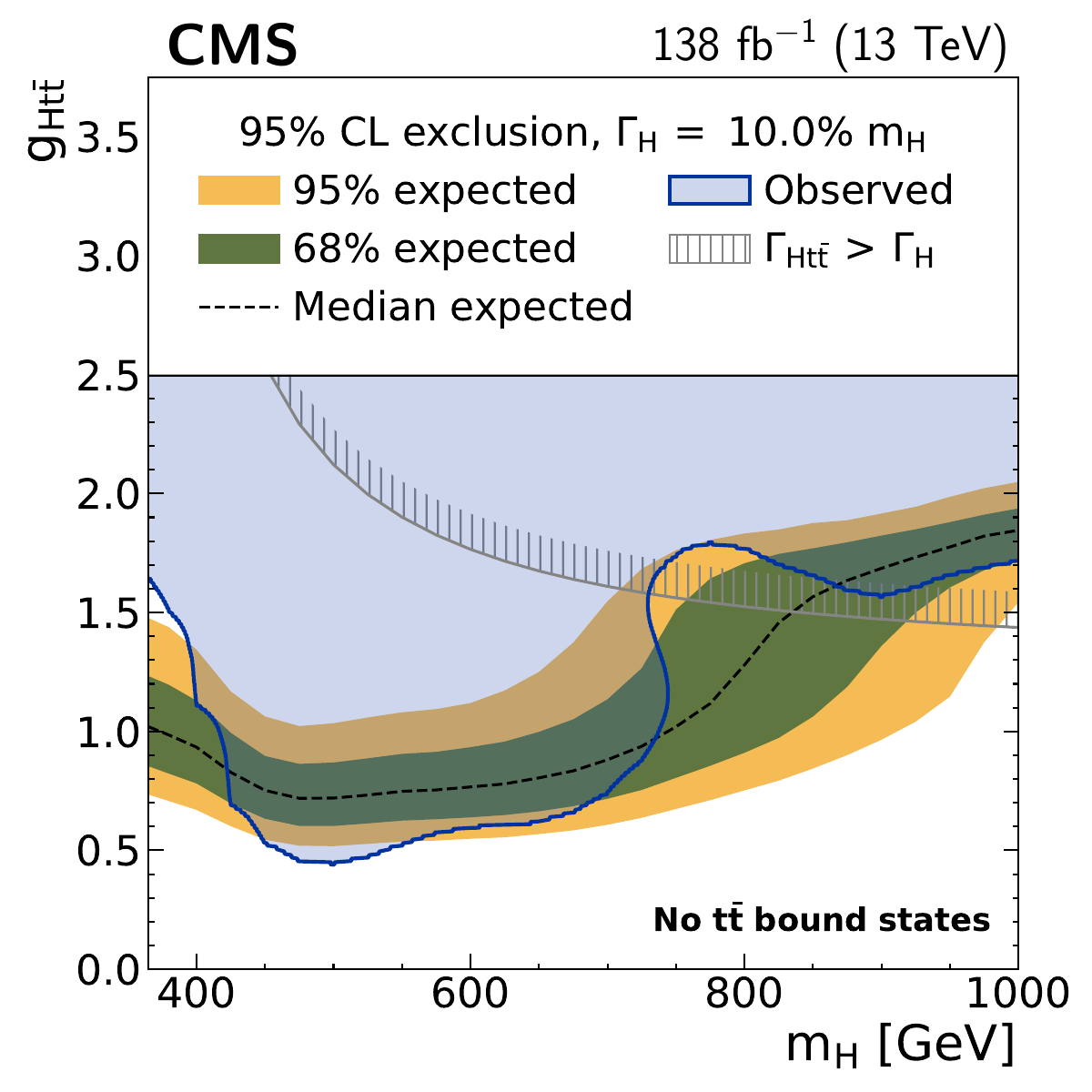}
\\
\includegraphics[width=0.42\textwidth]{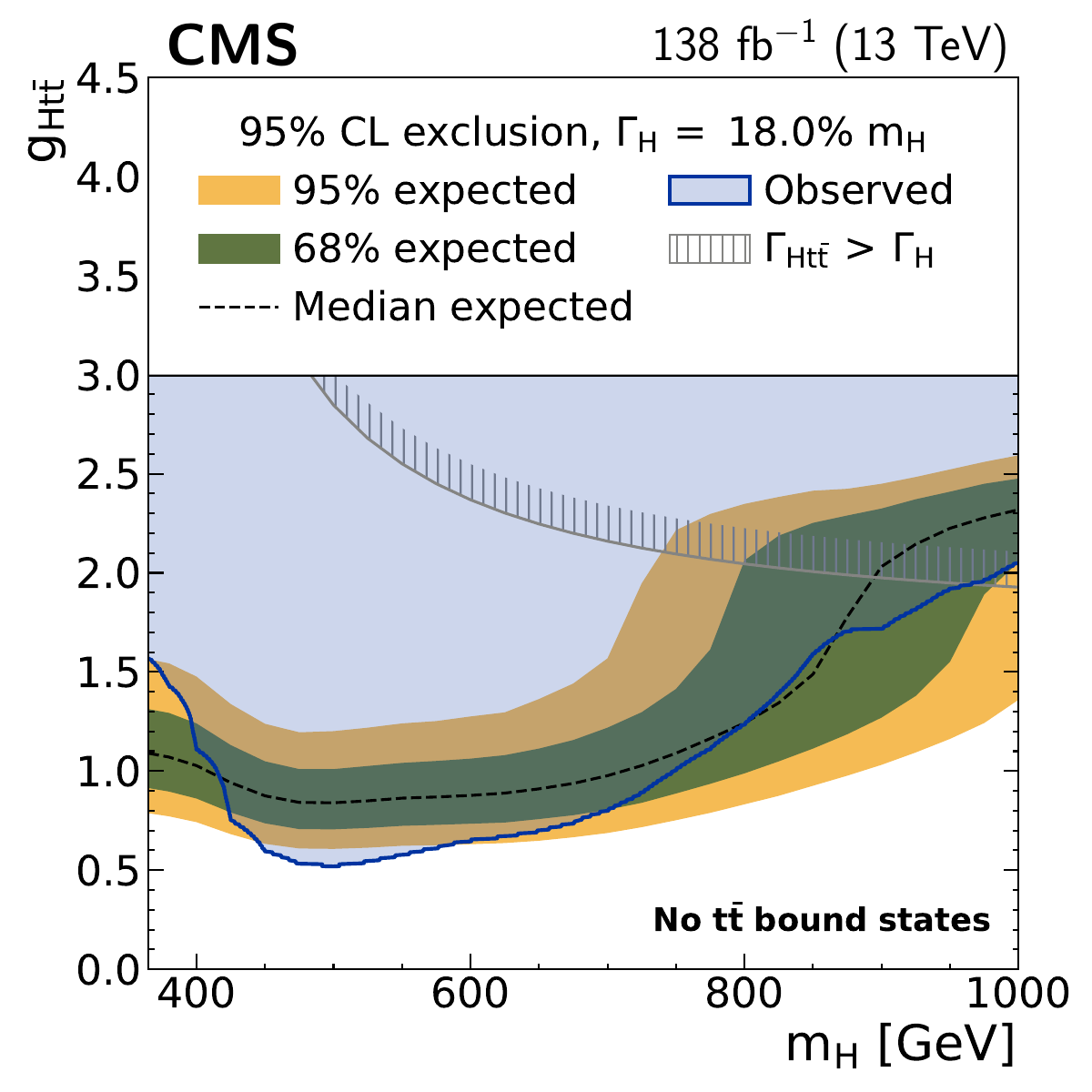}%
\hspace*{0.05\textwidth}%
\includegraphics[width=0.42\textwidth]{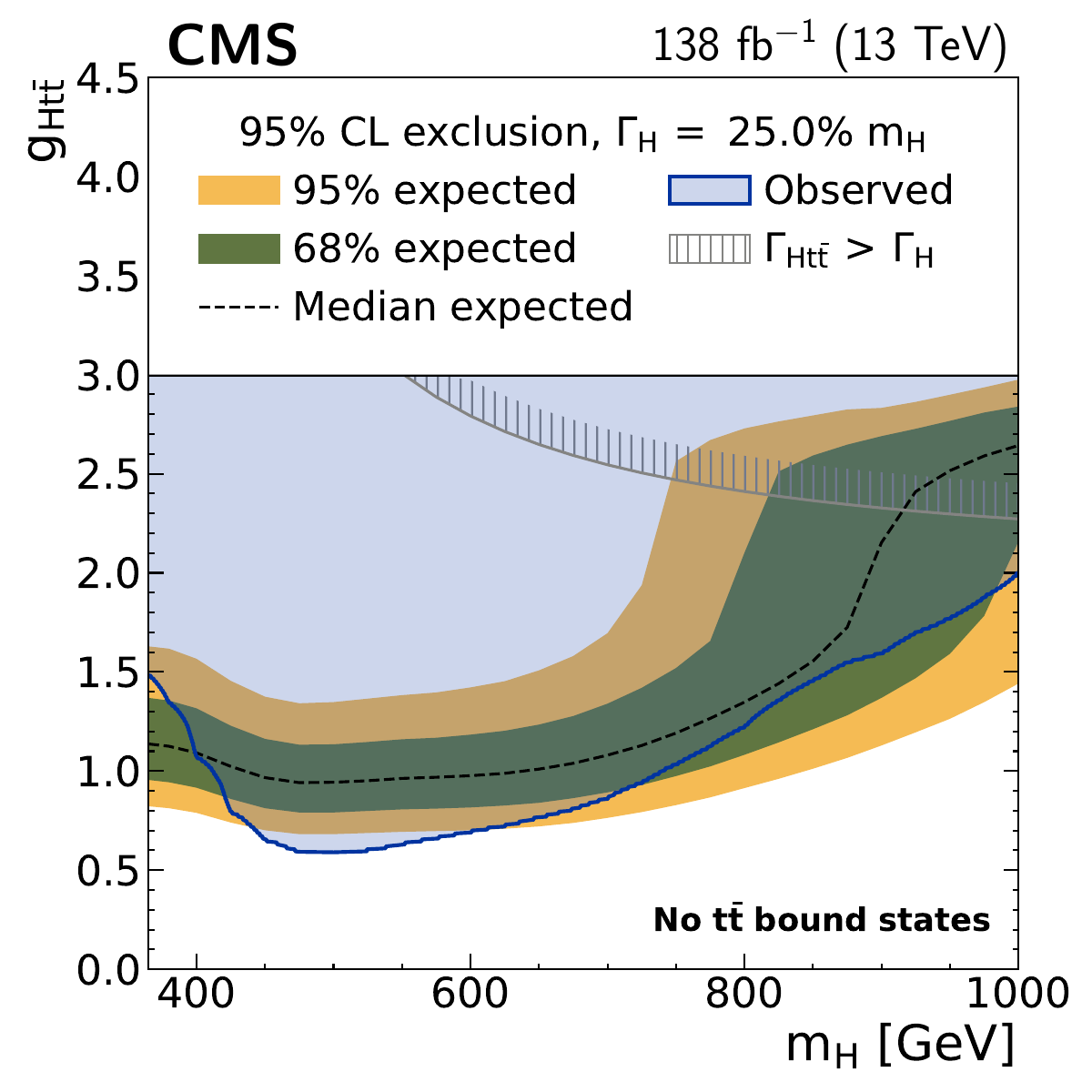}
\caption{%
    Model-independent constraints on \gphitt[\PSH] as functions of the \PSH boson mass in the background scenario without \etat contribution,
    shown in the same fashion as in Fig.~\ref{fig:limit_1D_a_smtt}.
}
\label{fig:limit_1D_h_smtt}
\end{figure*}

Combining the fit results for the 2D templates in $(\mtt,\acpTTT)$ of the \ttltj and \ttlfj channels (as discussed in Section~\ref{sec:singlelepanalysis}), with the results derived from the 3D templates in $(\mtt,\chel,\chan)$ of the \ttll channels (as discussed in Section~\ref{sec:dilepanalysis}), for all lepton flavors and eras, upper exclusion limits on \gAtt and \gHtt at the 95\% \CL are presented in Figs.~\ref{fig:limit_1D_a_smtt}--\ref{fig:limit_1D_h_smtt}, for the background scenario without \etat contribution, as functions of \mpphi for different assumptions on the \Gpphirel.
The expected constraints on \gphitt evolve in accordance with the signal cross section, as \pAorH boson mass and width values increase.
The relatively sharper decline in sensitivity for \pAorH bosons with $700<\mpphi<900\GeV$ and larger \Gpphi is due to cancellations in the cross sections for the resonance and interference signal components.

The expected constraints on \gphitt obtained in this analysis improve upon
the previous results presented in
Ref.~\cite{CMS:2019pzc}, which were based on a smaller data set and a simpler analysis strategy.
In the \ttlj channel, the addition of the three jets category increases the statistical power of the analysis.
In the \ttll channel, the addition of \chan as an observable improved sensitivity of the search, particularly for \PSH bosons.

These improvements also result in significantly stronger observed constraints on \gphitt compared to previous results, across most of the mass and width values in both \CP scenarios. Interestingly, there is a significant deviation between observed and expected limits at low \mpphi values, for both the \PSA and \PSH boson interpretations. The largest differences are for \PSA boson signal hypotheses with narrow widths. The best fit to the data is achieved for the \pntA{365}{2\%} signal hypothesis, corresponding to the lowest generated \PSA boson mass value,
with an observed local significance over the background-only hypothesis in the background scenario without \etat contribution of more than five SDs.
This hypothesis corresponds to the lowest \mpphi value probed in this analysis.

\subsection{Single \texorpdfstring{\PGF}{A/H} boson interpretation with \texorpdfstring{\etat}{toponium} in the background model}
\label{sec:ahetat_1D}

The same limit extraction as in Section~\ref{sec:ahsmtt_1D} is repeated assuming single \PGF boson production as signal, but now including
\etat production to the fit as additional background, with the normalization treated as an unconstrained nuisance parameter, as outlined in Section~\ref{sec:statfit}.

\begin{figure*}[!tp]
\centering
\includegraphics[width=0.42\textwidth]{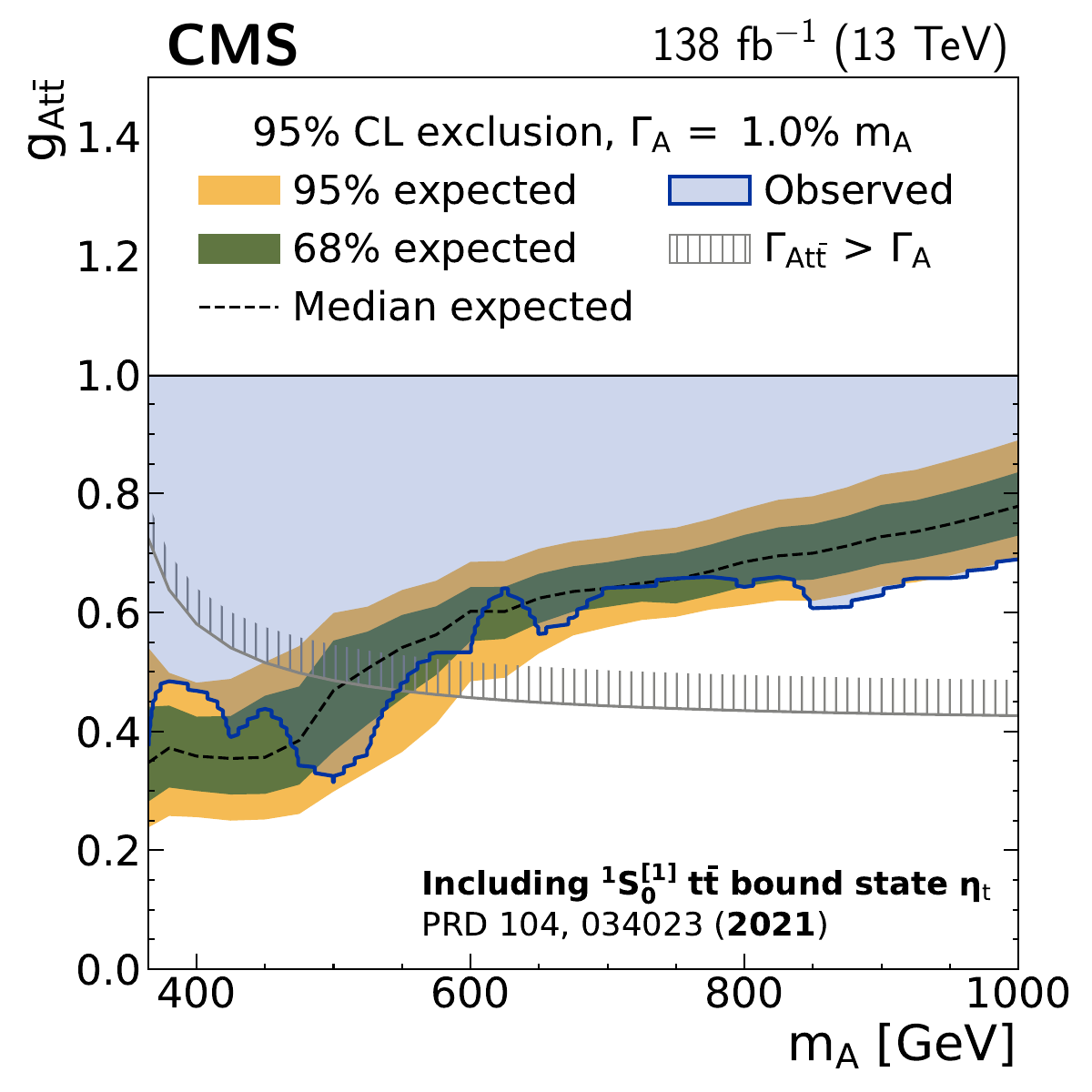}%
\hspace*{0.05\textwidth}%
\includegraphics[width=0.42\textwidth]{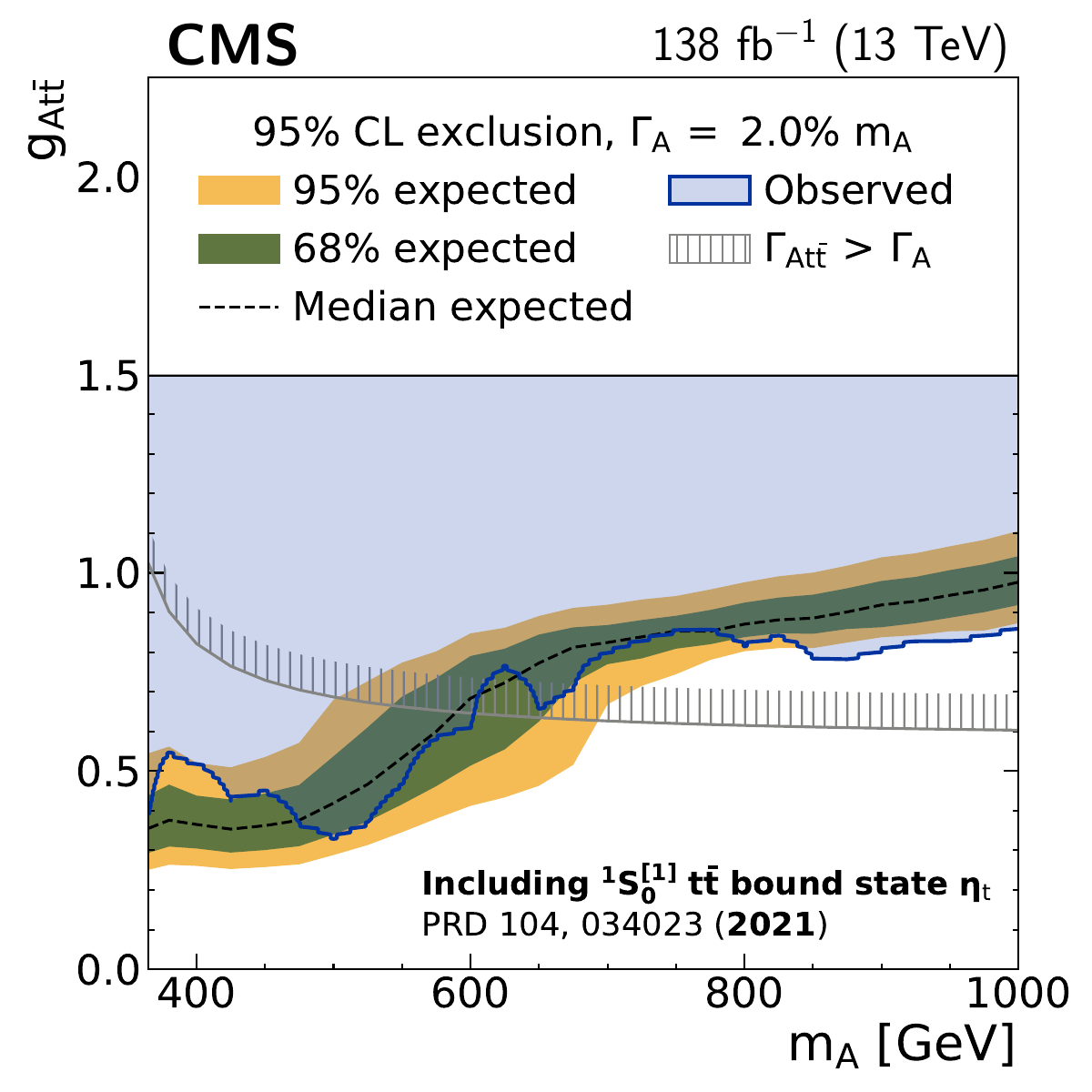} \\
\includegraphics[width=0.42\textwidth]{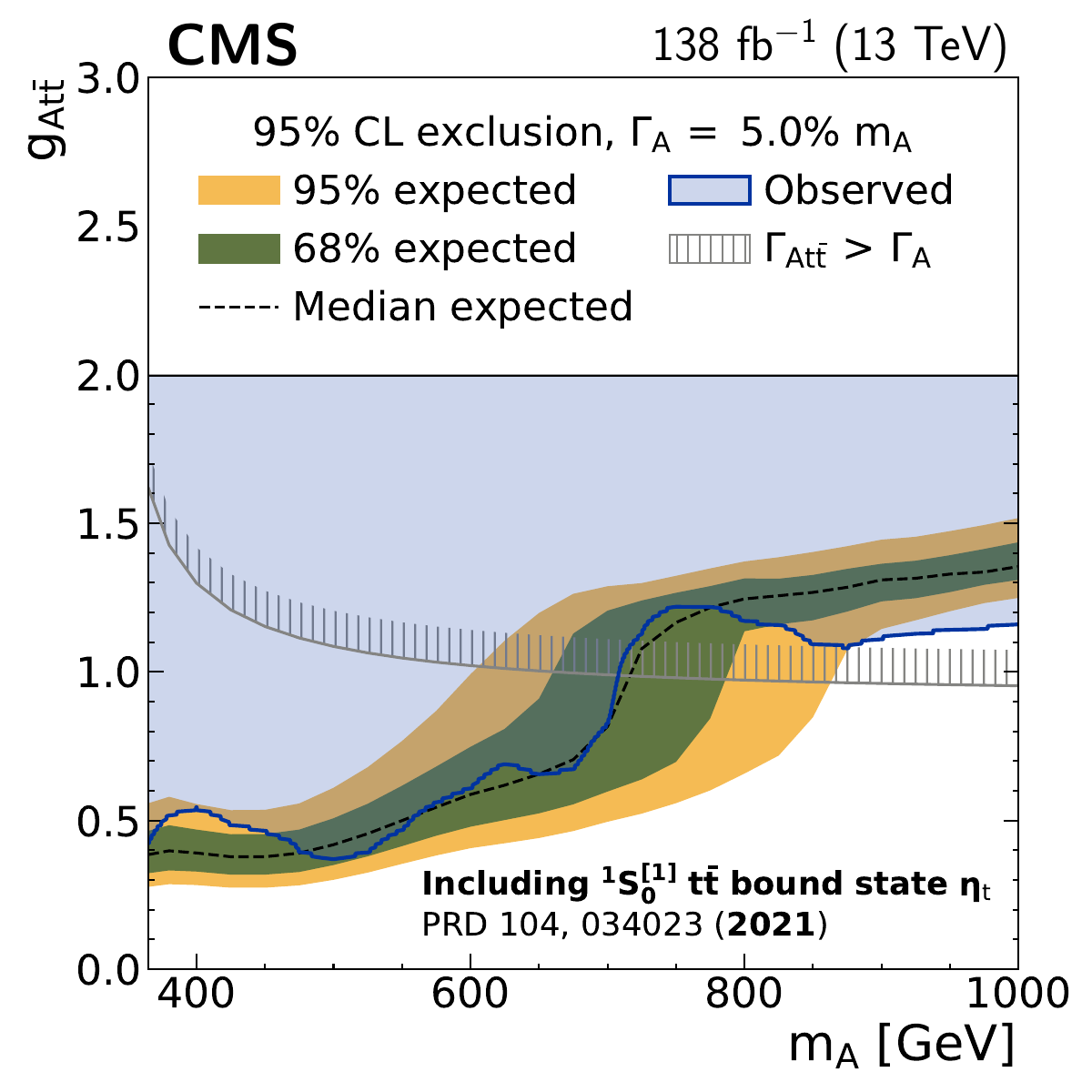}%
\hspace*{0.05\textwidth}%
\includegraphics[width=0.42\textwidth]{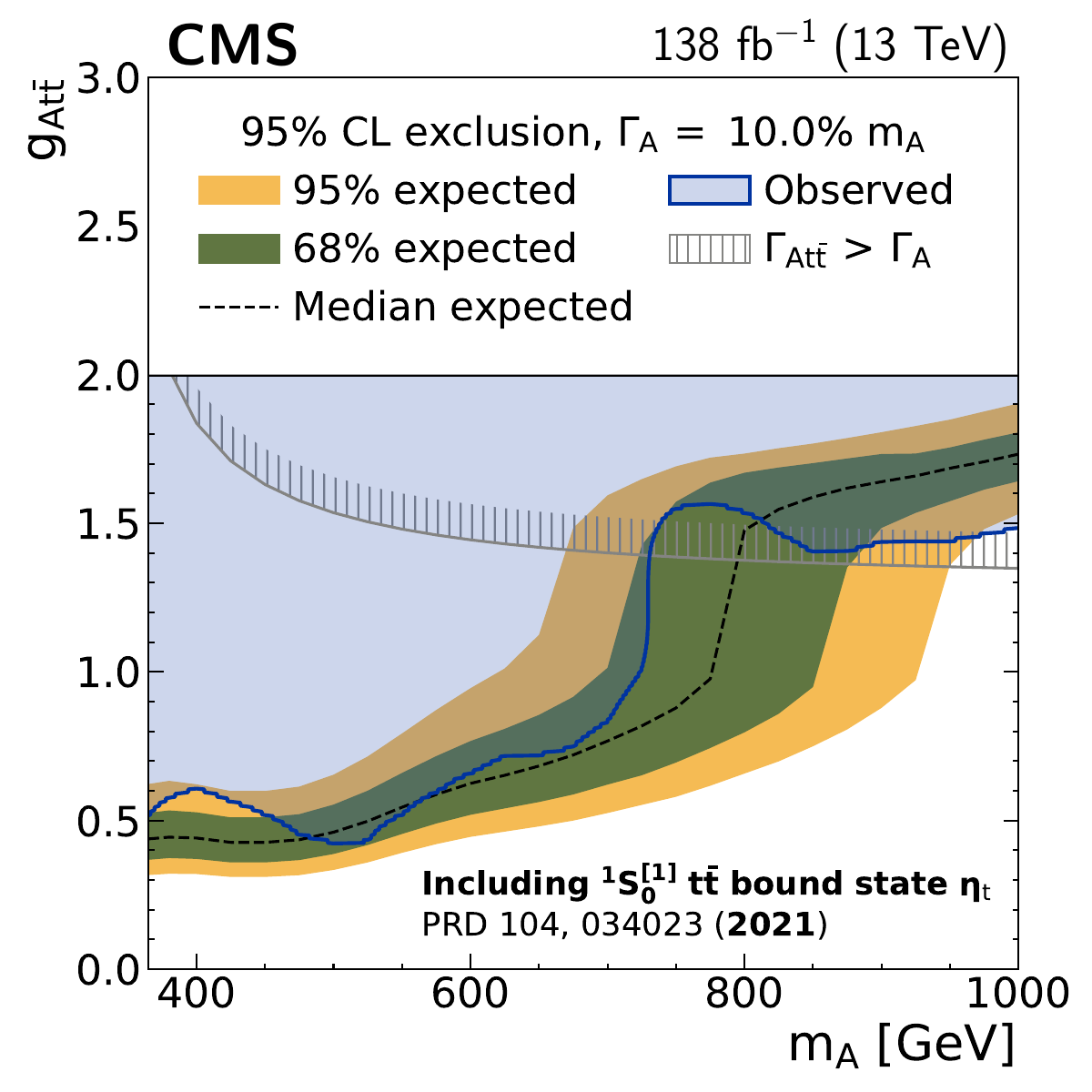} \\
\includegraphics[width=0.42\textwidth]{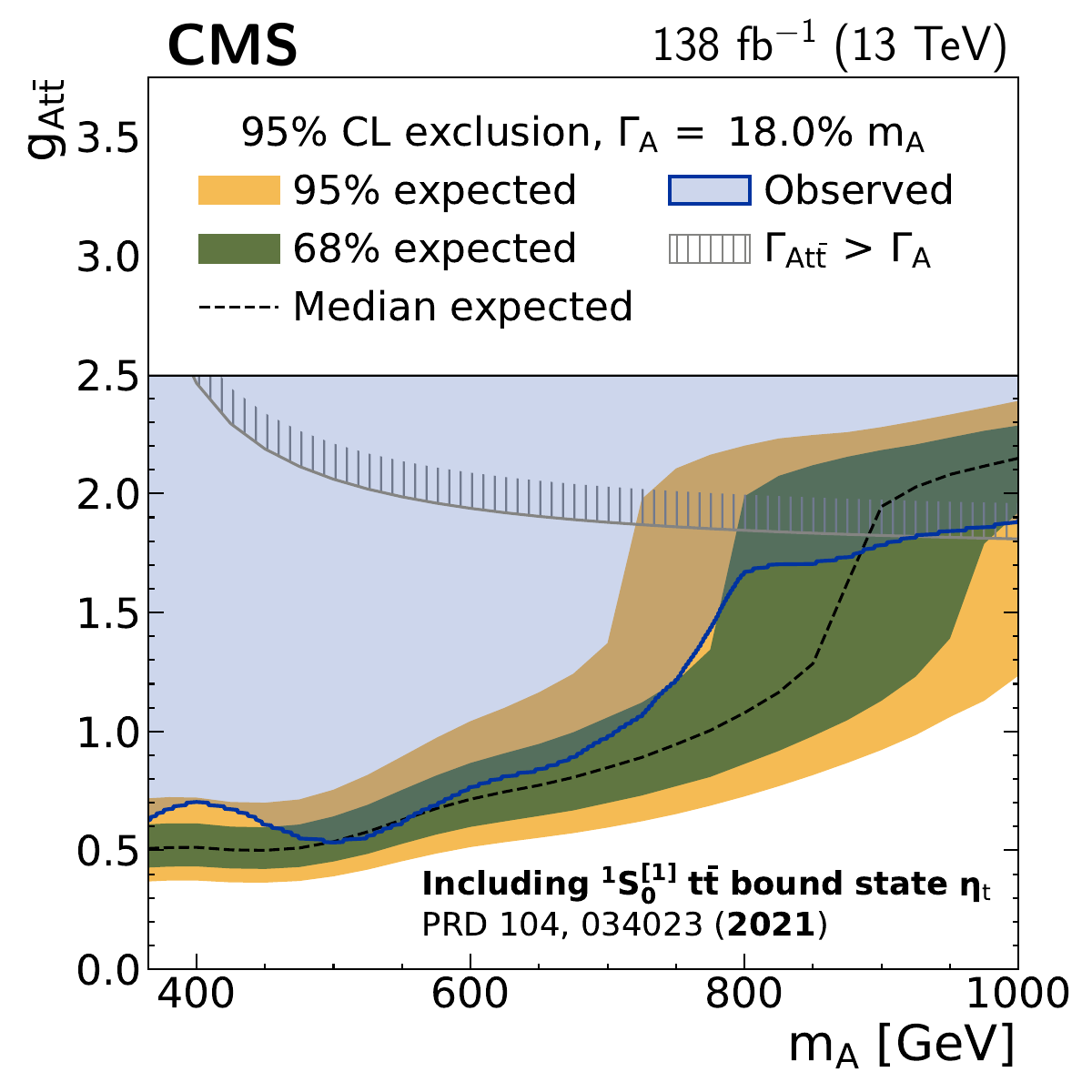}%
\hspace*{0.05\textwidth}%
\includegraphics[width=0.42\textwidth]{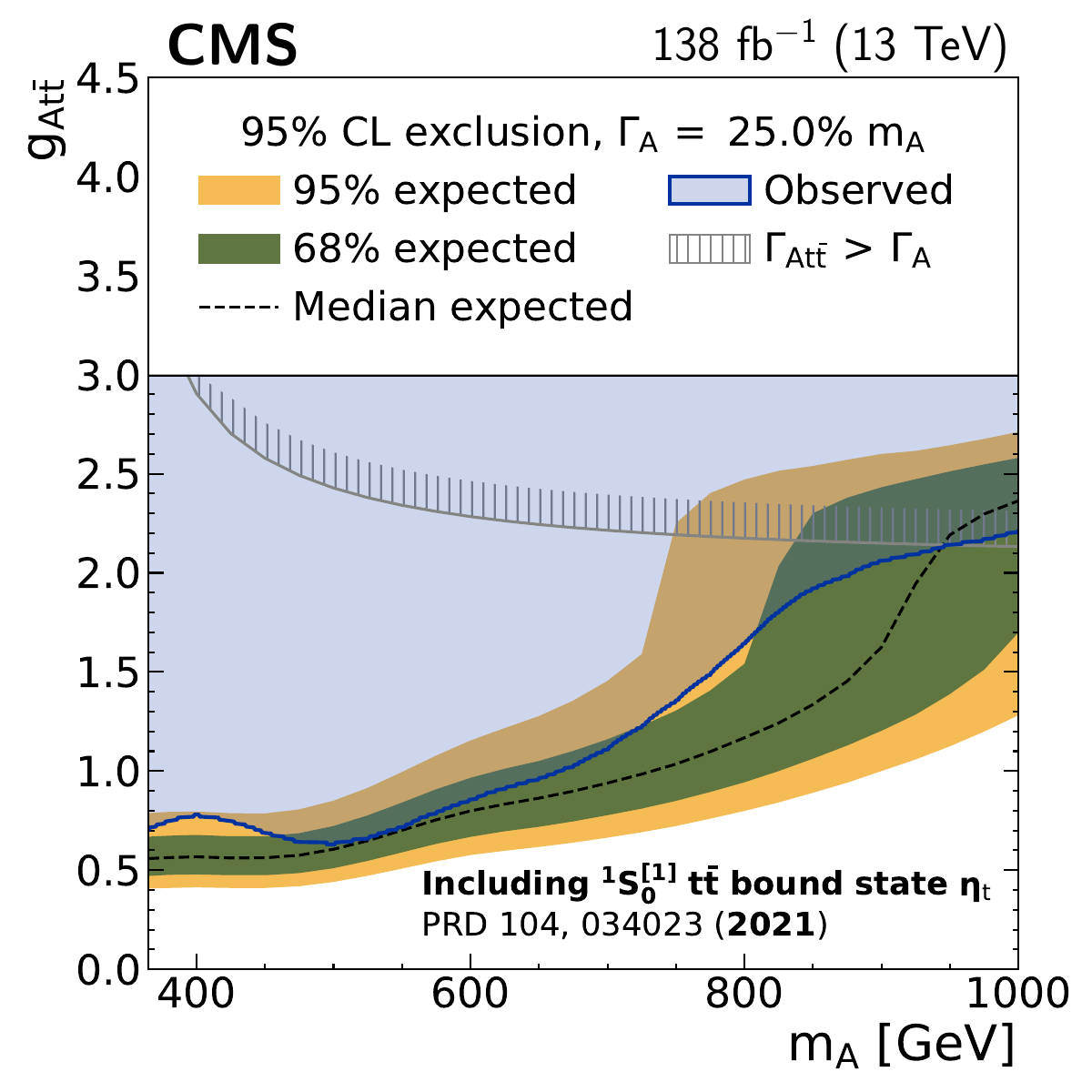}
\caption{%
    Model-independent constraints on \gphitt[\PSA] as functions of the \PSA boson mass in the background scenario with \etat contribution,
    shown in the same fashion as in Fig.~\ref{fig:limit_1D_a_smtt}.
}
\label{fig:limit_1D_a_etat}
\end{figure*}

\begin{figure*}[!tp]
\centering
\includegraphics[width=0.42\textwidth]{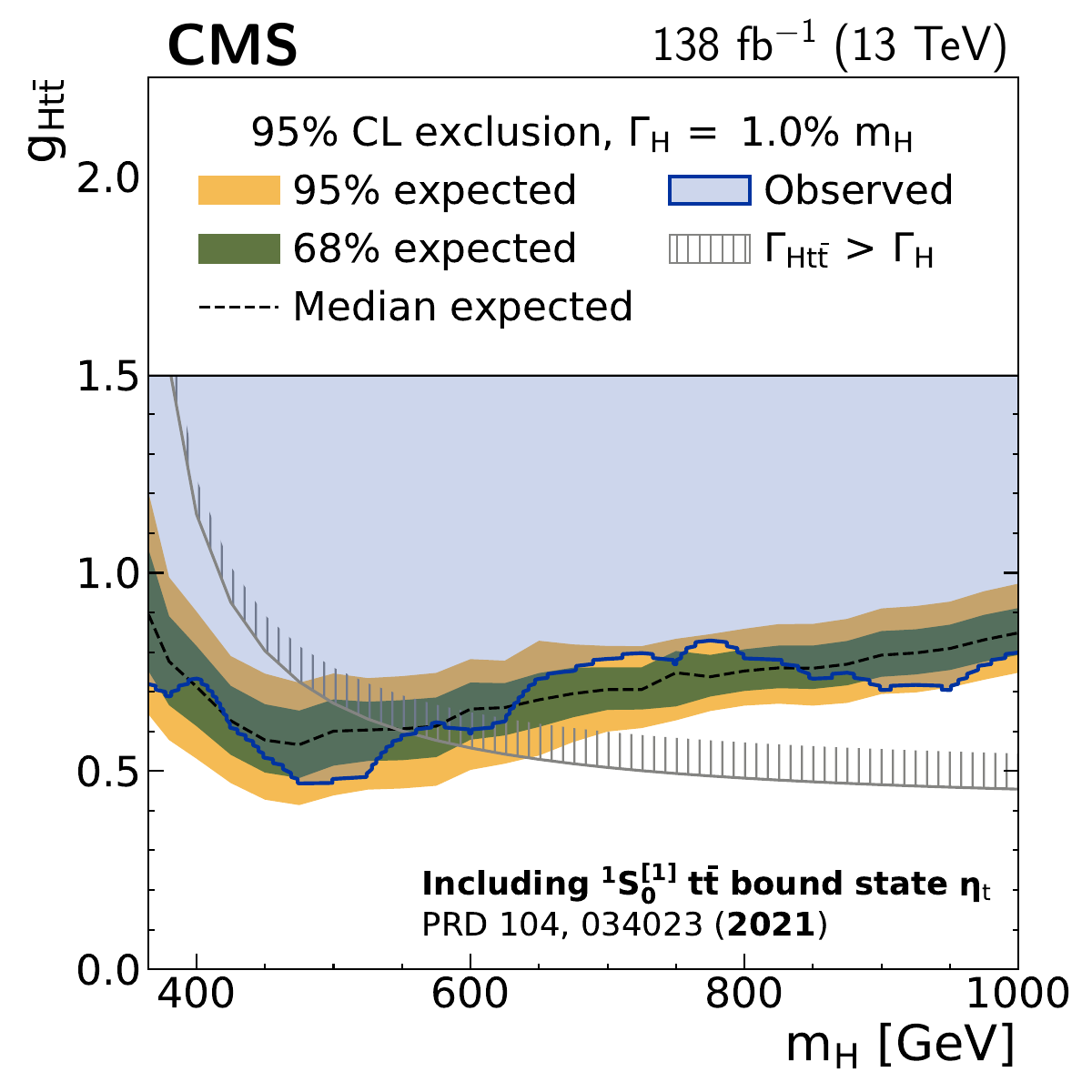}%
\hspace*{0.05\textwidth}%
\includegraphics[width=0.42\textwidth]{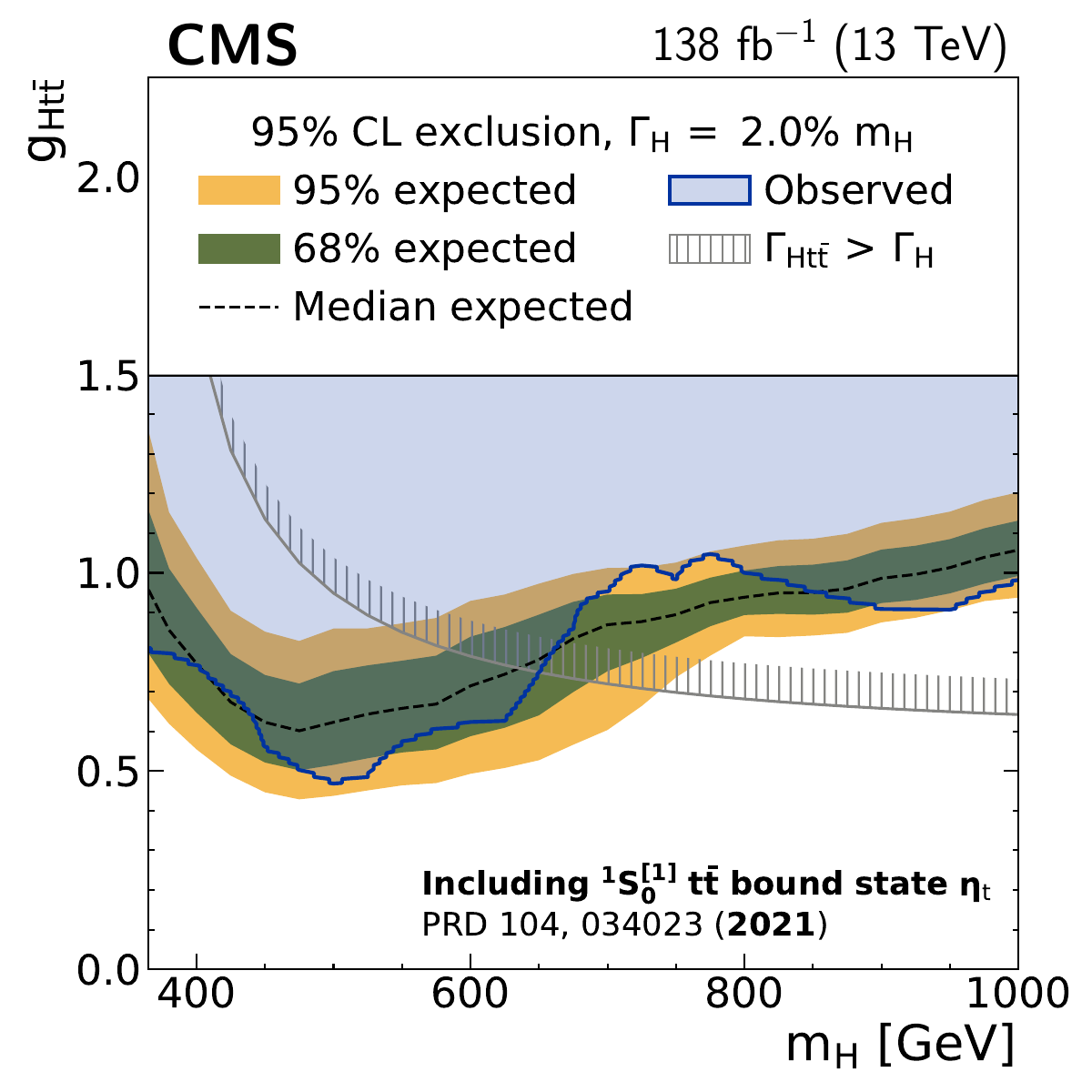} \\
\includegraphics[width=0.42\textwidth]{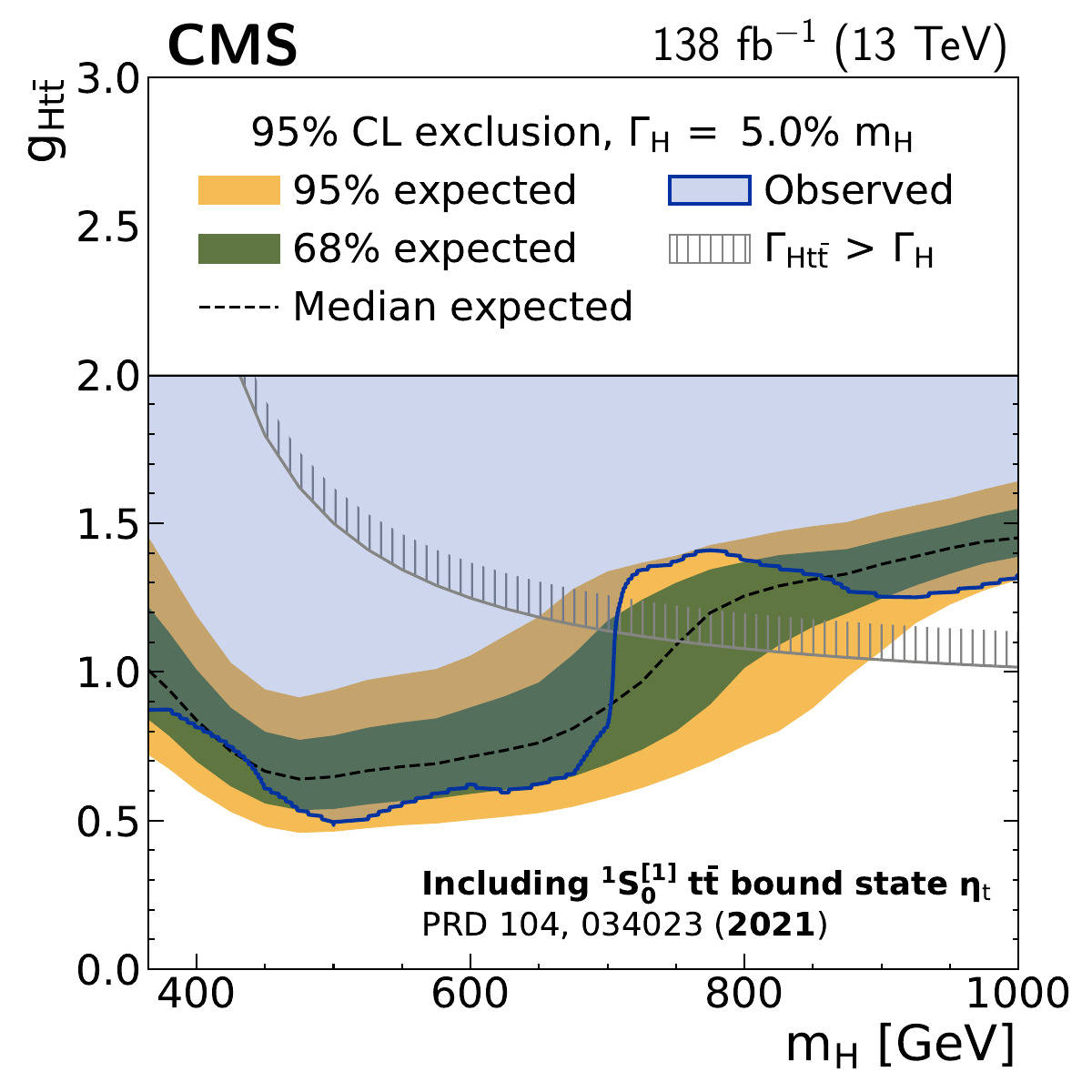}%
\hspace*{0.05\textwidth}%
\includegraphics[width=0.42\textwidth]{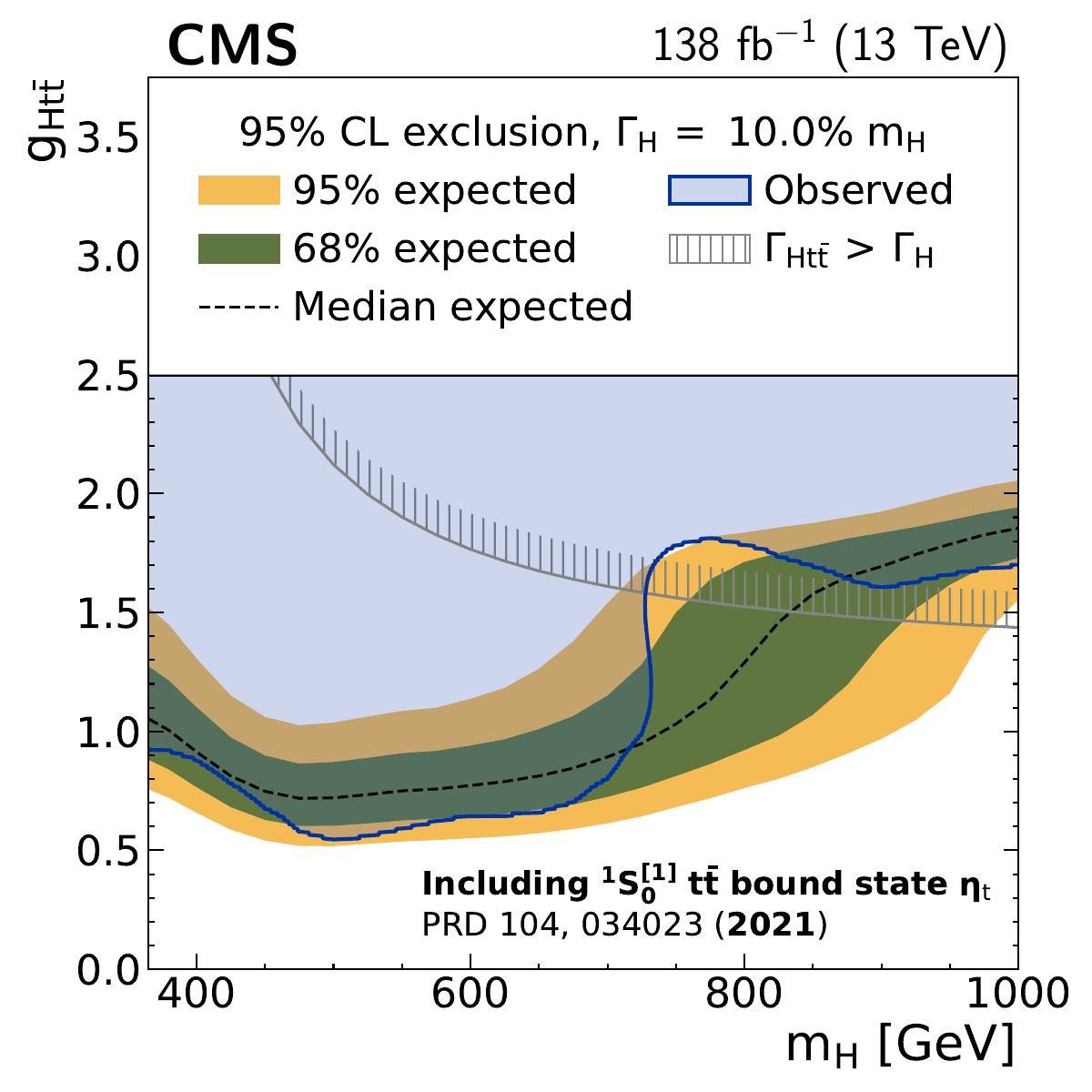} \\
\includegraphics[width=0.42\textwidth]{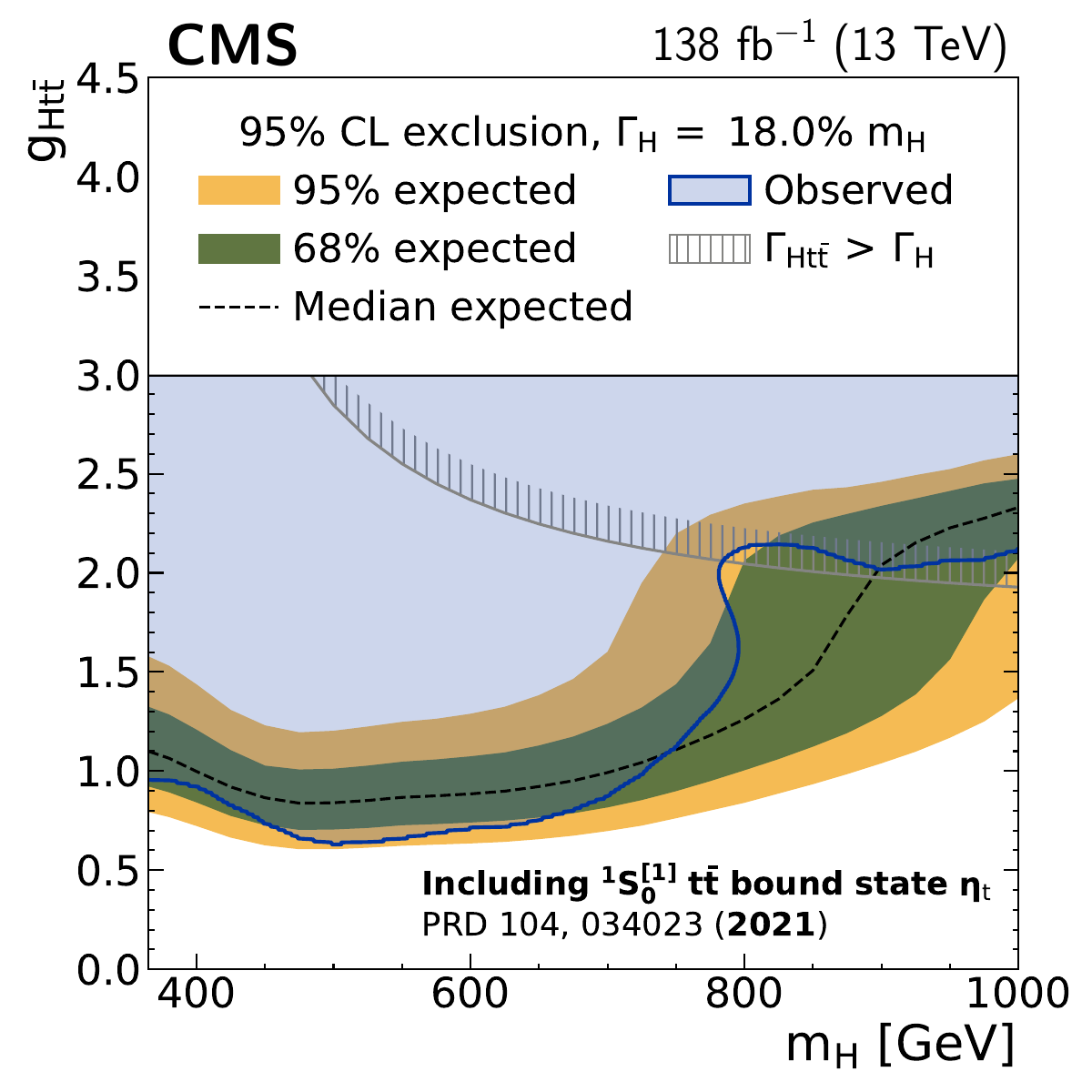}%
\hspace*{0.05\textwidth}%
\includegraphics[width=0.42\textwidth]{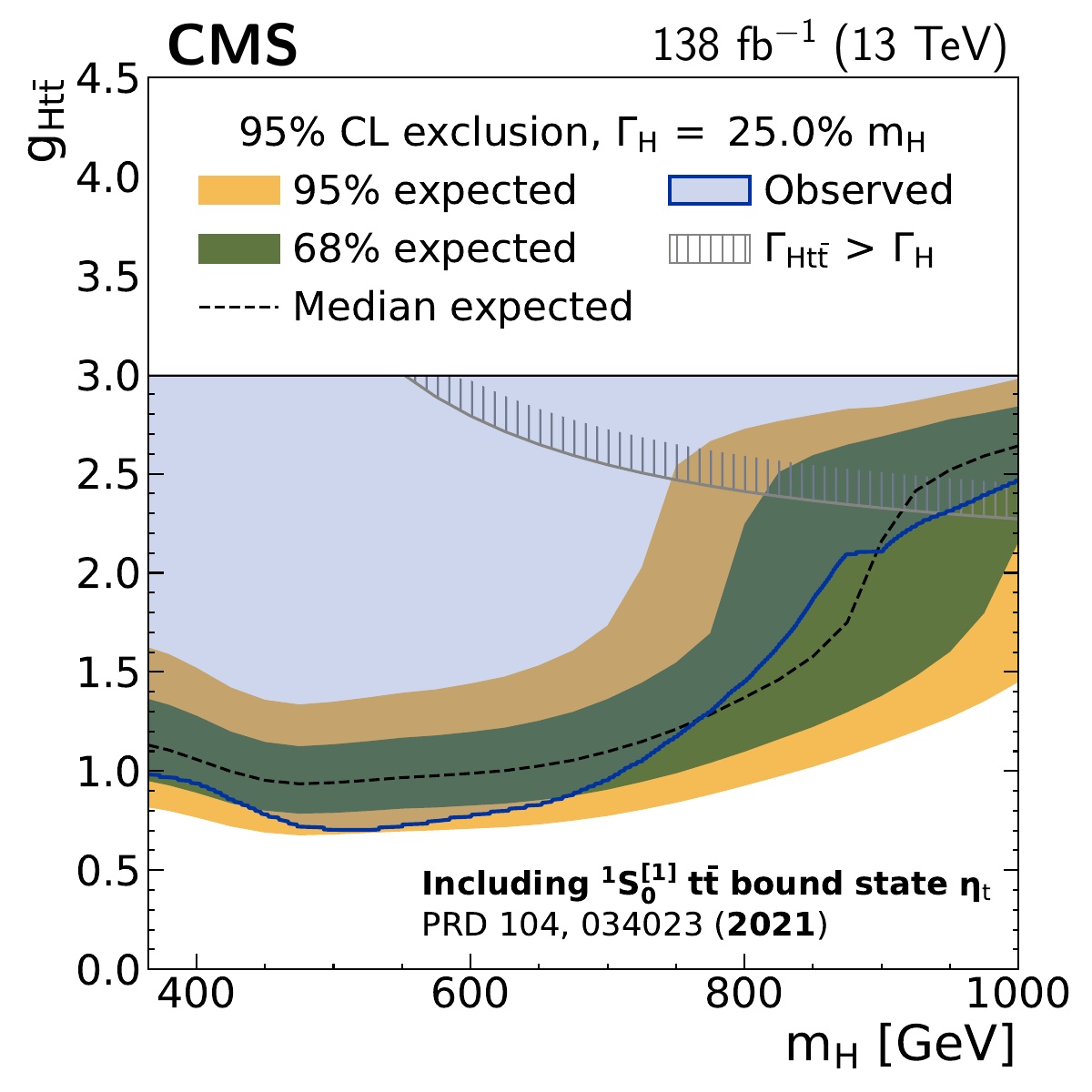}
\caption{%
    Model-independent constraints on \gphitt[\PSH] as functions of the \PSH boson mass in the background scenario with \etat contribution,
    shown in the same fashion as in Fig.~\ref{fig:limit_1D_a_smtt}.
}
\label{fig:limit_1D_h_etat}
\end{figure*}

The obtained 95\% \CL upper limits on \gphitt as a function of \mpphi are shown in Figs.~\ref{fig:limit_1D_a_etat}--\ref{fig:limit_1D_h_etat}.
The observed limits are consistent with the expected ones within two
SDs for both \CP scenarios, across all width values and the entire mass range. Notably, the excess at low masses seen in Figs.~\ref{fig:limit_1D_a_smtt}--\ref{fig:limit_1D_h_smtt}, where the background model without \etat contribution is assumed, has disappeared. This suggests that the data are well described when \etat production is included in the background model.
Moreover, a comparison of the exclusion regions in Figs.~\ref{fig:limit_1D_a_smtt}--\ref{fig:limit_1D_a_etat} at low masses indicates a slight preference for the
\etat hypothesis over the single \PSA boson production hypothesis for the lowest probed mass point at \pntA{365}{2\%}. However, the current
analysis has limited discriminatory power between these hypotheses based on their
\mtt lineshapes due to the limited experimental resolution, preventing a definitive preference for one explanation over the other.

\subsection{The \texorpdfstring{\pAandH}{A+H} boson interpretation}
\label{sec:ahetat_2D}

Many extensions of the Higgs sector, such as 2HDMs~\cite{Branco:2011iw}, predict the existence of both \PSA and \PSH bosons, with their masses and widths potentially falling within the range probed by this analysis.
To investigate this possibility, we perform a simultaneous \pAandH boson interpretation, considering various \pAorH boson pairs beyond the one analyzed in Section~\ref{sec:ahsmvseta_2D}, including the \etat contribution in the background scenario.

\begin{figure*}[!t]
\centering
\includegraphics[width=0.4\textwidth]{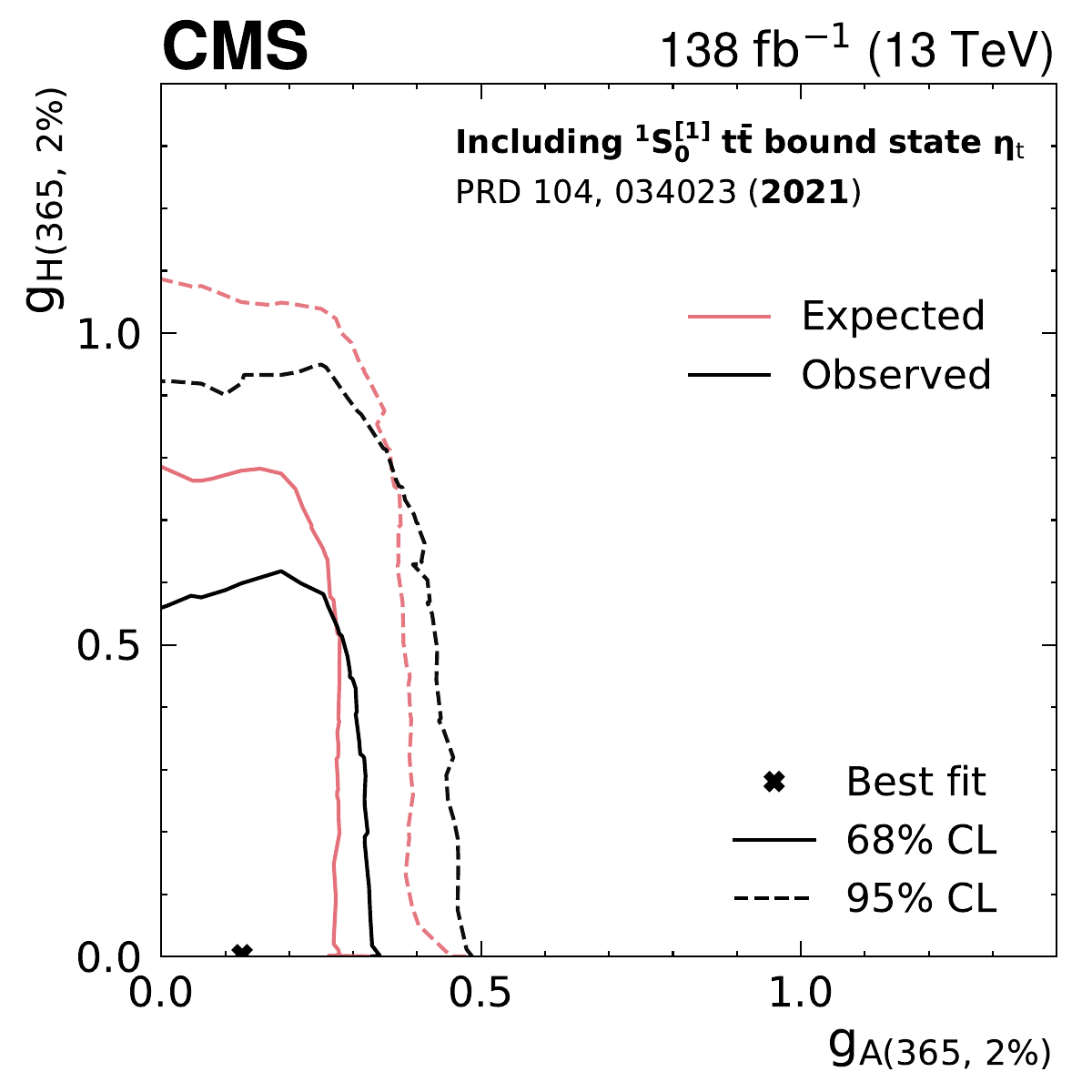}%
\hspace*{0.05\textwidth}%
\includegraphics[width=0.4\textwidth]{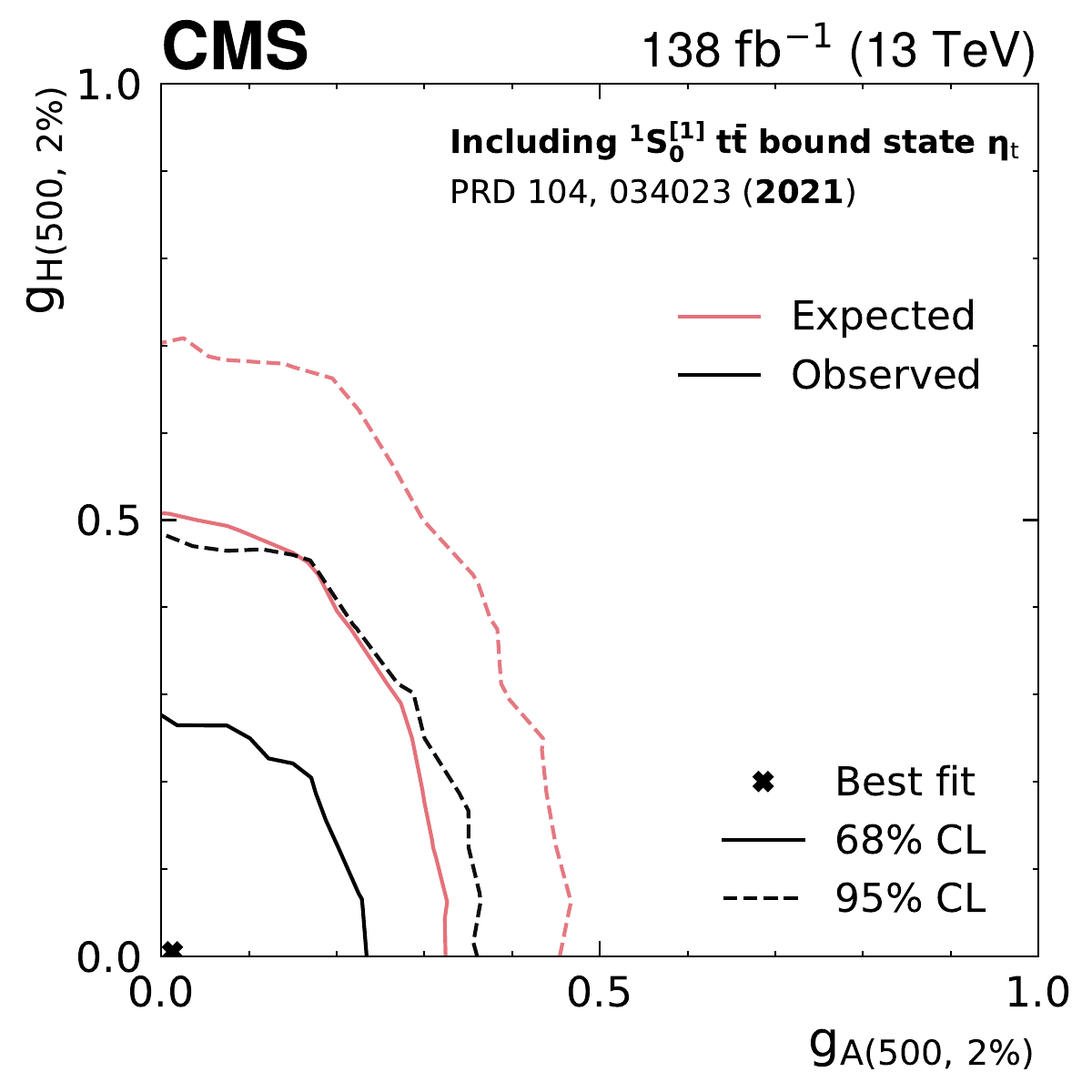} \\
\includegraphics[width=0.4\textwidth]{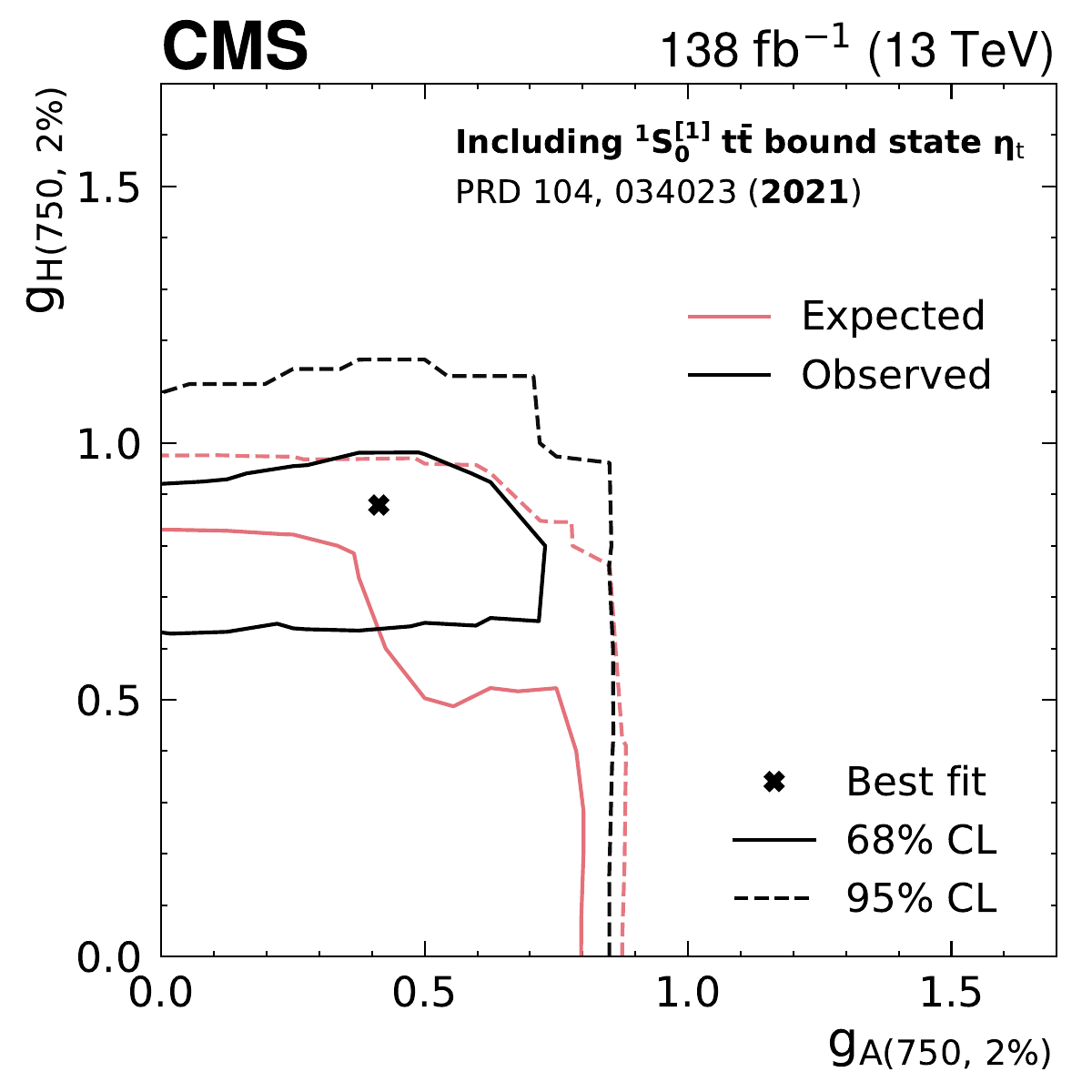}%
\hspace*{0.05\textwidth}%
\includegraphics[width=0.4\textwidth]{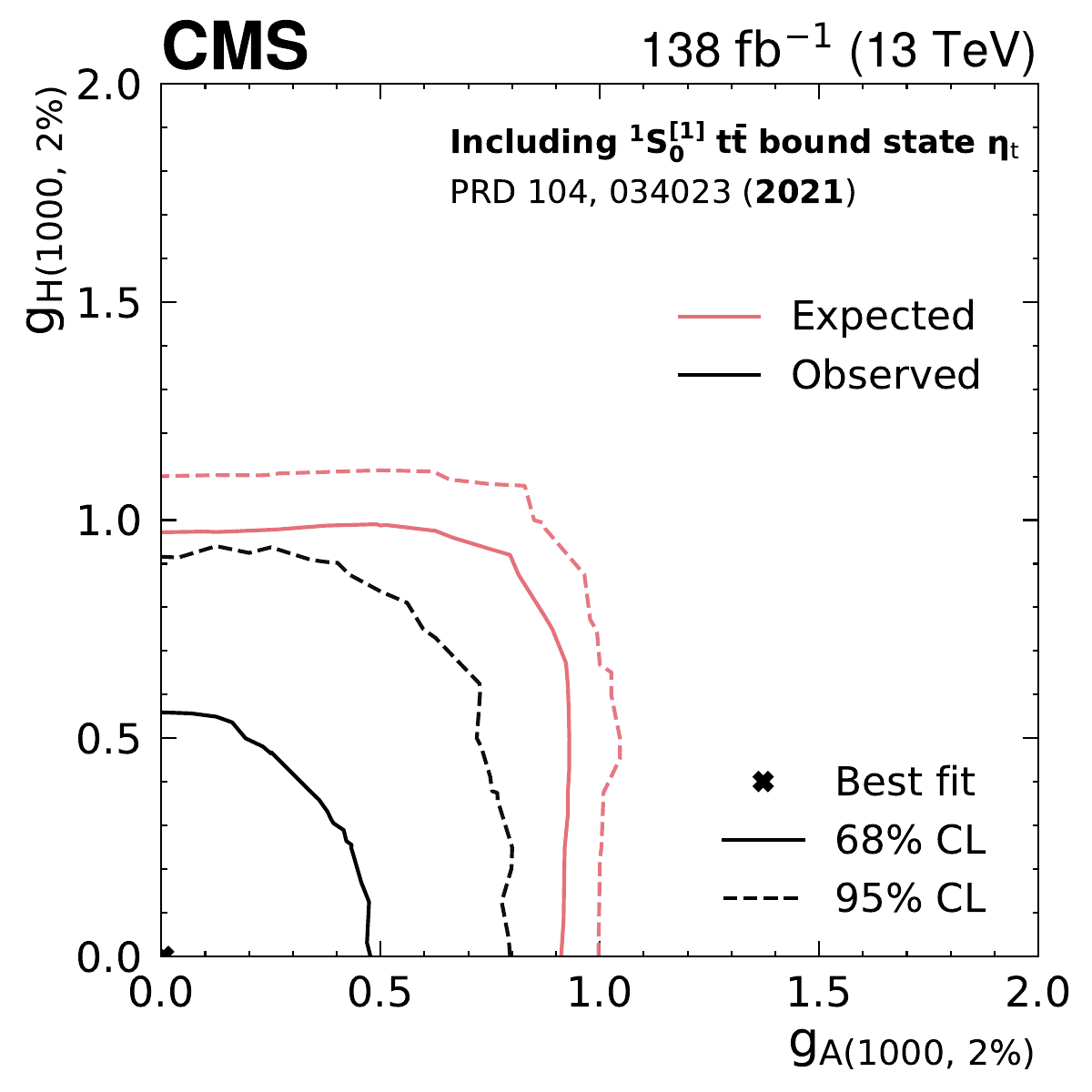}
\caption{%
    Frequentist 2D exclusion contours for \gphitt[\PSA] and \gphitt[\PSH] in the \pAandH boson interpretation for four different signal hypotheses with identical \PSA and \PSH boson masses of 365\GeV (upper left), 500\GeV (upper right), 750\GeV (lower left), and 1000\GeV (lower right), all assuming a relative width of 2\%. The expected and observed contours, evaluated with the Feldman--Cousins prescription~\cite{Feldman:1997qc,Cousins:1991qz}, are shown in pink and black, respectively, with the solid and dashed lines corresponding to exclusions at 68 and 95\% \CL. The regions outside of the contours are considered excluded. In all cases, \etat production is included in the background model.
}
\label{fig:limit_2D_ah_etat_0}
\end{figure*}

\begin{figure*}[!tp]
\centering
\includegraphics[width=0.4\textwidth]{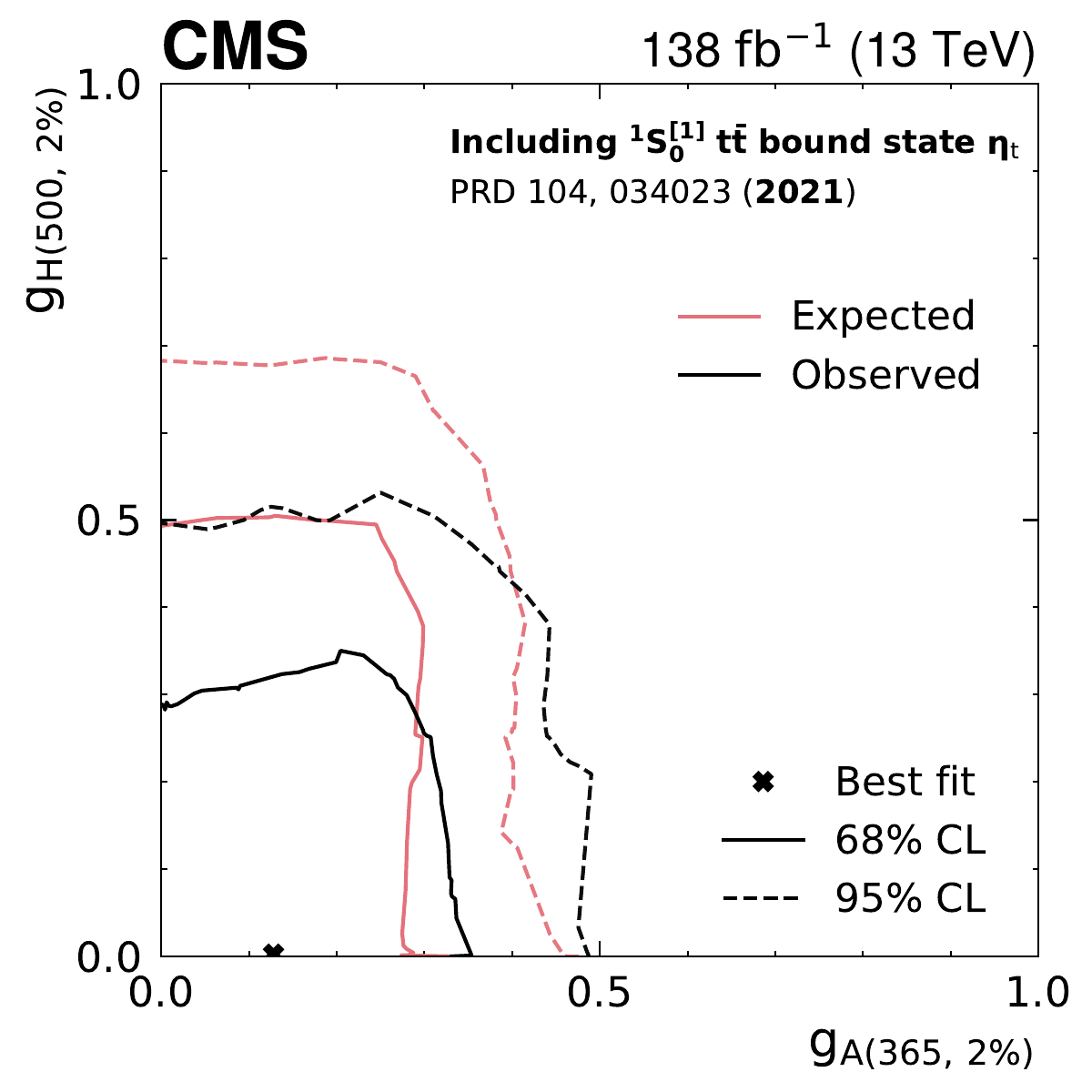}%
\hspace*{0.05\textwidth}%
\includegraphics[width=0.4\textwidth]{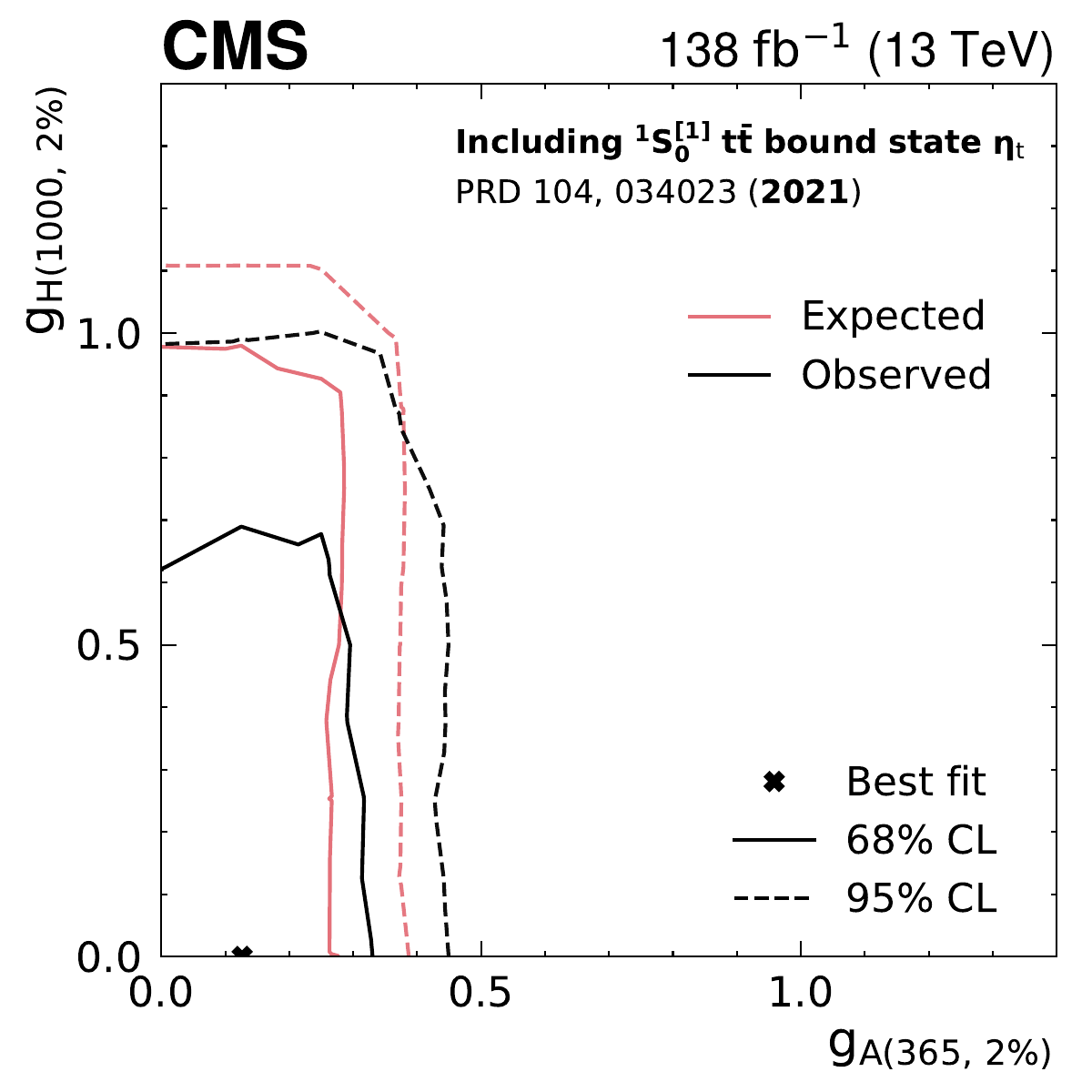} \\
\includegraphics[width=0.4\textwidth]{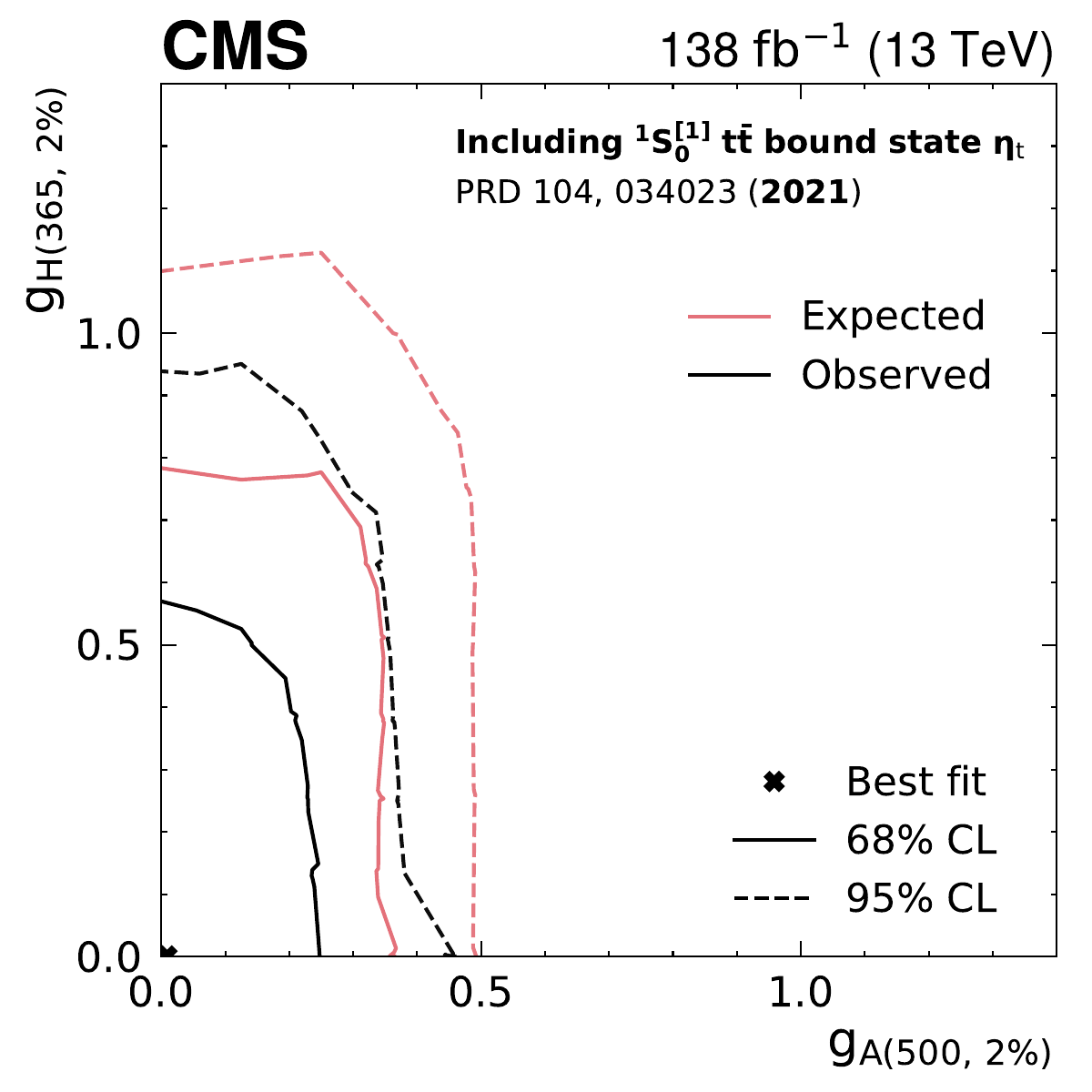}%
\hspace*{0.05\textwidth}%
\includegraphics[width=0.4\textwidth]{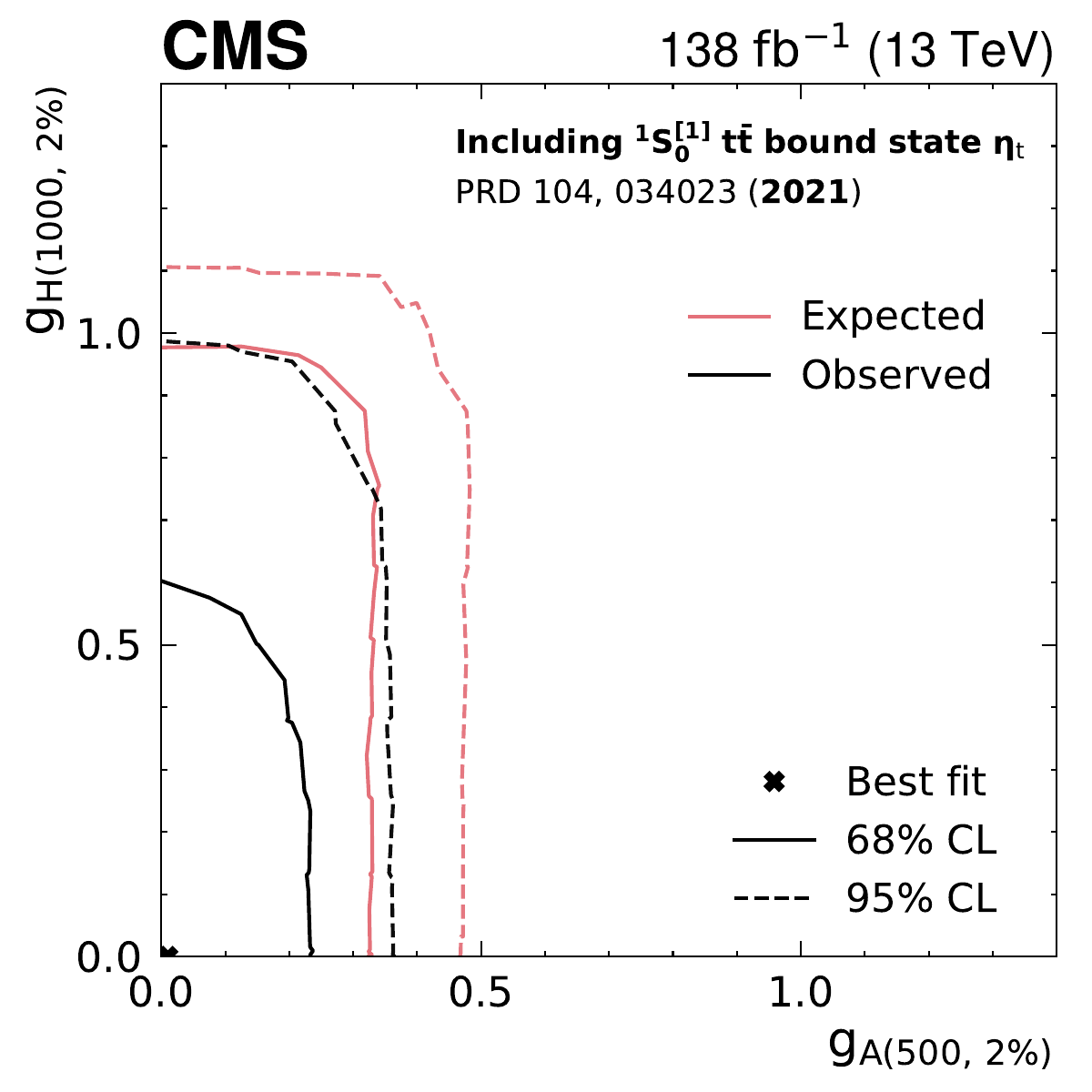} \\
\includegraphics[width=0.4\textwidth]{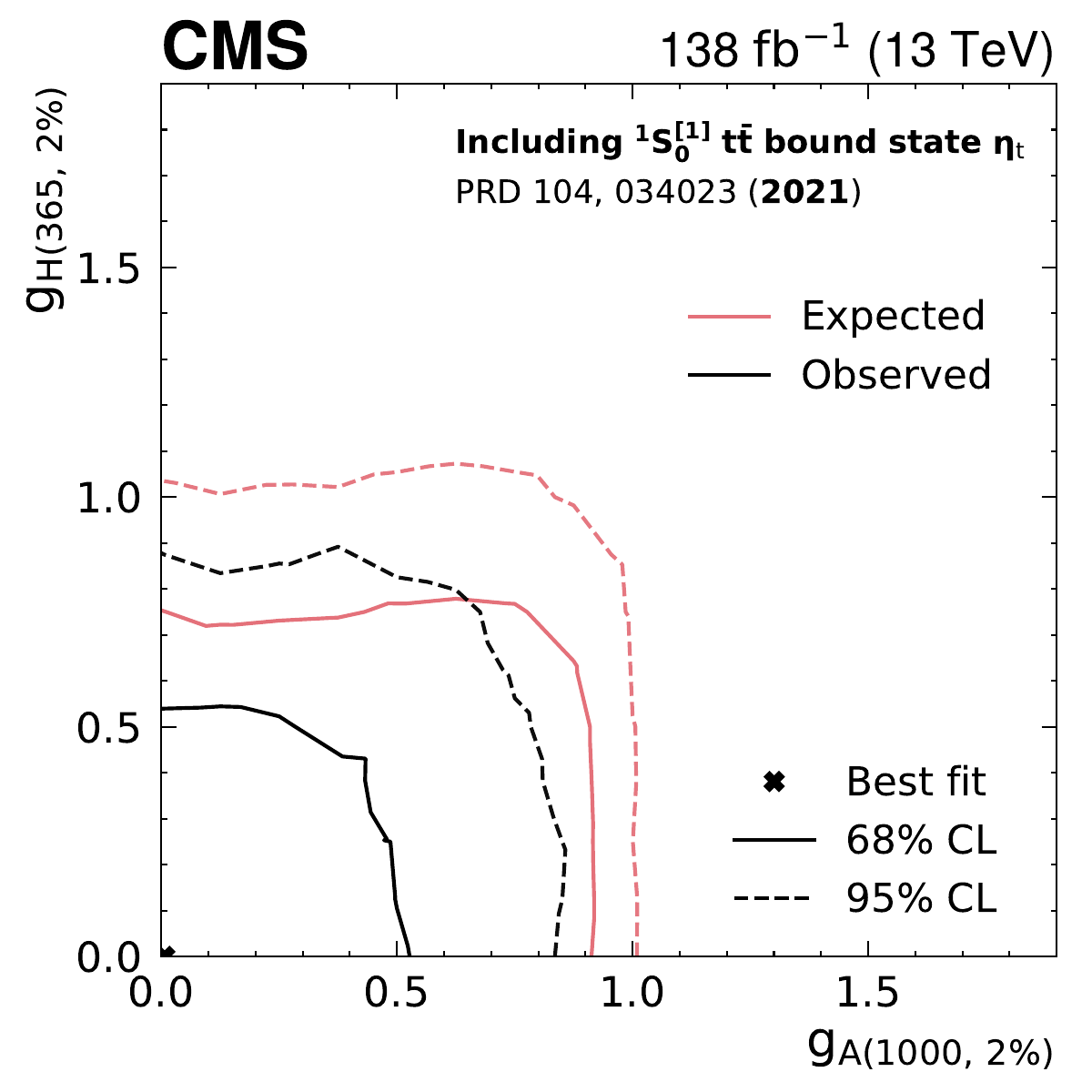}%
\hspace*{0.05\textwidth}%
\includegraphics[width=0.4\textwidth]{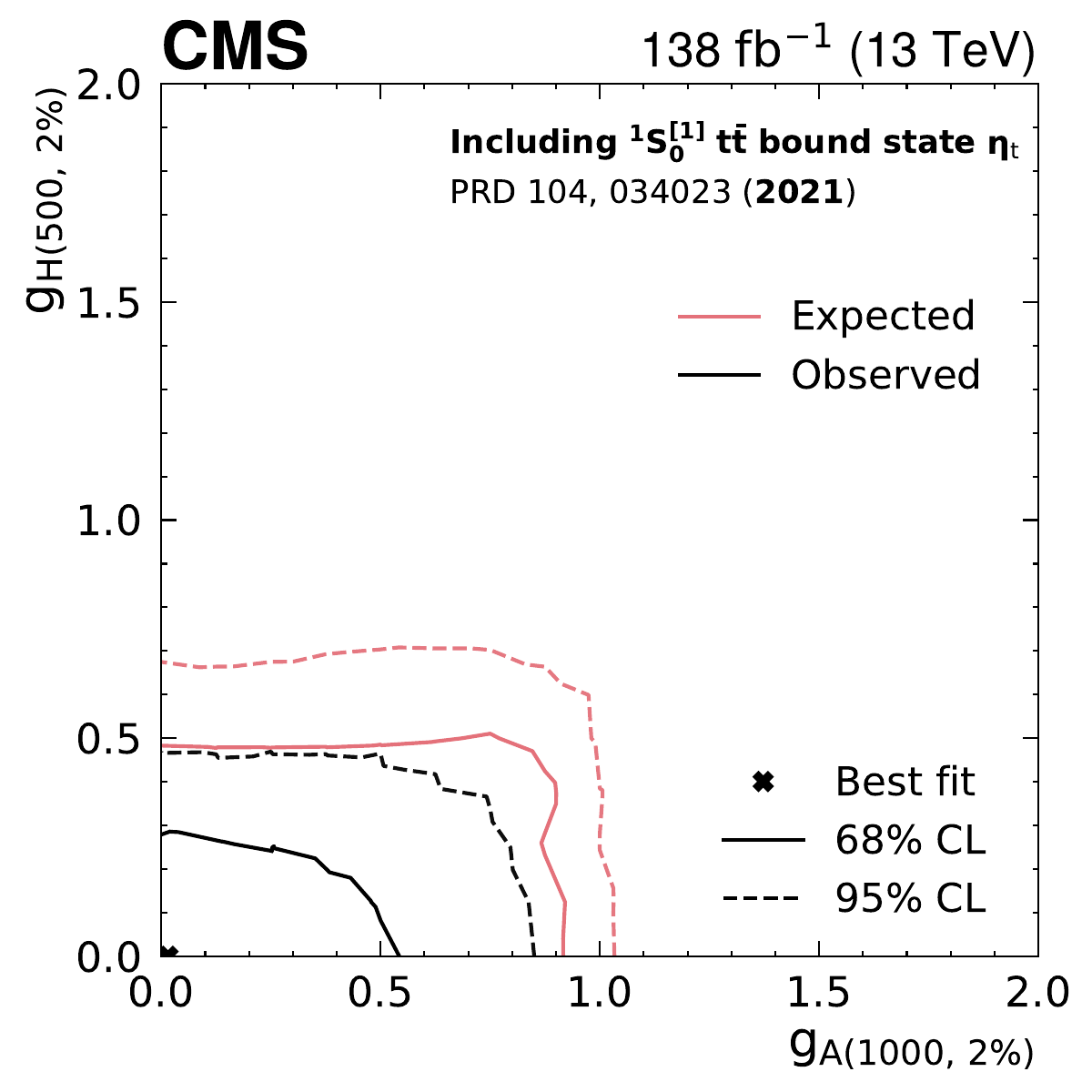}
\caption{%
    Frequentist 2D exclusion contours for \gphitt[\PSA] and \gphitt[\PSH] in the \pAandH boson interpretation for six different signal hypotheses with unequal \PSA and \PSH boson masses, corresponding to combinations of 365, 500, and 1000\GeV, all assuming a relative width of 2\%. The expected and observed contours, evaluated with the Feldman--Cousins prescription~\cite{Feldman:1997qc,Cousins:1991qz}, are shown in pink and black, respectively, with the solid and dashed lines corresponding to exclusions at 68 and 95\% \CL. The regions outside of the contours are considered excluded. In all cases, \etat production is included in the background model.
}
\label{fig:limit_2D_ah_etat_1}
\end{figure*}

The results are presented in Fig.~\ref{fig:limit_2D_ah_etat_0} for the case of identical \PSA and \PSH boson masses and in Fig.~\ref{fig:limit_2D_ah_etat_1} for differing masses, all assuming a width of 2\%.
In all cases, the observed exclusion contours are consistent with zero \pAandH boson contribution.
We note that the difference between expected and observed contours in Fig.~\ref{fig:limit_2D_ah_etat_0} (lower left) corresponds to a local tension at the level of 1--2~SDs for \mpH between 700 and 780\GeV and $\GpH/\mpH=2\%$, similar as in Fig.~\ref{fig:limit_1D_h_etat} (upper left).

\section{Summary}

A search has been presented for the production of pseudoscalar or scalar bosons in proton-proton collisions at $\sqrt{s}=13\TeV$, decaying into a top quark pair (\ttbar) in final states with one or two charged leptons.
The analysis uses data collected with the CMS detector at the LHC, corresponding to an integrated luminosity of 138\fbinv.
To discriminate the signal from the standard model \ttbar background, the search utilizes the invariant mass of the reconstructed \ttbar system along with angular observables sensitive to its spin and parity. The signal model accounts for both the resonant production of the new boson and its interference with the perturbative quantum chromodynamics (pQCD) \ttbar background.

A deviation from the background prediction, modeled using fixed-order (FO) pQCD, is observed near the \ttbar production threshold. This deviation is similar to the moderate excess previously reported by CMS using data corresponding to an integrated luminosity of 35.9\fbinv~\cite{CMS:2019pzc}. The local significance of the excess exceeds five standard deviations, with a strong preference for the pseudoscalar signal hypothesis over the scalar one.

Incorporating the production of a color-singlet \oneSone\ \ttbar quasi-bound state, \etat, within a simplified nonrelativistic QCD model, with an unconstrained normalization to the background, yields agreement with the observed data, eliminating the need for additional exotic pseudoscalar or scalar boson production.
However, the precision of the measurement is insufficient to clearly favor either the \etat  production model, or a new \PSA boson down to a mass of 365 GeV, or any potential mixture of the two.
A detailed analysis of the excess using the \ttbar quasi-bound-state interpretation is provided in Ref.~\cite{CMS:TOP-24-007}.

Exclusion limits at the 95\% confidence level are set on the coupling strength between top quarks and new bosons, covering mass ranges of 365--1000\GeV and relative widths of 0.5--25\%.
When the background model includes both FO pQCD \ttbar production and \etat production, stringent constraints are obtained for three scenarios: a new pseudoscalar boson, a new scalar boson, and the simultaneous presence of both. Coupling values as low as 0.4 (0.6) are excluded for the pseudoscalar (scalar) case. These limits are similar to the ATLAS results~\cite{ATLAS:2024vxm} in case of pseudoscalar production, and represent the most stringent limits on scalar resonances decaying into \ttbar over a wide range of mass and width values.

\bibliography{auto_generated}
\cleardoublepage \appendix\section{The CMS Collaboration \label{app:collab}}\begin{sloppypar}\hyphenpenalty=5000\widowpenalty=500\clubpenalty=5000\cmsinstitute{Yerevan Physics Institute, Yerevan, Armenia}
{\tolerance=6000
A.~Hayrapetyan, V.~Makarenko\cmsorcid{0000-0002-8406-8605}, A.~Tumasyan\cmsAuthorMark{1}\cmsorcid{0009-0000-0684-6742}
\par}
\cmsinstitute{Institut f\"{u}r Hochenergiephysik, Vienna, Austria}
{\tolerance=6000
W.~Adam\cmsorcid{0000-0001-9099-4341}, J.W.~Andrejkovic, L.~Benato\cmsorcid{0000-0001-5135-7489}, T.~Bergauer\cmsorcid{0000-0002-5786-0293}, M.~Dragicevic\cmsorcid{0000-0003-1967-6783}, C.~Giordano, P.S.~Hussain\cmsorcid{0000-0002-4825-5278}, M.~Jeitler\cmsAuthorMark{2}\cmsorcid{0000-0002-5141-9560}, N.~Krammer\cmsorcid{0000-0002-0548-0985}, A.~Li\cmsorcid{0000-0002-4547-116X}, D.~Liko\cmsorcid{0000-0002-3380-473X}, M.~Matthewman, I.~Mikulec\cmsorcid{0000-0003-0385-2746}, J.~Schieck\cmsAuthorMark{2}\cmsorcid{0000-0002-1058-8093}, R.~Sch\"{o}fbeck\cmsAuthorMark{2}\cmsorcid{0000-0002-2332-8784}, D.~Schwarz\cmsorcid{0000-0002-3821-7331}, M.~Shooshtari, M.~Sonawane\cmsorcid{0000-0003-0510-7010}, W.~Waltenberger\cmsorcid{0000-0002-6215-7228}, C.-E.~Wulz\cmsAuthorMark{2}\cmsorcid{0000-0001-9226-5812}
% ...
\par}
\vskip\cmsinstskip
\dag:~Deceased\\
$^{1}$Also at Yerevan State University, Yerevan, Armenia\\
$^{2}$Also at TU Wien, Vienna, Austria\\
% ...
\end{sloppypar}
\end{document}